\documentclass[12pt,a4paper,twoside,english]{report}
\usepackage[T1]{fontenc}
\usepackage[latin1]{inputenc}
\usepackage{babel}
\usepackage{graphics}

\makeatletter

\providecommand{\LyX}{L\kern-.1667em\lower.25em\hbox{Y}\kern-.125emX\@}

\usepackage{amsfonts}

\usepackage{fancyhdr}
\fancypagestyle{plain}{\fancyhf{} \fancyhead[LE,RO]{\thepage}}

\makeatother
\begin{document}
\noindent \begin{titlepage}

\noindent \vspace*{2,5cm}

{\noindent \centering {\LARGE The application of \( \star  \)-products
to noncommutative geometry and gauge theory}\LARGE \par}

\noindent \vspace*{2cm}

{\noindent \centering {\large Andreas Sykora}\large \par}

\noindent \vspace*{4cm}

\noindent \end{titlepage}

\renewcommand{\thepage}{\roman{page}}

\noindent \cleardoublepage

\vspace*{2,5cm}

{\noindent \centering {\LARGE The application of \( \star  \)-products
to noncommutative geometry and gauge theory}\LARGE \par}

\noindent \vspace*{2cm}

{\noindent \centering {\large Andreas Sykora}\large \par}

\noindent \vspace*{3cm}

{\noindent \centering Dissertation der Fakultät für Physik \\
der Ludwig Maximilian Universität München%
\footnote{Erstgutachter: Prof. Dr. Julius Wess

Zweitgutachter: Prof. Dr. Ivo Sachs

Tag der mündlichen Prüfung: 2.11.2004
}\par}

\noindent \vspace*{1cm}

{\noindent \centering Juni 2004 \par}

\noindent \cleardoublepage

\noindent \vspace*{0.5cm}

{\centering \textbf{\large Abstract}\large \par}

\bigskip

\noindent Due to the singularities arising in quantum field theory
and the difficulties in quantizing gravity it is often believed that
the description of spacetime by a smooth manifold should be given
up at small length scales or high energies. In this work we will replace
spacetime by noncommutative structures arising within the framework
of deformation quantization. The ordinary product between functions
will be replaced by a \( \star  \)-product, an associative product
for the space of functions on a manifold. 

We develop a formalism to realize algebras defined by relations on
function spaces. For this porpose we construct the Weyl-ordered \( \star  \)-product
and present a method how to calculate \( \star  \)-products with
the help of commuting vector fields.

Concepts developed in noncommutative differential geometry will be
applied to this type of algebras and we construct actions for noncommutative
field theories. In the classical limit these noncommutative theories
become field theories on manifolds with nonvanishing curvature. It
becomes clear that the application of \( \star  \)-products is very
fruitful to the solution of noncommutative problems. In the semiclassical
limit every \( \star  \)-product is related to a Poisson structure,
every derivation of the algebra to a vector field on the manifold.
Since in this limit many problems are reduced to a couple of differential
equations the \( \star  \)-product representation makes it possible
to construct noncommutative spaces corresponding to interesting Riemannian
manifolds.

Derivations of \( \star  \)-products makes it further possible to
extend noncommutative gauge theory in the Seiberg-Witten formalism
with covariant derivatives. The resulting noncommutative gauge fields
may be interpreted as one forms of a generalization of the exterior
algebra of a manifold. For the Formality \( \star  \)-product we
prove the existence of the abelian Seiberg-Witten map for derivations
of these \( \star  \)-products. We calculate the enveloping algebra
valued non abelian Seiberg-Witten map pertubatively up to second order
for the Weyl-ordered \( \star  \)-product. A general method to construct
actions invariant under noncommutative gauge transformations is developed.
In the commutative limit these theories are becoming gauge theories
on curved backgrounds.

We study observables of noncommutative gauge theories and extend the
concept of so called open Wilson lines to general noncommutative gauge
theories. With help of this construction we give a formula for the
inverse abelian Seiberg-Witten map on noncommutative spaces with nondegenerate
\( \star  \)-products.

\noindent \cleardoublepage

\noindent \tableofcontents{} \cleardoublepage

\textbf{\large Acknowledgements}{\large \par}

\bigskip

I am indebted to Julius Wess for letting me join his group in Munich
and drawing my attention to noncommutative gauge theory and \( \star  \)-products.
I thank Claudia Jambor and Wolfgang Behr for their fruitful collaboration
on parts of the material included in this work. I thank Fabian Bachmaier,
Christian Blohmann, Marija Dimitrijevic, Larisa Jonke, Branislav Jurco,
Florian Koch, John Madore, Frank Meyer, Dzo Mikulovic, Lutz M\"oller,
Harold Steinacker, Alexander Schmidt, Effrosini Tsouchnika, Hartmut
Wachter and Michael Wohlgenannt for many useful and inspiring discussions. 

\noindent \cleardoublepage

\chapter{Introduction}

\setcounter{page}{1}

\noindent \renewcommand{\thepage}{\arabic{page}}

All experiments in physics support the assumption that spacetime should
be described by a differential manifold and all successful theories
may be formulated as field theories on such manifolds. But in quantum
field theories there are some intrinsic difficulties at high energy
or short distances that can not be resolved. No hints are given by
experiment where and how these difficulties should be solved. But
there are other fomulations of successful theories like the algebraic
approch to quantum mechanics that leave the setting of differential
manifolds. 

In the early days of quantum field theory it was already suggested
by Heisenberg that spacetime might be modified at very short distances
by algebraic properties that could lead to uncertainty relations for
the space coordinates. The first one to write an entire article about
this subject was Snyder \cite{Snyder:1947qz}. The idea behind spacetime
noncommutativity is mainly inspired by quantum mechanics. Quantum
phase space is defined by replacing canonical variables \( q^{i},p_{j} \)
by hermitian operators which obey the Heisenberg commutation relations
\( [\hat{q}^{i},\hat{p}_{j}]=i\hbar \delta ^{i}_{j} \). Now the space
becomes smeared out and the notion of a point is replaced by a Planck
cell. In the limit \( \hbar \rightarrow 0 \) one can recover the
ordinary phase space. In its simplest form spacetime noncommutativity
can be described in the same way by replacing the commutative coordinate
functions \( x^{i} \) by operators \( \hat{x}^{i} \) of a general
algebra obeying the relations \[
[\hat{x}^{i},\hat{x}^{j}]=\hat{c}^{ij}.\]
 The righthand side of this equations should tend to zero in a certain
limit and one recovers in this way the classical commuting space.
Although this idea seemed quite promising the progress was very slow
due to the success of renormalization theory on the one hand and the
mathematical complexity of noncommutative structures on the other
hand. It took a long time until noncommutative geometry was mathematically
defined and physical models were formulated \cite{Connes:1994a,Landi:1997sh,Gracia-Bondia:2001tr,Madore:1999bi,Madore:2000aq}.

\medskip

Perhaps one reason for the slow progress is that postulating an uncertainty
relation between position measurements will lead to a nonlocal theory,
with all of the resulting difficulties. A secondary reason is that
noncommutativity of the spacetime coordinates generally conflicts
with Lorentz invariance. Although it is not implausible that a theory
defined using such coordinates could be effectively local on length
scales longer than that of the uncertainty, it is harder to believe
that the breaking of Lorentz invariance would be unobservable at these
scales.

\medskip

One big hope associated with the application of noncommutative geometry
in physics is a better description of quantized gravity. Quantum gravity
has an uncertainty principle which prevents one from measuring positions
to better accuracies than the Planck length: the momentum and energy
required to make such a measurement will itself modify the geometry
at these scales \cite{DeWitt:1962}. At least it should be possible
to construct effective actions where traces of this unknown theory
remain. If one believes that quantum gravity is in a sense a quantum
field theory, then its observables are operators on a Hilbert space
and therefore elements of an algebra. Some properties of this algebra
should be reflected in the noncommutative geometry the effective actions
are constructed on. As in this case the noncommutativity should be
induced by background gravitational fields, the classical limit of
the effective actions should reduce to actions on curved spacetimes
\cite{Madore:1997ta,Connes:1996gi}.

A related motivation is that quantum gravity might not be local in
the conventional sense. Nonlocality brings with it deep conceptual
and practical problems which have not been well understood, and one
might want to understand them in the simplest examples first, before
proceeding to a more realistic theory of quantum gravity. Further
there is an interesting similiarity in the gauge structure of general
relativity and noncommutative gauge theory. In the later gauge transformations
can be interpreted as a special subgroup of the group of diffeomorphisms.
Again with the growing understanding of noncommutative theories one
perhaps improves the knowledge about diffeomorphism invariant theories
like general relativity.

There are other reasons for introducing noncommutativity into physics.
One of the simplest is that it might improve the renormalizability
properties of a theory at short distances or even render it finite.
However it is known today that certain models develop new devergencies
absent in commutative theories \cite{Chaichian:1998kp,Matusis:2000jf}. 

\medskip

At the moment most of the applications of noncommutativity to physics
are done with noncommutative field theory \cite{Douglas:2001ba,Grosse:2000gd}.
As one thinks of these models as analogs to classical physics there
are also attempts to quantize these theories \cite{Szabo:2001kg,Chaichian:1999wy,Grosse:2001pr}.
A new approach is to treat noncommutative geometries as matrix models
and take advantage of the noncommutativity to quantize them \cite{Steinacker:2003sd,Landi:2003de}
since they have finite dimensional representations. Noncommutative
field theory is also known to appear naturally in condensed matter
physics. One example is the theory of electrons in a magnetic field
projected to the lowest Landau level, which is naturally thought of
as a noncommutative field theory. Thus these ideas are relevant to
the theory of the quantum Hall effect, and indeed, noncommutative
geometry has been found very useful in this context \cite{Bellissard:1994}.

\medskip


Symmtries have always played a very important role in physical models.
But noncommutative spaces mostly are not compatible with the symmetry
groups of their commutative counterparts. One way to circumvent these
problems are quantum groups. One does not only deform the space but
also the symmetry group acting on it. Beginning with the noncommutative
plane a large number of deformed spaces with deformed symmetries have
been constructed. Among others there are for example the \( q \)-deformed
Lie algebra of rotations \cite{Faddeev:1990ih} and \( q \)-deformed
Euclidean space \cite{Lorek:1997eh}, the \( q \)-deformed Lorentz
algebra \cite{Schmidke:1991mr} and \( q \)-deformed Minkowski space
\cite{Cerchiai:1998ee}, the \( q \)-deformed Poincare algebra \cite{Ogievetsky:1992pn},
\( \kappa  \)-deformed Poincare invariant space \cite{Majid:1994cy},
to name only a few. 

\medskip


Noncommutative geometry may be useful to describe effective field
theories derived from the low energy limit of loop quantum gravity.
Since here geometric objects are replaced by operators on a Hilbert
space \cite{Ashtekar:1998ak}, it would not be very unexpected if
noncommutative structures appeared in the continuum limit of this
theory. However the relation between loop quantum gravity and noncommutative
geometry has not been explored very well. Nevertheless there are hints
that an effective theory should be a noncomutative one. For example
there exists a nonperturbative quantization of gravity with an isolated
horizon as inner boundary within the formalism of loop quantum gravity.
The quantum geometry of the horizon looks like a noncommutative torus
\cite{Ashtekar:2000eq}.

\medskip


String theory made its first contact with noncommutative geometry
with a conjecture called M-theory. It was proposed that all known
string theories are the low energy limit of this theory. Further it
was conjectured that this M-theory may be formulated in the framework
of matrix quantum mechanics leading to the name M(atrix)-theory \cite{Konechny:2000dp}.
It was found in \cite{Connes:1998cr} that noncommutative geometry
arises very naturally in M(atrix)-theory. 

Noncommutative geometry entered string theory a second time with the
descriptions of open strings in a background \( B \)-field \cite{Chu:1998qz,Schomerus:1999ug}.
The \( D \)-brane is then a noncommutative space whose fluctuations
are governed by a noncommutative version of Yang-Mills theory \cite{Seiberg:1999vs}
and noncommutativity is induced by a so called \( \star  \)-product.
On a curved brane the \( B \)-field becomes position dependent \cite{Cornalba:2001sm}.
In the case of a constant \( B \)-field it has been shown quite soon
that there is an equivalent description in terms of ordinary gauge
theory. The two pictures are releated by a choice of regularization
\cite{Andreev:1999pv}. Therefore there must exist a field redefinition
mapping the one picture to the other, the Seiberg-Witten map \cite{Seiberg:1999vs}.

\bigskip


Most of the noncommutativity in this work will be formulated with
the help of \( \star  \)-products, i. e. with associative products
defined on function spaces. Throughout this work we will formulate
them with the help of differential operators\[
f\star g=fg+\frac{i\hbar }{2}\theta ^{ij}\partial _{i}f\, \partial _{j}g+\cdots \]
and assume that they may be expanded in some parameter of noncommutativity.
\( \star  \)-products first emerged from quantum mechanics. Due to
Weyl's quantization procedure \cite{Weyl:1927vd} one was able to
pull back noncommutativity to the classical phase space and the first
\( \star  \)-product was formulated \cite{Sternheimer:1998yg}, an
associative product between functions on phase space. In this formulation
the classical limit of quantum mechanics is very intuitive, the \( \star  \)-product
depends on \( \hbar  \) and for this parameter tending to zero it
becomes the ordinary product between functions. The Poisson bracket
can be obtained by looking at the first order deviation in \( \hbar  \).
With this in mind deformation quantization \cite{Bayen:1978a} was
formulated. One has to look for defomations of algebras of functions
of Poisson manifolds and realize quantum mechanics on this manifolds
in this way. A more abstract picture of \( \star  \)-products was
developed being now an associative product on the space of functions
on a manifold.

\medskip


The formulation of gauge theories in this work will be done with the
mentioned Seiberg-Witten map formalism emerging from string theory.
After its discovery the Seiberg-Witten map has been extensivly studied
and applied to noncommutative field theory. An interesting approch
is set within the Kontsevich \( \star  \)-product formalism \cite{Kontsevich:1997vb}.
Here the Seiberg-Witten map is found to be a part of the Formality
map \cite{Jurco:2000fb,Jurco:2000fs,Jurco:2001my,Jurco:2001kp}. In
particular these studies show that the Seiberg-Witten map is an integral
feature of any noncommutative geometry obtained through deformation
quantization of a Poisson manifold. Additionally the Seiberg-Witten
map was extended to nonabelian gauge groups. The noncommutative gauge
transformations are not longer Lie-algebra valued and have to be defined
on the enveloping algebra \cite{Jurco:2000ja}. 

On noncommutative \( \mathbb {R}^{N}_{\theta } \) which is characterized
by constant parameters \( \theta ^{ij} \) the Seiberg-Witten map
can be constructed using various techniques. The Seiberg-Witten equations
lead to a consistency condition which may be solved order by order
\cite{Jurco:2001rq}. Further it can be solved with a cohomological
approach within the BRST formalism \cite{Cerchiai:2002ss}. There
exist few Seiberg-Witten maps on other noncommutative spaces. On the
fuzzy sphere a Seiberg-Witten map was constructed up to second order
for a \( \star  \)-product that does not truncate the space of functions
and for the finite dimensional representations \( S^{2}_{N} \) \cite{Grimstrup:2003rd}.
On \( \kappa  \)-Minkowski spacetime it has been calculated in \cite{Dimitrijevic:2003wv,Dimitrijevic:2003pn}.
There are extensions of the constant case Seiberg-Witten map to supersymmetric
gauge theories \cite{Mikulovic:2003sq,Putz:2002ib}. Another remarkable
aspect of Seiberg-Witten gauge theory is that it is sensitive to the
representation of the gauge group. Due to this grand unified theories
do not have unique noncommutative analogs \cite{Aschieri:2002mc}.

\medskip

The first attempt to quantize noncommutative field theories in the
\( \star  \)-product representation was done in \cite{Filk:1996dm}.
This was done similar to the perturbative way interacting commutative
field theories are treated and Seiberg-Witten gauge theories are mostly
quantized using this method. However it is not quite clear how the
quantization of Seiberg-Witten gauge theory can be done in a consistent
way since the solution of the Seiberg-Witten equations is not unique
and other solutions are related by nonlocal field redefinitions \cite{Asakawa:1999cu}.
Nevertheless this was used in \cite{Bichl:2001cq} to show that noncommutative
abelian gauge theory on the \( \mathbb {R}^{N}_{\theta } \) in the
\( \star  \)-product representation is renormalizable. The same was
done for \( U(n) \) gauge groups up to one loop level in \cite{Buchbinder:2002fa}.

\bigskip


After this general introduction we begin to deal with \( \star  \)-products
and the representation of algebras by them. We begin with the definition
and first properties like the semiclassical limit and the equivalence
of \( \star  \)-products with respect to linear transformations on
function space. The semiclassical limit will be crucial to all applications
throughout this work. In this limit the \( \star  \)-products are
in one-to-one correspondence to Poisson structures up to the mentioned
linear transformations on function space. After that we start with
the representation of algebras defined by relations on function spaces
and calculate the Weyl-ordered \( \star  \)-product up to second
order. The Weyl-ordered \( \star  \)-product will be very important
for us to give explict formulas in noncommutative gauge theory. In
the end we give closed formulas for \( \star  \)-products for several
algebras, mainly quantum spaces like \( M(so_{q}(3)) \), \( M(so_{q}(1,3)) \)
and \( M(so_{q}(4)) \). For this we generalize the Moyal-Weyl product
with the help of commuting vector fields and give a method how to
calculate this type of \( \star  \)-representation for relation-defined
algebras. It will become clear that a big amount of the information
we have about the algebra is already contained in the Poisson structure
of the \( \star  \)-product.

\medskip

The purpose of chapter 3 is to relate noncommutative differential
geometry and \( \star  \)-product algebras. After a short introduction
to aspects of noncommutative differential geometry we need later,
we apply the \( \star  \)-product formalism to the commuting frame
formalism developed in \cite{Madore:2000aq}. We will see that in
the semiclassical limit, an algebra with nonconstant commutator and
therefore nonconstant Poisson structure will in general yield a curved
background. With the application of the commuting frame formalism
we are now able to construct noncomutative spaces with interesting
classical limit. The general considerations yield a system of partial
differential equations, which we can try to solve for certain interesting
geometries. In two dimensions we are quite successful and we are able
to construct algebras for all spaces of constant curvature. In four
dimensions this is not the case, since the mentioned system of partial
differential equations is getting more and more overdetermined in
higher dimensions. At the end of the chapter we give another very
interesting application for \( \star  \)-products in noncommutative
geometry. We calculate rotational invariant Poisson structures in
four dimensions and quantize them with the help of \( \star  \)-products.
On the resulting algebra we construct a first order differential calculus
having a frame for the Schwarzschild metric as classical limit.

\medskip

We will see in chapter 3 that derivations are very useful for formulating
noncommutative geometry on quantized Poisson manifolds. In chapter
4 therefore we make general considerations about derivations of \( \star  \)-products.
We again come to the conclusion that the important informations are
included in the semiclassical limit of the \( \star  \)-product.
Vector fields in a sense compatible with the Poisson structure of
the \( \star  \)-product and derivations are in one-to-one correspondence.
We apply our results to the Formality \( \star  \)-product and the
Weyl-ordered \( \star  \)-product from the second chapter. An alternative
definition of noncommutative forms will be later useful in combination
with Seiberg-Witten gauge theory. To make contact with physical application
we introduce traces on \( \star  \)-products at the end of the chapter.
With them we start to construct actions on noncommutative spaces having
field theories on curved backgrounds as classical limit. As an example
we give an noncommutative action being the deformation of \( \phi ^{4} \)-theory
on a space of constant curvature.

\medskip

Chapter 5 is dedicated to the application of \( \star  \)-products
to noncommutative gauge theory. We start with a introduction to noncommutative
gauge theory and the special case of Seiberg-Witten gauge theory,
a fomulation of noncommutative gauge theory only possible in the \( \star  \)-product
representation. Only in Seiberg-Witten gauge theory at the moment
it is possible to formulate noncommutative analogs to general nonabelian
gauge theories. Our main purpose in the following is the extension
of the Seiberg-Witten map to derivations of \( \star  \)-products.
Then we give a closed formula for the abelian Seiberg-Witten map for
the Fomality \( \star  \)-product. The Seiberg-Witten map for the
Weyl-ordered \( \star  \)-product is calculated up to second order.
We relate the resulting objects to the noncommutative forms introduced
in the chapter 4. Now we are able to construct actions on noncommutative
spaces invariant under noncommutative gauge transformations. The actions
have as classical limit a gauge theory on a curved background. We
give an example of a noncommutative version of electrodynamics on
a background with constant curvature. At the end we deal with observables
of noncommutative gauge theories. Most useful in this context are
the so called open Wilson lines introduced in the case of constant
commutator. We will generalize them to general \( \star  \)-product
algebras. With these observables we are able to give a formula for
the inverse abelian Seiberg-Witten map on symplectic manifolds.

\cleardoublepage

\chapter{\protect\( \star \protect \)-products}

The first \( \star  \)-product emerged from Weyl's quantization procedure
\cite{Weyl:1927vd}. Assume that \( f(q_{i},p_{j}) \) is a function
on a classical phase space and associate the following operator with
it\begin{equation}
\label{Weyl_quantisation_formula}
\hat{f}=\Omega (f)=\int d^{n}\xi \, d^{n}\eta \, \tilde{f}(\xi ,\eta )\, e^{\frac{i}{\hbar }(\hat{q}\cdot \xi +\hat{p}\cdot \eta )}.
\end{equation}
Here \( \tilde{f}(\xi ,\eta ) \) is the Fourier transform of \( f \),
the operators \( \hat{q}_{i} \) and \( \hat{p}_{j} \) should fulfill
the canonical commutation relations \( [\hat{q}_{i},\hat{p}_{j}]=i\hbar \delta _{ij} \)
. In this case it is possible to give an inverse operation\[
\Omega ^{-1}(\hat{f})=\int d^{n}\xi \, d^{n}\eta \, Tr\left( \hat{f}\, e^{-\frac{i}{\hbar }(\hat{q}\cdot \xi +\hat{p}\cdot \eta )}\right) \, e^{i(q\cdot \xi +p\cdot \eta )}.\]
 Here \( Tr \) is the trace on the Fock space representation of the
operator algebra. Now one can pull back the product between two operators
to a product between two functions on phase space\[
f\star _{M}g=\Omega ^{-1}(\Omega (f)\Omega (g)),\]
which yields the Moyal product on classical phase space. If \( P^{IJ}\partial _{I}\wedge \partial _{J} \)
(\( \partial _{I}=(\partial _{q_{i}},\partial _{p_{j}}) \) ) is the
Poisson structure of the classical phase space, i. e. \( \{f,g\}=P^{IJ}\partial _{I}f\partial _{J}g \)
and \( \{q_{i},p_{j}\}=\delta _{ij} \), it is possible to write down
an explicit formula\begin{equation}
\label{Moyal_product}
f\star _{M}g=\sum ^{\infty }_{n=0}\frac{(i\hbar )^{n}}{2^{n}n!}P^{I_{1}J_{1}}\cdots P^{I_{n}J_{n}}\partial _{I_{1}}\cdots \partial _{I_{n}}f\, \partial _{J_{1}}\cdots \partial _{J_{n}}g
\end{equation}
for the Moyal product. A good introduction to this is \cite{Sternheimer:1998yg}
and references therein. The generalization of the above construction
is called deformation quantization \cite{Bayen:1978a,Bayen:1978hb},
where one tries to quantize phase spaces by finding appropriate \( \star  \)-products
for functions on phase space.

\section{Definition and first properties}

The Moyal product (\ref{Moyal_product}) is a special case of a \( \star  \)-product.
To define more general \( \star  \)-products let \( M \) be an arbitrary
(sufficiently smooth) finite dimensional manifold. A \( \star  \)-product
on \( M \) is an associative, \( \mathbb {C} \)-linear product on
the space of functions (with values in \( \mathbb {C} \)) on \( M \)
given by \[
f\star g=fg+\frac{h}{2}B_{1}(f,g)+(\frac{h}{2})^{2}B_{2}(f,g)+\cdots \]
where \( f \) an \( g \) are two such functions and the \( B_{i} \)
are bidifferential operators on \( M \). The parameter \( h \) is
called deformation parameter. There is a natural gauge group acting
on \( \star  \)-products consisting of \( \mathbb {C} \)-linear
transformations on the space of functions\[
f\rightarrow f+hD_{1}(f)+h^{2}D_{2}(f)+\cdots \]
where the \( D_{i} \) are differential operators. They may be interpreted
as a generalization of coordinate transformations. If \( D \) is
such a linear transformation it maps a \( \star  \)-product to a
new \( \star  \)-product \begin{equation}
\label{linear_transformation_on_star_product}
f\star ^{\prime }g=D^{-1}(D(f)\star D(g)).
\end{equation}
If one expands this equation in \( h \) one sees that a linear transformation
\( D \) acting on \( \star  \) only affects the symmetric part of
\( B_{1} \)\[
B^{\prime }_{1}(f,g)=B_{1}(f,g)+D_{1}(fg)-fD_{1}(g)-D_{1}(f)g\]
 and one can show that the symmetric part of \( B_{1} \) may be canceled
by a linear transformation. For this we may assume \( B_{1} \) to
be antisymmetric. Since \( \star  \) is associative, the commutator\[
[f\stackrel{\star }{,}g]=f\star g-g\star f=hB_{1}(f,g)+\cdots \]
has to be a derivation\[
[f\star g\stackrel{\star }{,}h]=f\star [g\stackrel{\star }{,}h]+[f\stackrel{\star }{,}h]\star g.\]
Up to first order this means that the antisymmetric part of \( B_{1} \)
is a derivation with respect to both functions \( f \) and \( g \).
Additionally the Jakobi-identity is fulfilled for the commutator\[
[f\stackrel{\star }{,}[g\stackrel{\star }{,}h]]+[h\stackrel{\star }{,}[f\stackrel{\star }{,}g]]+[g\stackrel{\star }{,}[h\stackrel{\star }{,}f]]=0.\]
Up to second order this implies that \( B_{1} \) is a Poisson structure\[
\{f,\{g,h\}\}+\{h,\{f,g\}\}+\{g,\{h,f\}\}=0\]
where \( \{f,g\}=B_{1}(f,g) \). Therefore after a certain linear
transformation we can always write on a local patch of the manifold
(here locally \( \{f,g\}=\Pi ^{ij}\partial _{i}f\partial _{j}g \))
\[
f\star g=fg+\frac{ih}{2}\Pi ^{ij}\partial _{i}f\, \partial _{j}g+\cdots \]
 with\begin{equation}
\label{Poisson_condition_local}
\Pi ^{il}\partial _{l}\Pi ^{jk}+\Pi ^{kl}\partial _{l}\Pi ^{ij}+\Pi ^{jl}\partial _{l}\Pi ^{ki}=0
\end{equation}
 \bigskip

We have seen that \( \star  \)-products up to second order are classified
by Poisson structures on the manifold. On the other hand, if there
is a manifold with a given Poisson structure \( \{,\} \) on it, it
is possible to construct \( \star  \)-products with\[
f\star g=fg+\frac{ih}{2}\{f,g\}+\cdots .\]
 This was first done for symplectic manifolds (manifolds with invertible
\( \Pi ^{ij} \)) in \cite{DeWilde:1983b,Fedosov:1994a}. In \cite{Kontsevich:1997vb}
a general construction for arbitrary Poisson manifolds has been given
(see also \cite{Cattaneo:1999fm}). It makes use of the so called
formality map that we will use later for constructing noncommutative
gauge theories, too.

\section{Algebras and \protect\( \star \protect \)-products}

Suppose we are taking \( \mathbb {R}^{N} \) as the manifold and parametrize
it by \( N \) coordinates \( x^{i} \). Then \( \theta ^{ij}=const.\, (i,j=1\cdots N) \)
clearly fulfills the Poisson condition (\ref{Poisson_condition_local}).
With a view to the original Moyal product (\ref{Moyal_product}) we
can write down a \( \star  \)-product for this Poisson structure\begin{equation}
\label{definiton_moyal_weyl_star_product}
f\star g=\sum ^{\infty }_{n=0}\frac{(ih)^{n}}{2^{n}n!}\theta ^{i_{1}j_{1}}\cdots \theta ^{i_{n}j_{n}}\partial _{i_{1}}\cdots \partial _{i_{n}}f\, \partial _{j_{1}}\cdots \partial _{j_{n}}g
\end{equation}
where \( f \) and \( g \) are functions on \( \mathbb {R}^{N} \).
We will again call it Moyal-Weyl \( \star  \)-product. A proof that
it is really associative will be given in (\ref{Associativity_of_vector_star}).
Since \[
[x^{i}\stackrel{\star }{,}x^{j}]=ih\theta ^{ij},\]
the space of functions on \( \mathbb {R}^{N} \) together with the
\( \star  \)-product forms a realization of the algebra\[
\mathcal{A}=\mathbb {C}<\hat{x}^{1},\cdots ,\hat{x}^{N}>/([\hat{x}^{i},\hat{x}^{j}]-ih\theta ^{ij}).\]
In opposite to the representation on a Hilbert space we will call
it a \( \star  \)-product representation. Now the question arises
if we can do the same with other relation-defined algebras. We will
see that this is possible if we invent an ordering description. Other
possibilities for relations are Lie algebra structures with\[
[\hat{x}^{i},\hat{x}^{j}]=ihC^{ij}{}_{k}\hat{x}^{k},\hspace {1cm}h,C^{ij}{}_{k}\in \mathbb {C}\]
 and quantum space structures \cite{Wess:1999,Cerchiai:1998eg,Wess:1991vh,Schmidke:1991mr}
with\[
\hat{x}^{i}\hat{x}^{j}=qR^{ij}{}_{kl}\hat{x}^{k}\hat{x}^{l},\hspace {1cm}q=e^{h},R^{ij}{}_{kl}\in \mathbb {C}\]

Instead of considerering these special relations we will in the following
discuss a more general case. We assume that the algebra \( \mathcal{A} \)
is generated by \( N \) elements \( \hat{x}^{i} \) and relations

\[
[\hat{x}^{i},\hat{x}^{j}]=\hat{c}^{ij}(\hat{x})=ih\widetilde{\hat{c}}^{ij}(\hat{x})\]
where we assume that the right hand side of this formula is containing
a parameter \( h \), and is becoming in a sense small if this parameter
approches zero. Mathematically more correct we have to use a \( h \)-adic
expanded algebra \begin{equation}
\label{hadic_expanded_algabra}
\mathcal{A}=\mathbb {C}<\hat{x}^{1},\cdots ,\hat{x}^{N}>[[h]]/([\hat{x}^{i},\hat{x}^{j}]-ih\widetilde{\hat{c}}^{ij}(\hat{x}))
\end{equation}
where it is possible to work with formal power series in \( h \).
Note that this kind of algebras all have the Poincare-Birkhoff-Witt
property since a reordering of two \( \hat{x}^{i} \) never affects
the polynomials of same order in \( h \). An algebra with Poicare-Birkhoff-Witt
property possesses a basis of lexicographically ordered monomials.
For an algebra generated by two elements \( \hat{x} \) and \( \hat{y} \)
this means that the monomials \( \hat{x}^{n}\hat{y}^{m} \) constitute
a basis. For example the algebra defined by the relations \( [\hat{x},\hat{y}]=\hat{x}^{2}+\hat{y}^{2} \)
does not have this property. An example of an algebra that is not
included in the above three cases but fulfilling the property (\ref{hadic_expanded_algabra})
is given later in (\ref{rotational_invariant_algebra}).

\subsection{Algebra generator orderings}

Note that Weyl's quantization procedure (\ref{Weyl_quantisation_formula})
does not make reference to any algebra relation. So let us calculate
what Weyl is doing on an algebraic level with this formula. For this
let \[
f(p)=\int d^{n}x\, f(x)e^{ip_{i}x^{i}}\]
 be the Fourier transform of \( f \). Formally we get for a monomial
in \( \mathbb {R}^{N} \) \begin{eqnarray*}
\int d^{n}x\, x^{1}\cdots x^{m}\, e^{ip_{i}x^{i}} & =(-i\partial _{p_{i_{1}}})\cdots (-i\partial _{p_{i_{m}}})\delta (p).
\end{eqnarray*}
The Weyl operator associated to a function \( f \) is defined by\begin{equation}
\label{weyl_ordering_operator}
W(f)=\int \frac{d^{n}p}{(2\pi )^{n}}f(p)e^{-ip_{i}\hat{x}^{i}}
\end{equation}
 (see e. g. \cite{Madore:2000en}). For a monomial we get \begin{eqnarray*}
W(x^{i_{1}}\cdots x^{i_{m}}) & = & \frac{1}{m!}\partial _{p_{i_{1}}}\cdots \partial _{p_{i_{m}}}(p_{i}\hat{x}^{i})^{m}
\end{eqnarray*}
and therefore the Weyl operator really maps monomials to the corresponding
symmetrical ordered polynomial in the algebra, e. g. for three generators\[
W(x^{i}x^{j}x^{k})=\frac{1}{3!}(\hat{x}^{i}\hat{x}^{j}\hat{x}^{k}+\hat{x}^{i}\hat{x}^{k}\hat{x}^{j}+\hat{x}^{k}\hat{x}^{i}\hat{x}^{j}+\hat{x}^{j}\hat{x}^{i}\hat{x}^{k}+\hat{x}^{j}\hat{x}^{k}\hat{x}^{i}+\hat{x}^{k}\hat{x}^{j}\hat{x}^{i}).\]
A similar calculation may be done for normal ordering with the result
that\[
N(f)=\int \frac{d^{n}p}{(2\pi )^{n}}f(p)e^{-ip_{1}\hat{x}^{1}}\cdots e^{-ip_{n}\hat{x}^{n}.}\]
 In the end we see that for calculating a \( \star  \)-product like
Weyl we need a ordering description \( \Omega  \) that maps monomials
in the coordinates \( x^{i} \) to polynomials in the algebra generators
\( \hat{x}^{i} \). Then the \( \star  \)-product is defined by\begin{equation}
\label{definition_ordered_star_product}
\Omega (f\star _{\Omega }g)=\Omega (f)\Omega (g)
\end{equation}
 for two functions \( f \) and \( g \). If we have used another
ordering description \( \Omega ^{\prime } \), the resulting \( \star  \)-product
is gauge equivalent to this \( \star  \)-product by the linear transformation\[
D=\Omega ^{-1}\Omega ^{\prime }\]
since\[
f\star _{\Omega ^{\prime }}g=D^{-1}(D(f)\star _{\Omega }D(g)).\]
The choice of different ordering descriptions is equivalent to taking
a different gauge of \( \star  \)-product. 

For calculating the \( \star  \)-product in the constant and Lie
algebra case with the Weyl-ordering operator see e. g. \cite{Madore:2000en}.
There a normal ordered \( \star  \)-product is calculated for the
Manin plane, too. For the influence of ordering descriptions in the
constant case see e. g. \cite{Meyer:2003nu}. Many examples of \( \star  \)-products
for algebras and corresponding ordering descriptions are given in
\cite{Schraml:2000}.

\subsection{\label{wely_star_construction}Weyl-ordered \protect\( \star \protect \)-products}

In this section we will calculate a \( \star  \)-product generated
by symmetric ordering (\ref{weyl_ordering_operator}) of the generators
of the algebra (\ref{hadic_expanded_algabra}), the Weyl-ordered \( \star  \)-product
\cite{Behr:2003qc}. The algebra of functions equipped with the Weyl-ordered
\( \star  \)-product is isomorphic by construction to the noncommutative
algebra it is based on. 

\bigskip

With look to (\ref{definition_ordered_star_product}) we start with\[
f\star g=\int \frac{d^{n}k}{(2\pi )^{n}}\int \frac{d^{n}p}{(2\pi )^{n}}f(k)g(p)\, W^{-1}(e^{-ik_{i}\hat{x}^{i}}e^{-ip_{i}\hat{x}^{i}})\]
where we have used\[
W(e^{ik_{i}x^{i}})=e^{ik_{i}\hat{x}^{i}}.\]
We are therefore able to write down the \( \star  \)-product of the
two functions if we know the form of the last expression. For this
we expand it in terms of commutators. We use\[
e^{\hat{A}}e^{\hat{B}}=e^{\hat{A}+\hat{B}}R(\hat{A},\hat{B})\]
with

\begin{eqnarray*}
R(\hat{A},\hat{B}) & = & 1+\frac{1}{2}[\hat{A},\hat{B}]\nonumber \\
 & - & \frac{1}{6}[\hat{A}+2\hat{B},[\hat{A},\hat{B}]]+\frac{1}{8}[\hat{A},\hat{B}][\hat{A},\hat{B}]+\mathcal{O}(3).
\end{eqnarray*}
If we set \( \hat{A}=-ik_{i}\hat{x}^{i} \) and \( \hat{B}=-ip_{i}\hat{x}^{i} \)
the above mentioned expression becomes\[
W^{-1}(e^{-ik_{i}\hat{x}^{i}}e^{-ip_{i}\hat{x}^{i}})=\]
 \begin{eqnarray*}
 &  & e^{-i(k_{i}+p_{i})x^{i}}+\frac{1}{2}(-ik_{i})(-ip_{j})W^{-1}(e^{-i(k_{i}+p_{i})\hat{x}^{i}}[\hat{x}^{i},\hat{x}^{j}])\nonumber \\
 &  & -\frac{1}{6}(-i)(k_{m}+2p_{m})(-ik_{i})(-ip_{j})W^{-1}(e^{-i(k_{i}+p_{i})\hat{x}^{i}}[[\hat{x}^{m},[\hat{x}^{i},\hat{x}^{j}]])\\
 &  & +\frac{1}{8}(-ik_{m})(-ip_{n})(-ik_{i})(-ip_{j})W^{-1}(e^{-i(k_{i}+p_{i})\hat{x}^{i}}[\hat{x}^{m},\hat{x}^{n}][\hat{x}^{i},\hat{x}^{j}])\nonumber \\
 &  & +\mathcal{O}(3).\nonumber 
\end{eqnarray*}
If we assume that the commutators of the generators are written in
Weyl ordered form\[
\hat{c}^{ij}=W(c^{ij}),\]
we see that\[
[\hat{x}^{m},[\hat{x}^{i},\hat{x}^{j}]]=W(c^{ml}\partial _{l}c^{ij})+\mathcal{O}(3),\]
\[
[\hat{x}^{m},\hat{x}^{n}][\hat{x}^{i},\hat{x}^{j}]=W(c^{mn}c^{ij})+\mathcal{O}(3).\]
Further we can derive\begin{eqnarray*}
W^{-1}(e^{iq_{i}\hat{x}^{i}}W(f)) & = & W^{-1}\left( \int \frac{d^{n}p}{(2\pi )^{n}}f(p)e^{-i(q_{i}+p_{i})\hat{x}^{i}}R(-iq_{i}\hat{x}^{i},-ip_{i}\hat{x}^{i})\right) \nonumber \\
 & = & e^{-iq_{i}x^{i}}\left( f+\frac{1}{2}(-iq_{i})c^{ij}\partial _{j}f\right) +\mathcal{O}(2).
\end{eqnarray*}
Putting all this together yields

\begin{eqnarray*}
W^{-1}(e^{-ik_{i}\hat{x}^{i}}e^{-ip_{i}\hat{x}^{i}}) & = & e^{-i(k_{i}+p_{i})x^{i}}\left( 1+\frac{1}{2}c^{ij}(-ik_{i})(-ip_{j})\right. \nonumber \\
 & + & \frac{1}{8}c^{mn}c^{ij}(-ik_{m})(-ip_{n})(-ik_{i})(-ip_{j})\\
 & + & \left. \frac{1}{12}c^{ml}\partial _{l}c^{ij}(-i)(k_{m}-p_{m})(-ik_{i})(-ip_{j})\right) \nonumber \\
 & + & \mathcal{O}(3),\nonumber 
\end{eqnarray*}
and we can write down the Weyl ordered \( \star  \)-product up to
second order for an arbitrary algebra\begin{eqnarray}
f\star g & = & fg+\frac{1}{2}c^{ij}\partial _{i}f\partial _{j}g\nonumber \\
 & + & \frac{1}{8}c^{mn}c^{ij}\partial _{m}\partial _{i}f\partial _{n}\partial _{j}g\label{weyl_ordered_star_product} \\
 & + & \frac{1}{12}c^{ml}\partial _{l}c^{ij}(\partial _{m}\partial _{i}f\partial _{j}g-\partial _{i}f\partial _{m}\partial _{j}g)+\mathcal{O}(3).\nonumber 
\end{eqnarray}

\bigskip

Let us collect some properties of the just calculated \( \star  \)-product.
First\[
[x^{i}\stackrel{\star }{,}x^{j}]=c^{ij}\]
is the Weyl ordered commutator of the algebra. Further, if there is
a conjugation on the algebra and if we assume that the noncommutative
coordinates are real \( \overline{\hat{x}^{i}}=\hat{x}^{i} \), then
the Weyl ordered monomials are real, too. This is also true for the
monomials of the commutative coordinate functions. Therefore this
\( \star  \)-product respects the ordinary complex conjugation\[
\overline{f\star g}=\overline{g}\star \overline{f}.\]
On the level of the Poisson tensor this means\[
\overline{c^{ij}}=-c^{ij}.\]

\bigskip

It is very instructive to calculate the action of a linear transformation
(\ref{linear_transformation_on_star_product}) \begin{eqnarray*}
\Omega  & = & e^{\Omega ^{m}\partial _{n}+\Omega ^{mn}\partial _{m}\partial _{n}+\cdots }\\
 & = & 1+(\Omega ^{m}+\frac{1}{2}\Omega ^{n}\partial _{n}\Omega ^{m})\partial _{m}+(\Omega ^{mn}+\frac{1}{2}\Omega ^{m}\Omega ^{n})\partial _{m}\partial _{n}\\
 &  & +\cdots \\
\Omega ^{-1} & = & e^{-\Omega ^{m}\partial _{n}-\Omega ^{mn}\partial _{m}\partial _{n}-\cdots }\\
 & = & 1-(\Omega ^{m}-\frac{1}{2}\Omega ^{n}\partial _{n}\Omega ^{m})\partial _{m}-(\Omega ^{mn}-\frac{1}{2}\Omega ^{m}\Omega ^{n})\partial _{m}\partial _{n}\\
 &  & +\cdots 
\end{eqnarray*}
on the Weyl ordered \( \star  \)-product. We find\begin{eqnarray*}
\Omega ^{-1}(\Omega (f)\star \Omega (g)) & = & f\star g\\
 &  & +\frac{1}{2}(c^{in}\partial _{n}\Omega ^{j}-c^{jn}\partial _{n}\Omega ^{i}-\Omega ^{n}\partial _{n}c^{ij})\, \partial _{i}f\, \partial _{j}g\\
 &  & -2\Omega ^{ij}\, \partial _{i}f\, \partial _{j}g\\
 &  & +\cdots .
\end{eqnarray*}
The first deviation is the Lie derivative of the vector field \( \Omega ^{i}\partial _{i} \)
for \( c^{ij} \). Later we will compare the Weyl ordered \( \star  \)-product
to another one and give in this case an explicit formula for the transformation
\( \Omega  \).

\subsection{\label{example_so_a(n)}Example: \protect\( M(so_{a}(n))\protect \) }

Here we will start the example of a quantum space introduced in \cite{Majid:1994cy}.
Although this quantum space is covariant under the quantum group \( SO_{a}(n) \),
we will never use this property. We have taken it because of its simple
relations. Further it has a nontrivial center and there exist outer
derivations that will below serve as a useful example. 

Since we are using the \( n \)-dimensional generalisation introduced
in \cite{Dimitrijevic:2003pn,Dimitrijevic:2003wv} we will simply
call it \( SO_{a}(n) \) covariant quantum space or abreviated \( M(so_{a}(n)) \).
The relations of this quantum space are\[
[\widehat{x}^{0},\widehat{x}^{i}]=ia\widehat{x}^{i}\; \; \; \mbox {for}\; \; \; i\neq 0,\]
 with \( a \) a real number. The \( \hat{x}^{i} \) simply commute
with each other. In the following of the example Greek indices will
run from \( 0 \) to \( n-1 \), whereas Latin indices will run from
\( 1 \) to \( n-1 \). It is easy to see that the Poisson tensor
corresponding to the algebra is\[
c^{\mu \nu }=iax^{i}(\delta ^{\mu }_{0}\delta _{i}^{\nu }-\delta ^{\nu }_{0}\delta _{i}^{\mu }).\]
Since we are dealing here with the case of a Lie algebra we surely
have \( W(c^{\mu \nu })=[\hat{x}^{\mu },\hat{x}^{\nu }] \). In this
case the Weyl ordered \( \star  \)-product takes the following form
(compare \cite{Dimitrijevic:2003pn})\begin{eqnarray}
f\star g & = & fg+\frac{ia}{2}\left( \partial _{o}f\, x^{i}\partial _{i}g-x^{i}\partial _{i}f\, \partial _{o}g\right) \nonumber \\
 &  & -\frac{a^{2}}{8}\left( \partial ^{2}_{0}f\, x^{i}x^{j}\partial _{i}\partial _{j}g-2\partial _{0}x^{i}\partial _{i}f\, \partial _{0}x^{i}\partial _{i}g+x^{i}x^{j}\partial _{i}\partial _{j}f\, \partial ^{2}_{0}g\right) \label{SO_a(n)_star_product} \\
 &  & -\frac{a^{2}}{12}\left( \partial ^{2}_{0}f\, x^{i}\partial _{i}g-\partial _{0}f\partial _{0}\, x^{i}\partial _{i}g-\partial _{0}x^{i}\partial _{i}f\, \partial _{o}g+x^{i}\partial _{i}f\, \partial ^{2}_{0}g\right) \nonumber \\
 &  & +\mathcal{O}(a^{3}).\nonumber 
\end{eqnarray}
We will continue the example when we have derivations of the \( \star  \)-product
algebra.

\section{\protect\( \star \protect \)-products with commuting vector fields}

The \( \star  \)-products of the last section are only given up to
second order and we have not been able to derive closed formulas.
Here we present a closed formula for \( \star  \)-products that generalise
the Moyal-Weyl-\( \star  \)-product (\ref{definiton_moyal_weyl_star_product})
in a simple way. \cite{Jambor:2003qc} We only have to replace the
partial derivatives in the formula by commuting vector fields, since
they have the same algebraic properties. We will prove the associativity
of this \( \star  \)-product and make considerations of how to get
desired algebra relations. After that we calculate \( \star  \)-products
for some examples like the \( so(3) \) Lie algebra and several quantum
spaces.

\subsection{Definition and proof of associativity\label{Associativity_of_vector_star}}

Let \( X \) be a vector field. Then it is easy to show that\[
X^{i}(x)\frac{\partial }{\partial x^{i}}\left( f(x)\, g(x)\right) =\left. (X^{i}(y)\frac{\partial }{\partial y^{i}}+X^{i}(z)\frac{\partial }{\partial z^{i}})\left( f(y)g(z)\right) \right| _{y\rightarrow x,z\rightarrow x}.\]
To write the last formula in a more compact way we introduce the following
notation\[
X_{1}f_{1}g_{1}=\left. (X_{2}+X_{3})f_{2}g_{3}\right| _{2\rightarrow 1,3\rightarrow 1}.\]
With this we can derive a kind of Leibniz rule\begin{eqnarray*}
X^{l}_{1}f_{1}g_{1} & = & \left. (X_{2}+X_{3})^{l}f_{2}g_{3}\right| _{2\rightarrow 1,3\rightarrow 1}\\
P(X_{1})f_{1}g_{1} & = & \left. P(X_{2}+X_{3})f_{2}g_{3}\right| _{2\rightarrow 1,3\rightarrow 1}
\end{eqnarray*}
where \( P \) is a polynomial in \( X \). The last equation can
also be written in the form \begin{equation}
\label{eq: vector}
P(X_{1})\left( \left. f_{1}g_{1}\right| _{2\rightarrow 1,3\rightarrow 1}\right) =\left. P(X_{2}+X_{3})f_{2}g_{3}\right| _{2\rightarrow 1,3\rightarrow 1}.
\end{equation}
Let now be \( X_{a}=X^{i}_{a}\partial _{i} \) \( n \) commuting
vector fields, i. e. \( [X_{a},X_{b}]=0 \). Note that then locally
always a coordinate system \( y^{a}(x) \) may be found with \( X_{a}=\partial _{y^{a}} \).
Globally this does not have to be the case. Further let \( \sigma ^{ab} \)
be a constant matrix. Then we define a \( \star  \)-product via\begin{equation}
\label{def: vector_star_product}
\left. (f\star g)\right| _{1}=\left. e^{\sigma ^{ab}X_{a2}X_{b3}}f_{2}g_{3}\right| _{2=3=1}.
\end{equation}
 This \( \star  \) product is associative since\begin{eqnarray*}
\left. (f\star (g\star h))\right| _{1} & = & \left. e^{\sigma ^{ab}X_{a2}X_{b3}}f_{2}\left( \left. e^{\sigma ^{cd}X_{c4}X_{d5}}g_{4}h_{5}\right| _{4\rightarrow 3,5\rightarrow 3}\right) \right| _{2\rightarrow 1,3\rightarrow 1}\\
 & = & \left. e^{\sigma ^{ab}X_{a2}(X_{b4}+X_{b5})}f_{2}e^{\sigma ^{cd}X_{c4}X_{d5}}g_{4}h_{5}\right| _{4\rightarrow 3,5\rightarrow 3,2\rightarrow 1,3\rightarrow 1}\\
 & = & \left. e^{\sigma ^{ab}X_{a1}X_{b2}+\sigma ^{ab}X_{a1}X_{b3}}e^{\sigma ^{cd}X_{c2}X_{d3}}f_{1}g_{2}h_{3}\right| _{2\rightarrow 1,3\rightarrow 1}
\end{eqnarray*}
and\begin{eqnarray*}
\left. ((f\star g)\star h)\right| _{1} & = & \left. e^{\sigma ^{ab}X_{a1}X_{b2}}_{2}\left( \left. e^{\sigma ^{cd}X_{c3}X_{d4}}f_{3}g_{4}\right| _{3\rightarrow 1,4\rightarrow 1}\right) h_{2}\right| _{2\rightarrow 1}\\
 & = & \left. e^{\sigma ^{ab}(X_{a3}+X_{a4})X_{b2}}e^{\sigma ^{cd}X_{c3}X_{d4}}f_{3}g_{4}h_{2}\right| _{3\rightarrow 1,4\rightarrow 1,2\rightarrow 1}\\
 & = & \left. e^{\sigma ^{ab}X_{a1}X_{b3}+\sigma ^{ab}X_{a2}X_{b3}}e^{\sigma ^{cd}X_{c1}X_{d2}}f_{1}g_{2}h_{3}\right| _{2\rightarrow 1,3\rightarrow 1}
\end{eqnarray*}
where in the second step we used the relation (\ref{eq: vector}).
The two expressions are equal since the vector fields commute.

For future use we calculate the \( \star  \)-commutator\begin{eqnarray*}
[f\stackrel{\star }{,}g] & = & \left. \left( e^{\sigma ^{ab}X_{a1}X_{b2}}-e^{\sigma ^{ab}X_{a2}X_{b1}}\right) f_{1}g_{2}\right| _{2\rightarrow 1}\\
 & = & \left. 2\sinh (\sigma ^{ab}X_{a1}X_{b2})f_{1}g_{2}\right| _{2\rightarrow 1}.
\end{eqnarray*}
The last line is only valid for an antisymmetric matrix \( \sigma . \)

For the case of two vector fields, which we call \( X_{1}=X \) and
\( X_{2}=Y \), we write down the explicit formula for \( \sigma ^{12}=h,\sigma ^{21}=0 \)\begin{equation}
\label{def: asym star}
f\star g=\sum ^{\infty }_{n=0}\frac{h^{n}}{n!}(X^{n}f)\, (Y^{n}g)
\end{equation}
the asymmetric \( \star  \)-product and for \( \sigma ^{12}=\frac{h}{2},\sigma ^{21}=-\frac{h}{2} \)
\begin{equation}
\label{def: antisym star}
f\star g=\sum ^{\infty }_{n=0}\frac{h^{n}}{2^{n}n!}\sum ^{n}_{i=0}(-1)^{i}{n\choose i}(X^{n-i}Y^{i}f)\, (X^{i}Y^{n-i}g)
\end{equation}
which yields the antisymmetric \( \star  \)-product. Both \( \star  \)-products
have the same Poisson tensor \( \Pi =X\wedge Y \).

\subsection{Linear transformations}

If we have a \( \star  \)-product, we have seen that we simply can
produce a new \( \star  \)-product by a linear transformation on
the space of functions (\ref{linear_transformation_on_star_product}).
Suppose that \( D \) is such an invertible operator and that its
expansion in derivatives starts with 1. Additionally we now assume,
that \( D \) is of the form\begin{eqnarray*}
D=e^{\tau (X_{a})} & , & D^{-1}=e^{-\tau (X_{a})}
\end{eqnarray*}
where \( \tau  \) is a polynomial of the vector fields \( X_{a} \).
Then for the \( \star  \)-product (\ref{def: vector_star_product})
we see that \begin{eqnarray*}
f\star 'g & = & D^{-1}(D(f)\star D(g))\\
 & = & e^{-\tau (X_{a1})}\left( \left. e^{\sigma ^{ab}X_{a2}X_{b3}}e^{\tau (X_{a2})}f_{2}e^{\tau (X_{a3})}g_{3}\right| _{2\rightarrow 1,3\rightarrow 1}\right) \\
 &  & =\left. e^{-\tau (X_{a2}+X_{a3})+\sigma ^{ab}X_{a2}X_{b3}+\tau (X_{a2})+\tau (X_{a3})}f_{2}g_{3}\right| _{2\rightarrow 1,3\rightarrow 1}.
\end{eqnarray*}
For \( \tau  \) only quadratic in the \( X_{a} \) (note that \( \tau _{2}^{ab} \)
is symmetric) \begin{eqnarray*}
\tau  & = & \tau _{1}^{a}X_{a}+\frac{1}{2}\tau _{2}^{ab}X_{a}X_{b}
\end{eqnarray*}
we have\[
\tau (X_{a1})+\tau (X_{a2})-\tau (X_{a1}+X_{a2})=-\tau _{2}^{ab}X_{a1}X_{b2}\]
and the new \( \star  \)-product becomes\[
f\star 'g=\left. e^{(\sigma ^{ab}-\tau ^{ab}_{2})X_{a1}X_{b2}}\right| _{2\rightarrow 1}\]
So we see that the antisymmetric \( \star  \)-product (\ref{def: antisym star})
and the asymmetric \( \star  \)-product (\ref{def: asym star}) are
related by a linear transformation in function space.

\bigskip

As already mentioned locally commuting vector fields can be represented
by a coordinate transformation. It is very important that this need
not to be the case globally. This is the reason that the algebras
resulting from the \( \star  \)-product are not isomorphic to the
constant case algebra. We will see explicit examples for this later.

\subsection{Differences to other \protect\( \star \protect \)-products}

The Poisson tensor of the above defined \( \star  \)-product (\ref{def: vector_star_product})
is \( \Pi ^{ij}=\sigma ^{ab}X^{i}_{a}X^{j}_{b} \) with \( \sigma ^{ab} \)
antisymmetric. This we can plug into the formula of the Weyl ordered
\( \star  \)-product (\ref{weyl_ordered_star_product}) and make
a linear transformation. We can compare the result to the \( \star  \)-product
(\ref{def: vector_star_product}). After some calculations we get\[
e^{-\rho }\left( e^{\rho }(f)\star _{\textrm{Weyl}}e^{\rho }(g)\right) =f\star _{\sigma }g\]
with \begin{eqnarray*}
\rho  & = & 1+\frac{1}{16}\sigma ^{ab}\sigma ^{cd}(X^{m}_{a}\partial _{m}X^{i}_{c})(X^{n}_{b}\partial _{n}X^{j}_{d})\partial _{i}\partial _{j}\\
 &  & +\frac{1}{24}\sigma ^{ab}\sigma ^{cd}(X^{i}_{a}X^{j}_{c}X^{n}_{b}\partial _{n}X^{k}_{d}+X^{k}_{a}X^{i}_{c}X^{n}_{b}\partial _{n}X^{j}_{d}+X^{j}_{a}X^{k}_{c}X^{n}_{b}\partial _{n}X^{i}_{d})\partial _{i}\partial _{j}\partial _{k}\\
 &  & +\mathcal{O}(\sigma ^{3}).
\end{eqnarray*}
Therefore this two \( \star  \)-products are equivalent at least
up to second order. 

\bigskip

Later we will define the Kontsevich \( \star  \)-product. This \( \star  \)-product
can be constructed on every Poisson manifolds and proves that one
is able to find a \( \star  \)-product for every Poisson structure.
We were not able to show, that there is a equivalence between the
Kontsevich \( \star  \)-product and the \( \star  \)-product constructed
by commuting vector fields. There may be obstructions since the equivalence
is dependent of the Poisson cohomology of the Poisson manifold.

\subsection{Some examples in two dimensions}

We calculate some examples in two dimensions with the asymmetric \( \star  \)-product
(\ref{def: asym star}).

\bigskip

\subsubsection{\protect\( X=ax\partial _{x}\protect \) , \protect\( Y=\partial _{y}\protect \)}

We get \[
[x\stackrel{\star }{,}y]=ax,\]
 the algebra of two dimensional \( a \)-euclidian space. The algebra
relations follow from\begin{eqnarray*}
x\star x & = & x^{2},\\
x\star y & = & xy+ax,\\
y\star x & = & xy,\\
y\star y & = & y^{2}.
\end{eqnarray*}

\bigskip

\subsubsection{\protect\( X=(a+bx)\partial _{x}\protect \) , \protect\( Y=(c+dy)\partial _{y}\protect \)}

This is the general linear case. We get \[
x\star y=e^{bd}(y+\frac{c}{d})\star (x+\frac{a}{b}),\]
which follows from \begin{eqnarray*}
x\star x & = & x^{2},\\
x\star y & = & e^{bd}(y+\frac{c}{d})(x+\frac{a}{b}),\\
y\star x & = & xy,\\
y\star y & = & y^{2}.
\end{eqnarray*}

\bigskip

\subsubsection{\protect\( X=\frac{a}{\sqrt{x^{2}+y^{2}}}(x\partial _{x}+y\partial _{y})\protect \)
, \protect\( Y=x\partial _{y}-y\partial _{x}\protect \)}

These are the derivatives \( \partial _{r} \) and \( \partial _{\theta } \)
of the coordinate transformation \( x=r\cos \theta ,\, y=r\sin \theta  \).
We get \[
[x\stackrel{\star }{,}y]=a\sqrt{x\star x+y\star y},\]
 which follows from \begin{eqnarray*}
x\star x & = & x^{2}-a\frac{xy}{r},\\
x\star y & = & xy+a\frac{x^{2}}{r},\\
y\star x & = & xy-a\frac{y^{2}}{r},\\
y\star y & = & y^{2}+a\frac{xy}{r}.
\end{eqnarray*}

\bigskip

\subsubsection{\protect\( X=a(x\partial _{x}+y\partial _{y})\protect \) , \protect\( Y=x\partial _{y}-y\partial _{x}\protect \)}

This is a simplification of the previous case. We get \[
[x\stackrel{\star }{,}y]=(\tan a)(x\star x+y\star y),\]
which follows from \begin{eqnarray*}
x\star x & = & \cos a\, x^{2}-\sin a\, xy,\\
x\star y & = & \cos a\, xy+\sin a\, y^{2},\\
y\star x & = & \cos a\, xy-\sin a\, y^{2},\\
y\star y & = & \cos a\, y^{2}+\sin a\, xy,
\end{eqnarray*}
 \begin{eqnarray*}
x\star x+y\star y & = & \cos a\, (x^{2}+y^{2}),\\
x\star y-y\star x & = & \sin a\, (x^{2}+y^{2}).
\end{eqnarray*}
It is interesting that this algebra does not have the Poincare-Birkhoff-Witt
property for \( \tan a=1 \).

\bigskip

We have seen that even in this simple cases very rich structures surface.
But it is not quite clear what happens if we try to replace the formal
parameter in the \( \star  \)-product expansion by a number. In the
last case we see, that in this case higher order relations can arise.

\subsection{Realization of algebras}

If we want to represent an algebra with the help of a \( \star  \)-product,
we have seen that this is possible if we use an ordering description.
In this section we propose an other method of how to calculate a \( \star  \)-product
with the property, that it reproduces the algebra relations of some
desired algebra. With this second approach no ordering description
is needed. It is even not quite clear in the end, if there would be
an ordering description that would yield the same \( \star  \)-product
with the first approach.

We know that the \( \star  \)-commutator of a \( \star  \)-product
is a Poisson tensor up to first order\[
[f\stackrel{\star }{,}g]=h\{f,g\}+\mathcal{O}(h^{2})=h\Pi (f,g)+\mathcal{O}(h^{2}).\]
 where \( \Pi  \) is the Possion-bivector of the Poisson structure.
If we would have a \( \star  \)-product that reproduces the algebra
relations, the right hand side of the previous equation would be a
polynomial in the generators of the algebra, i. e. \[
[x^{i}\stackrel{\star }{,}x^{j}]=hc^{ij}_{\star }(x).\]
The \( \star  \) in the index of \( c_{\star }^{ij} \) indicates
that all products between the coordinate functions in it are \( \star  \)-products.
To calculate the leading order of \( c^{ij}_{\star }(x) \) it is
not necessary to know the explicit form of the \( \star  \)-product,
since it always starts with the ordinary product of functions. We
can conclude that\begin{equation}
\label{Poisson_like_algebra_comm}
\{x^{i},x^{j}\}=\Pi ^{ij}=c^{ij}(x),
\end{equation}
For the special case for the \( \star  \)-products (\ref{def: vector_star_product})
it is\[
\Pi =\sigma ^{ab}X_{a}\wedge X_{b}.\]
If we are able to write a general Poisson bivector in this form, we
can try to reconstruct the algebra relations with the help of the
\( \star  \)-products (\ref{def: vector_star_product}). For this
let \( f \) be a function and \( X_{f}=\{f,\cdot \} \) the Hamiltonian
vector field associated to \( f \). Then the commutator of vector
fields is\[
[X_{f},X_{g}]=X_{\{f,g\}},\]
due to the Jakobi identity of the Poisson bracket. If we can find
functions with \begin{eqnarray*}
\{f_{i},g_{j}\}=\delta _{ij}, & \{f_{i},f_{j}\}=0, & \{g_{i},g_{j}\}=0,
\end{eqnarray*}
this implies that all commutators between the associated Hamiltonian
vector fields vanish. Now one can deduce from the splitting theorem
for Poisson manifolds \cite{Weinstein:1983} that this is possible
in a neighborhood of a point if the rank of the Poisson tensor is
constant around this point. Since we do not want to find a \( \star  \)-product
on \( \mathbb {R}^{N} \), but a \( \star  \)-product with certain
commutation relations, we can reduce \( \mathbb {R}^{N} \) by the
set of points where the rank of the Poisson tensor jumps and we have
a good chance to find functions with the desired properties on the
new manifold. In this case we can write the Poisson tensor as\[
\Pi =\sum _{i}X_{f_{i}}\wedge X_{g_{i}}.\]
In the following we will find functions \( f_{i} \) anf \( g_{i} \)
for Poisson tensors of several algebras and will use the corresponding
Hamiltonian vector fields in the \( \star  \)-products (\ref{def: vector_star_product}).
We will calculate the resulting algebra relations from the \( \star  \)-product
and compare them to the original algebra relations.

\subsection{The quantum space \protect\( M(so_{a}(n))\protect \)}

We will start our examples by giving a closed formula for a second
\( \star  \)-product for the quantum space introduced in (\ref{example_so_a(n)}).
It is closely related to the \( \star  \)-product for the two dimensional
\( a \)-euclidean space given above. As manifold we take \( \mathbb {R}^{N} \)
with coordinates \( x^{0} \) and \( x^{i} \) with \( i=1,\dots ,N-1 \)
and use the asymmetric \( \star  \)-product (\ref{def: asym star})
with the two vectorfields\begin{eqnarray*}
X=iax^{i}\partial _{i}, &  & Y=\partial _{o}.
\end{eqnarray*}
With this we get \[
[x^{i}\stackrel{\star }{,}x^{0}]=iax^{i},\]
 the algebra of \( M(so_{a}(n)) \). The algebra relations follow
from\begin{eqnarray*}
x^{i}\star x^{j} & = & x^{i}x^{j},\\
x^{i}\star x^{0} & = & x^{i}x^{0}+iax^{i},\\
x^{0}\star x^{i} & = & x^{i}x^{0},\\
y\star y & = & y^{2}.
\end{eqnarray*}

\bigskip

To show the usefulness of the approach proposed in the last section
we now make a generalization of the above defined algebra. The new
relations are \[
[\hat{x}^{\alpha },\hat{x}^{\beta }]=i(a^{\alpha }\hat{x}^{\beta }-a^{\beta }\hat{x}^{\alpha })\]
where \( a^{\alpha } \) are now \( n \) deformation parameters.
For this relations to be consistent the Jacobi identities have to
be fulfilled, which easily can be proofed. In this case the commuting
vector fields can not so easy guessed like in the special case.

Since the right hand side of the relation is linear and we are therefore
dealing with a Lie algebra the Poisson tensor associated with the
algebra is simple\[
\{x^{\alpha },x^{\beta }\}=a^{\alpha }x^{\beta }-a^{\beta }x^{\alpha }.\]
If we want to find commuting vector fields that reproduce this Poisson
tensor we now follow the way outlined in the previous section. The
rank of this matrix is \( 2 \). Therefore we have to find two functions
fulfilling \( \{f,g\}=1 \). We make a guess and define\begin{eqnarray*}
f=a^{\alpha }x^{\alpha } &  & \tilde{x}^{\alpha }=x^{\alpha }-\frac{a^{\alpha }a^{\beta }}{a^{2}}x^{\beta }
\end{eqnarray*}
with \( a^{2}=a^{\alpha }a^{\alpha } \). These functions have commutation
relations very similar to the special case of \( M(so_{a}(n)) \).\begin{eqnarray*}
\{f,\tilde{x}^{\alpha }\}=a^{2}\tilde{x}^{\alpha } &  & \{\tilde{x}^{\alpha },\tilde{x}^{\beta }\}=0
\end{eqnarray*}
If we define \( g=\frac{1}{a^{2}}ln\, \sqrt{\tilde{x}^{\alpha }\tilde{x}^{\alpha }} \)
we see that\begin{equation}
\{f,g\}=1
\end{equation}
and the desired functions are found. The commuting vector fields are
now easy calculated\begin{eqnarray*}
X=\{f,\cdot \} & = & a^{2}x^{\beta }\partial _{\beta }-(a^{\alpha }x^{\alpha })a^{\beta }\partial _{\beta }\\
Y=\{\cdot ,g\} & = & -\frac{1}{a^{2}}a^{\beta }\partial _{\beta }
\end{eqnarray*}
 In this case we are lucky since no singularities have shown up and
the \( \star  \)-product can be defined on whole \( \mathbb {R}^{n} \).
Again we may use the asymmetric \( \star  \)-product (\ref{def: asym star})
and see that the algebra relations are reproduced.

\subsection{\protect\( q\protect \)-deformed Heisenberg algebra}

If we take the \( q \)-deformed Heisenberg algebra \cite{Wess:1999}
in two dimensions \[
\hat{x}\hat{y}=q\hat{y}\hat{x}+\theta \]
 we very easily can calculate a \( \star  \)-product in \( h=ln\, q \)
and \( \theta  \). The Poisson tensor \( \Pi  \) is\[
\Pi =(xy+\frac{\theta }{h})\partial _{x}\wedge \partial _{y}.\]
We see that with \( f=\ln (xy+\frac{\theta }{h}) \) and \( g=\ln y \)
\[
\{f,g\}=1.\]
The Hamiltonian vector fields are\begin{eqnarray*}
X=X_{f}=y\partial _{y}-x\partial _{x}, &  & Y=X_{g}=-(x+\frac{\theta }{hy})\partial _{x}.
\end{eqnarray*}
To calculate the \( \star  \)-products we note\begin{eqnarray*}
X^{n}(x)=(-1)^{n}x, &  & X^{n}(y)=y,\\
Y^{n}(x)=(-1)^{n}(x+\frac{\theta }{hy})\textrm{ for }n>0, &  & Y^{n}(y)=\delta ^{n,0}y.
\end{eqnarray*}
For the asymmetric \( \star  \)-product (\ref{def: asym star}) this
yields\begin{eqnarray*}
x\star y & = & xy,\\
y\star x & = & e^{-h}xy+(e^{-h}-1)\frac{\theta }{h}.
\end{eqnarray*}
For the antisymmetric \( \star  \)-product (\ref{def: antisym star})
we get\begin{eqnarray*}
x\star y & = & e^{\frac{h}{2}}xy+(e^{+\frac{h}{2}}-1)\frac{\theta }{h},\\
y\star x & = & e^{-\frac{h}{2}}xy+(e^{-\frac{h}{2}}-1)\frac{\theta }{h}.
\end{eqnarray*}
Both \( \star  \)-products have therefore the algebra relation \[
x\star y=e^{h}y\star x+(e^{h}-1)\frac{\theta }{h}\]
and by a redefinition of \( \theta  \) the original alegebra relations
are reproduced.

\subsection{The Lie algebra \protect\( so(3)\protect \)}

\subsubsection{First try}

The algebra relations of the eveloping algebra of \( so(3) \) are
\( [\hat{x}^{i},\hat{x}^{j}]=i\epsilon ^{ijk}\hat{x}^{k} \). The
corresponding poisson tensor is \[
\Pi =ix\partial _{y}\wedge \partial _{z}+iy\partial _{z}\wedge \partial _{x}+iz\partial _{x}\wedge \partial _{y}.\]
For \( f=-i\arctan \frac{x}{y} \) we have \( \{z,f\}=1 \). The Hamiltonian
vector fields are \begin{eqnarray*}
X=X_{f} & = & -\partial _{z}+\frac{z}{x^{2}+y^{2}}(x\partial _{x}+y\partial _{y}),\\
Y=X_{z} & = & i(y\partial _{x}-x\partial _{y}).
\end{eqnarray*}
We calculate (\( \rho =\sqrt{x^{2}+y^{2}} \)) \begin{eqnarray*}
Y^{n}(z)=\delta ^{n0}z, &  & X^{n}(z)=\delta ^{n0}z+\delta ^{n1},\\
Y^{2n}(x)=(-1)^{n}x, &  & Y^{2n+1}(x)=-(-1)^{n}y,\\
Y^{2n}(y)=(-1)^{n}y, &  & Y^{2n+1}(y)=(-1)^{n}x,\\
X(x)=-x\frac{z}{\rho ^{2}}, &  & Y(y)=-y\frac{z}{\rho ^{2}},\\
X^{n}(x)=x(1+\frac{z^{2}}{\rho ^{2}})f_{n}(\rho ,z), &  & X^{n}(y)=y(1+\frac{z^{2}}{\rho ^{2}})f_{n}(\rho ,z),\\
f_{2}=-\frac{1}{\rho ^{2}}, &  & f_{n+1}=\frac{z}{\rho ^{2}}f_{n}+\partial _{z}f_{n}-\frac{z}{\rho }\partial _{\rho }f_{n},\\
f_{3}=\frac{z}{\rho ^{4}}, &  & f_{4}=\frac{1}{\rho ^{4}}(1+\frac{5z^{2}}{\rho ^{2}}).
\end{eqnarray*}
If we transform to the \( x^{+},x^{-} \)coordinate system we are
now able to calculate the commutator of \( x^{+} \)and \( x^{-} \)and
get \begin{eqnarray*}
[x^{+}\stackrel{\star }{,}x^{-}] & = & 2\sum ^{\infty }_{n=0}\frac{h^{2n+1}}{(2n+1)!}\sum ^{\infty }_{i=0}{n\choose i}(X^{i}x^{+})(X^{n-i}x^{-})\\
 & = & -2hz+\frac{h^{3}}{6}(1+\frac{z^{2}}{\rho ^{2}})(\frac{4z}{\rho ^{2}})+\mathcal{O}(h^{5}).
\end{eqnarray*}
In this case the original algebra relations are not reproduced.

\subsubsection{Second try}

After a linear basis transformation we now can start with the algebra
relations

\begin{eqnarray*}
[\hat{z},\hat{x}^{+}]=\hat{x}^{+}, & [\hat{z},\hat{x}^{-}]=-\hat{x}^{-}, & [\hat{x}^{+},\hat{x}^{-}]=\hat{z}.
\end{eqnarray*}
With \( f=\ln x^{-} \)we have \( \{f,z\}=1 \). The Hamiltonian vector
fields become now

\begin{eqnarray*}
X=X_{f}=\partial _{z}-\frac{z}{x^{-}}\partial _{+}, &  & Y=X_{z}=x^{+}\partial _{+}-x^{-}\partial _{x^{-}}.
\end{eqnarray*}
 Therefore\begin{eqnarray*}
X^{n}(z)=\delta ^{n0}z+\delta ^{n1}, &  & Y^{n}(z)=\delta ^{n0}z,\\
X^{n}(x^{+})=-\delta ^{n1}\frac{z}{x^{-}}+\delta ^{n0}x^{+}-\delta ^{n2}\frac{1}{x^{-}}, &  & Y^{n}(x^{+})=x^{+},\\
X^{n}(x^{-})=\delta ^{n0}x^{-}, &  & Y^{n}(x^{-})=(-1)^{n}x^{-}.
\end{eqnarray*}
and with the asymmetric \( \star  \)-product (\ref{def: asym star})
we get \begin{eqnarray*}
z\star x^{+}=zx^{+}+hx^{+}, &  & x^{+}\star z=x^{+}z,\\
z\star x^{-}=zx^{-}-hx^{-}, &  & x^{-}\star z=x^{-}z,\\
x^{+}\star x^{-}=x^{+}x^{-}+hz-\frac{h^{2}}{2}, &  & x^{-}\star x^{+}=x^{+}x^{-}.
\end{eqnarray*}
and therefore\begin{eqnarray*}
[z\stackrel{\star }{,}x^{+}]=hx^{+}, & [z\stackrel{\star }{,}x^{-}]=-hx^{-}, & [x^{+}\stackrel{\star }{,}x^{-}]=hz-\frac{h^{2}}{2}.
\end{eqnarray*}
With \( \tilde{z}=z-\frac{h}{2} \) now the correct algebra relations
are reproduced.

\subsection{The quantum space \protect\( M(so_{q}(3))\protect \)}

Here we give as a second example the \( \star  \)-product for the
quantum space \-\( M(so_{q}(3)) \) invariant under the quantum group
\( SO_{q}(3). \) \cite{Lorek:1997eh} The algebra relations in the
basis adjusted to the quantum group terminology are \begin{eqnarray*}
\hat{z}\hat{x}^{+}=q^{2}\hat{x}^{+}\hat{z}, & \hat{z}\hat{x}^{-}=q^{-2}\hat{x}^{-}\hat{z}, & [\hat{x}^{-},\hat{x}^{+}]=(q-q^{-1})\hat{z}^{2}.
\end{eqnarray*}
For the commutators we get\begin{eqnarray*}
[\hat{z},\hat{x}^{+}]=(q^{2}-1)\hat{x}^{+}\hat{z}, & [\hat{z},\hat{x}^{-}]=(q^{2}-1)\hat{x}^{-}\hat{z}, & [\hat{x}^{-},\hat{x}^{+}]=(q-\frac{1}{q})\hat{z}^{2}
\end{eqnarray*}
and therefore the Poisson brackets are\begin{eqnarray*}
\{z,x^{+}\}=2zx^{+}, & \{z,x^{-}\}=-2zx^{-}, & \{x^{-},x^{+}\}=2z^{2}.
\end{eqnarray*}
For \( f=\ln x^{-} \)and \( g=\frac{1}{2}\ln z \) we have \( \{f,g\}=1 \)
and the Hamiltonian vector fields become\begin{eqnarray*}
X_{f}=2z\partial _{z}+\frac{2z^{2}}{x^{-}}\partial _{+}, &  & X_{g}=x^{+}\partial _{+}-x^{-}\partial _{-}.
\end{eqnarray*}
For the \( \star  \)-product we try now\begin{eqnarray*}
X=z\partial _{z}+\frac{\alpha z^{2}}{x^{-}}\partial _{+}, &  & Y=x^{+}\partial _{+}-x^{-}\partial _{-}.
\end{eqnarray*}
We have\begin{eqnarray*}
Y^{n}(x^{+})=x^{+}, & Y^{n}(x^{-})=(-1)^{n}x^{-}, & Y^{n}(z)=\delta ^{n,0}z,\\
 & X^{n}(x^{-})=\delta ^{n,0}x^{-}, & X^{n}(z)=z,\\
X^{n}(x^{+})=\alpha 2^{n-1}\frac{z^{2}}{x^{-}}\textrm{ } & \textrm{for }n>0. & 
\end{eqnarray*}
\bigskip

\noindent For the asymmetric \( \star  \)-product (\ref{def: asym star})
we calculate \begin{eqnarray*}
x^{+}\star z=x^{+}z, &  & z\star x^{+}=e^{h}x^{+}z,\\
x^{-}\star z=x^{-}z, &  & z\star x^{-}=e^{-h}x^{-}z,\\
x^{+}\star x^{-}=x^{+}x^{-}+\frac{\alpha }{2}(e^{-2h}-1)z^{2}, &  & x^{-}\star x^{+}=x^{+}x^{-},\\
z\star z=z^{2} &  & 
\end{eqnarray*}
and get algebra relations\begin{eqnarray*}
z\star x^{+}=e^{h}x^{+}\star z, & z\star x^{-}=e^{-h}x^{-}\star z, & [x^{+}\stackrel{\star }{,}x^{-}]=\frac{\alpha }{2}(e^{-2h}-1)z\star z.
\end{eqnarray*}
If we set\begin{eqnarray*}
q=e^{\frac{h}{2}}, &  & \alpha =\frac{2q^{2}}{q+\frac{1}{q}}
\end{eqnarray*}
this reproduces the algebra relations. 

\noindent \bigskip

\noindent For the antisymmetric \( \star  \)-product (\ref{def: antisym star})
we calculate \begin{eqnarray*}
x^{+}\star z=e^{-\frac{h}{2}}x^{+}z, &  & z\star x^{+}=e^{\frac{h}{2}}x^{+}z,\\
x^{-}\star z=e^{\frac{h}{2}}x^{-}z, &  & z\star x^{-}=e^{-\frac{h}{2}}x^{-}z,\\
x^{+}\star x^{-}=x^{+}x^{-}+\frac{\alpha }{2}(e^{-h}-1)z^{2}, &  & x^{-}\star x^{+}=x^{+}x^{-}+\frac{\alpha }{2}(e^{h}-1)z^{2},\\
z\star z=z^{2}. &  & 
\end{eqnarray*}
The algebra relations now are\begin{eqnarray*}
z\star x^{+}=e^{h}x^{+}\star z, & x^{-}z\star x^{-}=e^{-h}x^{-}\star z, & [x^{+}\stackrel{\star }{,}x^{-}]=-\frac{\alpha }{2}(e^{h}-e^{-h})z\star z.
\end{eqnarray*}
Here we can reproduce the algebra relations when we set\begin{eqnarray*}
q=e^{\frac{h}{2}}, &  & \alpha =-\frac{2}{q+\frac{1}{q}}.
\end{eqnarray*}

\subsection{The quantum space \protect\( M(so_{q}(1,3))\protect \)}

We try to generalize the previous example and start with the more
general commuting vector fields\begin{eqnarray*}
X=z\partial _{z}+\frac{1}{x^{-}}(\alpha z^{2}+\beta z)\partial _{+}, &  & Y=x^{+}\partial _{+}-x^{-}\partial _{-}.
\end{eqnarray*}
The only relation, that is changed is\[
X^{n}(x^{+})=\frac{1}{x^{-}}(2^{n-1}\alpha z^{2}+\beta z)\]
and we calculate with the antisymmetric \( \star  \)-product (\ref{def: antisym star})
\begin{eqnarray*}
x^{+}\star z=e^{-\frac{h}{2}}x^{+}z, &  & z\star x^{+}=e^{\frac{h}{2}}x^{+}z,\\
x^{-}\star z=e^{\frac{h}{2}}x^{-}z, &  & z\star x^{-}=e^{-\frac{h}{2}}x^{-}z,\\
z\star z=z^{2}, &  & 
\end{eqnarray*}
\[
x^{+}\star x^{-}=x^{+}x^{-}+\frac{\alpha }{2}(e^{-h}-1)z^{2}+\beta (e^{-\frac{h}{2}}-1)z,\]
\[
x^{-}\star x^{+}=x^{+}x^{-}+\frac{\alpha }{2}(e^{h}-1)z^{2}+\beta (e^{\frac{h}{2}}-1)z.\]
The relations become\begin{eqnarray*}
z\star x^{+}=e^{h}x^{+}\star z, &  & z\star x^{-}=e^{-h}x^{-}\star z,
\end{eqnarray*}
\[
[x^{+}\stackrel{\star }{,}x^{-}]=-\frac{\alpha }{2}(e^{h}-e^{-h})z\star z+-\beta (e^{\frac{h}{2}}-e^{-\frac{h}{2}})z.\]
The algebra relation of \( M(so_{q}(1,3)) \) are \cite{Lorek:1997eh,Cerchiai:1998ee}
\[
[\hat{t},\hat{x}^{i}]=0,\]
\[
[\hat{x}^{-},\hat{x}^{+}]=(q-\frac{1}{q})(\hat{z}^{2}-\hat{t}\hat{z}),\]
\[
\hat{z}\hat{x}^{+}=q^{2}\hat{x}^{+}\hat{z}+(1-q^{2})\hat{t}\hat{x}^{+},\]
\[
\hat{z}\hat{x}^{-}=q^{-2}\hat{x}^{-}\hat{z}+(1-q^{-2})\hat{t}\hat{x}^{-}.\]
We can define \( \tilde{\hat{z}}=\hat{z}-\hat{t} \) and get new relations\[
[\hat{t},\hat{x}^{i}]=0,\]
\[
[\hat{x}^{-},\hat{x}^{+}]=(q-\frac{1}{q})(\tilde{\hat{z}}^{2}+\hat{t}\tilde{\hat{z}}),\]
\[
\tilde{\hat{z}}\hat{x}^{+}=q^{2}\hat{x}^{+}\tilde{\hat{z}},\]
\[
\tilde{\hat{z}}\hat{x}^{-}=q^{-2}\hat{x}^{-}\tilde{\hat{z}}.\]
These relations are reproduced by the \( \star  \)-product if we
set\begin{eqnarray*}
q=e^{\frac{h}{2}} & \alpha =-\frac{2}{q+\frac{1}{q}}, & \beta =-1.
\end{eqnarray*}

\subsection{The quantum space \protect\( M(so_{q}(4))\protect \)}

\noindent The algebra relations of \( M(so_{q}(4)) \) are \cite{Faddeev:1990ih,Ocampo:1996}
\begin{eqnarray*}
\hat{x}_{1}\hat{x}_{2}=q\hat{x}_{2}\hat{x}_{1}, &  & \hat{x}_{1}\hat{x}_{3}=q\hat{x}_{3}\hat{x}_{1},\\
\hat{x}_{3}\hat{x}_{4}=q\hat{x}_{4}\hat{x}_{3}, &  & \hat{x}_{2}\hat{x}_{4}=q\hat{x}_{4}\hat{x}_{2},\\
\hat{x}_{2}\hat{x}_{3}=\hat{x}_{3}\hat{x}_{2}, &  & [\hat{x}_{4},\hat{x}_{1}]=(q-\frac{1}{q})\hat{x}_{2}\hat{x}_{3}.
\end{eqnarray*}
The Poisson brackets are\begin{eqnarray*}
\{x_{1},x_{2}\}=x_{1}x_{2}, &  & \{x_{2},x_{4}\}=x_{2}x_{4},\\
\{x_{1},x_{3}\}=x_{1}x_{3}, &  & \{x_{3},x_{4}\}=x_{3}x_{4},\\
\{x_{2},x_{3}\}=0, &  & \{x_{4},x_{1}\}=2x_{2}x_{3}.
\end{eqnarray*}
Since the Poisson tensor has two Casimir functions, two vector fields
will suffice. We take\begin{eqnarray*}
f=\ln x_{2}, &  & g=\ln x_{4},\\
\{f,g\}=1, &  & \\
X=X_{f}=x_{4}\partial _{4}-x_{1}\partial _{1}, &  & Y=X_{g}=-(x_{2}\partial _{2}+x_{3}\partial _{3})+2\frac{x_{2}x_{3}}{x_{4}}\partial _{1}.
\end{eqnarray*}
Therefore\begin{eqnarray*}
X^{n}(x_{1})=(-1)^{n}x_{1}, &  & Y^{n}(x_{1})=\delta ^{n0}x_{1}+\delta ^{ni}(-2)^{i}\frac{x_{2}x_{3}}{x_{4}},\\
X^{n}(x_{2})=\delta ^{n0}x_{2}, &  & Y^{n}(x_{2})=(-1)^{n}x_{2},\\
X^{n}(x_{3})=\delta ^{n0}x_{3}, &  & Y^{n}(x_{3})=(-1)^{n}x_{3},\\
X^{n}(x_{4})=x_{4}, &  & Y^{n}(x_{4})=\delta ^{n0}x_{4}.
\end{eqnarray*}
\bigskip

\noindent For the asymmetric \( \star  \)-product (\ref{def: asym star})
we get\begin{eqnarray*}
x_{1}\star x_{2}=e^{h}x_{1}x_{2}, &  & x_{2}\star x_{1}=x_{1}x_{2},\\
x_{1}\star x_{3}=e^{h}x_{1}x_{3}, &  & x_{3}\star x_{1}=x_{1}x_{3},\\
x_{1}\star x_{4}=x_{1}x_{4}, &  & x_{4}\star x_{1}=x_{1}x_{4}+(e^{-2h}-1)x_{2}x_{3},\\
x_{2}\star x_{3}=x_{2}x_{3}, &  & x_{3}\star x_{2}=x_{2}x_{3},\\
x_{2}\star x_{4}=x_{2}x_{4}, &  & x_{4}\star x_{2}=e^{-h}x_{2}x_{4},\\
x_{3}\star x_{4}=x_{3}x_{4}, &  & x_{4}\star x_{3}=e^{-h}x_{3}x_{4}
\end{eqnarray*}
which yields the algebra relations\begin{eqnarray*}
x_{1}\star x_{2}=e^{h}x_{2}\star x_{1}, &  & x_{1}\star x_{3}=e^{h}x_{3}\star x_{1},\\
x_{3}\star x_{4}=e^{h}x_{4}\star x_{3}, &  & x_{2}\star x_{4}=e^{h}x_{4}\star x_{2},\\
x_{2}\star x_{3}=x_{3}\star x_{2}, &  & [x_{1}\stackrel{\star }{,}x_{4}]=(e^{-2h}-1)x_{2}\star x_{3}.
\end{eqnarray*}
\bigskip

\noindent For the antisymmetric \( \star  \)-product (\ref{def: antisym star})
we calculate \begin{eqnarray*}
x_{1}\star x_{2}=e^{\frac{h}{2}}x_{1}x_{2}, &  & x_{2}\star x_{1}=e^{-\frac{h}{2}}x_{1}x_{2},\\
x_{1}\star x_{3}=e^{\frac{h}{2}}x_{1}x_{3}, &  & x_{3}\star x_{1}=e^{-\frac{h}{2}}x_{1}x_{3},\\
x_{1}\star x_{4}=x_{1}x_{4}+(e^{-h}-1)x_{2}x_{3}, &  & x_{4}\star x_{1}=x_{1}x_{4}+(e^{h}-1)x_{2}x_{3},\\
x_{2}\star x_{3}=x_{2}x_{3}, &  & x_{3}\star x_{2}=x_{2}x_{3},\\
x_{2}\star x_{4}=e^{\frac{h}{2}}x_{1}x_{2}, &  & x_{4}\star x_{2}=e^{-\frac{h}{2}}x_{1}x_{2},\\
x_{3}\star x_{4}=e^{\frac{h}{2}}x_{1}x_{3}, &  & x_{4}\star x_{3}=e^{-\frac{h}{2}}x_{1}x_{3}
\end{eqnarray*}
and we get the relations \begin{eqnarray*}
x_{1}\star x_{2}=e^{h}x_{2}\star x_{1}, &  & x_{1}\star x_{3}=e^{h}x_{3}\star x_{1},\\
x_{3}\star x_{4}=e^{h}x_{4}\star x_{3}, &  & x_{2}\star x_{4}=e^{h}x_{4}\star x_{2},\\
x_{2}\star x_{3}=x_{3}\star x_{2}, &  & [x_{1}\stackrel{\star }{,}x_{4}]=(e^{h}-e^{-h})x_{2}\star x_{3}.
\end{eqnarray*}
In this case the relations are exactly reproduced.

\subsection{Fourdimensional \protect\( q\protect \)-deformed Fock space }

The algebra relations are \cite{Kempf:1992pr,Kempf:1992re}\begin{eqnarray*}
\hat{x}_{1}\hat{x}_{2}=\frac{1}{q}\hat{x}_{2}\hat{x}_{1}, &  & \hat{y}_{1}\hat{y}_{2}=q\hat{y}_{2}\hat{y}_{1},\\
\hat{y}_{1}\hat{x}_{2}=q\hat{x}_{2}\hat{y}_{1}, &  & \hat{y}_{2}\hat{x}_{1}=q\hat{x}_{1}\hat{y}_{2},\\
\hat{y}_{1}\hat{x}_{1}=q^{2}\hat{x}_{1}\hat{y}_{1}+\theta , &  & \hat{y}_{2}\hat{x}_{2}=q^{2}\hat{x}_{2}\hat{y}_{2}+(q^{2}-1)\hat{x}_{1}\hat{y}_{1}+\theta .
\end{eqnarray*}
The Poisson tensor becomes\begin{eqnarray*}
\{x_{1},x_{2}\}=-x_{1}x_{2}, &  & \{y_{1},y_{2}\}=y_{1}y_{2},\\
\{x_{2},y_{1}\}=-x_{2}y_{1}, &  & \{x_{1},y_{2}\}=-x_{1}y_{2},\\
\{y_{1},x_{1}\}=2x_{1}y_{1}+\theta , &  & \{y_{2},x_{2}\}=2(x_{1}y_{1}+x_{2}y_{2})+\theta .
\end{eqnarray*}
After some calculations we find the desired functions\begin{eqnarray*}
f_{1}=-\ln x_{1}, &  & f_{2}=\frac{1}{2}\ln (2x_{1}y_{1}+\theta ),\\
g_{1}=f_{2}-\ln x_{2}, &  & g_{2}=\frac{1}{2}\ln \frac{2(x_{1}y_{1}+x_{2}y_{2})+\theta }{2x_{1}y_{1}+\theta }
\end{eqnarray*}
with \( \{f_{1},f_{2}\}=1,\, \{g_{1},g_{2}\}=1 \), the other brackets
vanish. The Hamiltonian vector fields are\begin{eqnarray*}
X_{1}=X_{f_{1}}=x_{2}\partial _{x_{2}}+y_{2}\partial _{y_{2}}+\frac{2x_{1}y_{1}+\theta }{x_{1}}\partial _{y_{1}}, &  & Y_{1}=X_{f_{2}}=x_{1}\partial _{x_{1}}-y_{1}\partial _{y_{1}},\\
X_{2}=X_{g_{1}}=\frac{2(x_{1}y_{1}+x_{2}y_{2})+\theta }{x_{2}}\partial _{y_{2}}, &  & Y_{2}+X_{g_{2}}=x_{2}\partial _{x_{2}}-y_{2}\partial _{y_{2}}.
\end{eqnarray*}
We calculate\begin{eqnarray*}
X^{n}_{1}(x_{1})=\delta ^{n0}x_{1}, &  & Y^{n}_{1}(x_{1})=x_{1},\\
X^{n}_{1}(x_{2})=x_{2}, &  & Y^{n}_{1}(x_{2})=\delta ^{n0}x_{2},\\
X^{n}_{1}(y_{1})=\frac{2^{n}}{x_{1}}(x_{1}y_{1}+\frac{\theta }{2})\, \, (n>0), &  & Y^{n}_{1}(y_{1})=(-1)^{n}y_{1},\\
X^{n}_{1}(y_{2})=y_{2}, &  & Y^{n}_{1}(y_{2})=\delta ^{n0}y_{2},
\end{eqnarray*}
\begin{eqnarray*}
X^{n}_{2}(x_{1})=\delta ^{n0}x_{1}, &  & Y^{n}_{2}(x_{1})=\delta ^{n0}x_{1},\\
X^{n}_{2}(x_{2})=\delta ^{n0}x_{2}, &  & Y^{n}_{2}(x_{2})=x_{2},\\
X^{n}_{2}(y_{1})=\delta ^{n0}y_{1}, &  & Y^{n}_{2}(y_{1})=\delta ^{n0}y_{1},\\
X^{n}_{2}(y_{2})=\frac{2^{n}}{x_{2}}(x_{1}y_{1}+x_{2}y_{2}+\frac{\theta }{2})\, \, \textrm{for }n>0, &  & Y^{n}_{2}(y_{2})=(-1)^{n}y_{2}.
\end{eqnarray*}
Since we now have four vector fields we use a generalization of the
asymmetric \( \star  \)-product\[
f\star g=\sum _{n=0,m=0}^{\infty }\frac{h^{n+m}}{n!m!}\left( X^{n}_{1}X^{m}_{2}f\right) \left( Y^{n}_{1}Y^{m}_{2}g\right) \]
and get\begin{eqnarray*}
x_{1}\star x_{2}=x_{1}x_{2}, &  & x_{2}\star x_{1}=e^{h}x_{1}x_{2},\\
y_{1}\star y_{2}=y_{1}y_{2}, &  & y_{2}\star y_{1}=e^{-h}y_{1}y_{2},\\
y_{1}\star x_{2}=y_{1}x_{2}, &  & x_{2}\star x_{1}=e^{-h}x_{2}y_{1},\\
y_{2}\star x_{1}=e^{h}x_{1}y_{2}, &  & x_{1}\star y_{2}=x_{1}y_{2},\\
x_{1}\star y_{1}=x_{1}y_{1}, &  & x_{2}\star y_{2}=x_{2}y_{2},
\end{eqnarray*}
\[
y_{1}\star x_{1}=e^{2h}x_{1}y_{1}+\frac{e^{2h}-1}{2}\theta ,\]
\[
y_{2}\star x_{2}=e^{2h}x_{2}y_{2}+(e^{2h}-1)x_{1}y_{1}+\frac{e^{2h}-1}{2}\theta .\]
The algebra relations for this \( \star  \)-product are\begin{eqnarray*}
x_{1}\star x_{2}=e^{-h}x_{2}\star x_{1}, &  & y_{1}\star y_{2}=e^{h}y_{2}\star y_{1},\\
y_{1}\star x_{2}=e^{h}x_{2}\star y_{1}, &  & y_{2}\star x_{1}=e^{h}x_{1}\star y_{2},
\end{eqnarray*}
\[
y_{1}\star x_{1}=e^{2h}x_{1}\star y_{1}+\frac{e^{2h}-1}{2}\theta ,\]
\[
y_{2}\star x_{2}=e^{2h}x_{2}\star y_{2}+(e^{2h}-1)x_{1}\star y_{1}+\frac{e^{2h}-1}{2}\theta .\]
And we get the same relations as in the original algebra if we set\begin{eqnarray*}
q=e^{h}, &  & \theta ^{\prime }=\frac{2}{q^{2}-1}\theta .
\end{eqnarray*}

\noindent \cleardoublepage

\chapter{Geometry}

To study physics in the noncommutative realm, one replaces the commutative
algebra of functions on a space with a noncommutative algebra. Such
a replacement is generally controlled by a parameter so that in some
limit we can get back a commutative space. The same we expect from
theories built an a noncommutative space: In the commutative limit
they should reduce to a meaningful commutative theory. \( \star  \)-products
have shown to be very useful tools for constructing such deformations
since their classical limit is very easily calculated. In this chapter
we apply \( \star  \)-products to the commuting frame formalism developed
in \cite{Madore:2000aq}. For a noncommutative space where the commutator
of the coordinates is constant, the commutative limit of this formalism
is the usual flat spacetime. For noncommutative spaces with more complicated,
non-constant commutators this limit can be a curved manifold. 

After a short introduction to noncommutative differential geometry
where we fix our notation, we will calculate the semi-classical limit
of the commuting frame formalism. In this limit we will see that the
construction of a Poisson tensor for a given frame reduces to solving
a couple of differential equations. The deformation quantization of
the Poisson tensor gives us a \( \star  \)-product and we have constructed
a noncommutative space with desired classical limit. We give some
examples and we will see that the formalism works well in two dimensions,
but has its restrictions in four dimensions. In the end we will construct
Poisson structures having the same symmetries as the Schwarzschild
metric. Here we are able to give a first order differential calculus
with the desired classical limit.

\section{Noncommutative differential geometry}

Locally every manifold can be described by \( N \) coordinates \( x^{i} \).
The set of all derivations acting on functions on the manifold forms
a module over the algebra of functions. The partial derivatives \( \partial _{i} \)
form a basis for all these derivations. Dual to the space of derivations
is the space of one forms. The differentials \( dx^{i} \) form a
basis of this space and they are dual to the partial derivatives\[
dx^{i}(\partial _{j})=\delta _{j}^{i}.\]
 With help of the differentials one is able to introduce the de Rham
differential mapping functions to one-forms\[
df=dx^{i}\partial _{i}f.\]
If one introduces higher order forms with the rule\[
dx^{i}dx^{j}=-dx^{j}dx^{i},\]
one can extend the de Rham differential to a nilpotent graded derivation.
The differential \( d \) and all higher order forms are the exterior
algebra of the manifold. One can show that the whole topology of the
manifold is encoded in the properties of \( d \) or the exterior
algebra respectively.

\bigskip

In noncommutative geometry one replaces the commutative algebra of
functions on the manifold by an noncommutative algebra. Here we again
restrict ourself to algebras defined by relations\begin{equation}
\label{relational_defined_algebra}
\mathcal{A}=\mathbb {C}<\hat{x}^{1},\cdots ,\hat{x}^{N}>/\mathcal{R}.
\end{equation}
To find something similar to differential geometry one can go on and
construct differential calculi to these type of algebras. Just like
in the commutatitve case a differential calculus on \( \mathcal{A} \)
is a \( \mathbb {Z} \)-graded algebra \[
\Omega (\mathcal{A})=\bigoplus _{r\geq 0}\Omega ^{r}(\mathcal{A})\]
where the spaces \( \Omega ^{r}(\mathcal{A}) \) are \( \mathcal{A} \)-bimodules
with \( \Omega ^{0}(\mathcal{A})=\mathcal{A} \). The elements of
\( \Omega ^{r}(\mathcal{A}) \) are called \( r \)-forms. There is
a linear map \[
\hat{d}:\vspace {0.5cm}\Omega ^{r}(\mathcal{A})\, \rightarrow \, \Omega ^{r+1}(\mathcal{A})\]
with the same properties as the commutative differential. It is nilpotent
\[
\hat{d}^{2}=0\]
 and graded\begin{equation}
\label{differential_Leibniz_rule}
\hat{d}(\omega _{1}\omega _{2})=(\hat{d}\omega _{1})\omega _{2}+(-1)^{r}\omega _{1}\hat{d}\omega _{2}
\end{equation}
where \( \omega _{1}\in \Omega ^{r}(\mathcal{A}) \) and \( \omega _{2}\in \Omega (\mathcal{A}) \).
Additionally we assume that \( \hat{d} \) generates the spaces \( \Omega ^{r}(\mathcal{A}) \)
for \( r>0 \) in the sense that\[
\Omega ^{r+1}(\mathcal{A})=\mathcal{A}\cdot \hat{d}\Omega ^{r}(\mathcal{A})\cdot \mathcal{A}.\]
Using the Leibniz rule (\ref{differential_Leibniz_rule}), every element
of \( \Omega ^{r}(\mathcal{A}) \) can be written as a linear combination
of monomials \( f(\hat{x})\hat{d}\hat{x}^{i_{1}}\hat{d}\hat{x}^{i_{2}}\cdots \hat{d}\hat{x}^{i_{r}} \).
The action of \( \hat{d} \) is determined by \[
\hat{d}(f(\hat{x})\hat{d}\hat{x}^{i_{1}}\hat{d}\hat{x}^{i_{2}}\cdots \hat{d}\hat{x}^{i_{r}})=\hat{d}(f(\hat{x}))\hat{d}\hat{x}^{i_{1}}\hat{d}\hat{x}^{i_{2}}\cdots \hat{d}\hat{x}^{i_{r}}.\]

\bigskip

To construct a differential calculus on the algebra \( \mathcal{A} \)
(\ref{relational_defined_algebra}) one starts with a first oder differential
calculus, that means one restricts to the \( 1 \)-forms and the differential\[
\hat{d}:\vspace {0.5cm}\mathcal{A}\rightarrow \Omega ^{1}(\mathcal{A}).\]
 The Leibniz rule (\ref{differential_Leibniz_rule}) and the relations
\( \mathcal{R} \) of the algebra have to be consistent with the bimodule
structure of \( \Omega ^{1}(\mathcal{A}) \). In the following all
relations will be given in terms of commutators \( [\hat{x}^{i},\hat{x}^{j}]=c^{ij}(\hat{x}) \),
therefore\[
[\hat{d}\hat{x}^{i},\hat{x}^{j}]+[\hat{x}^{i},\hat{d}\hat{x}^{j}]=\hat{d}c^{ij}(\hat{x}).\]
For the higher order differential calculus one has to go on in the
same way. The relations of the bimodule structure again have to be
consistent with \( \hat{d}^{2}=0 \) and the Leibniz rule.

\section{Commuting frame formalism}

Surely in commutative differential geometry one is not forced to use
the partial derivatives of the coordinates as basis for the space
of derivations. One can also use a comoving frame\begin{equation}
\label{definition_comoving_frame}
e_{a}=e_{a}{}^{i}\partial _{i},
\end{equation}
where \( e_{a}{}^{i} \) is an invertible matrix. Here \( a=1\cdots N \)
is an index numbering the derivations of the frame. The dual frame
is therefore (\( e_{\nu }{}^{a}e_{a}{}^{\mu }=\delta _{\nu }^{\mu } \))
\[
\theta ^{a}=e^{a}{}_{\mu }(x)dx^{\mu }.\]
The differential can be written only with this new basis elements\[
df=\theta ^{a}e_{a}(f).\]
These formulas all have global extensions to the whole manifold. To
go on we can restrict ourselves to special differential calculi related
to derivations of the algebra. The set of all derivations on the algebra
is not any more a module. But we can take a special set of linear
independent derivations \( \hat{e}_{a} \) and introduce a first order
differential calculus in the following way. The space of one forms
should be a bimodule over the algbra generated by \( \hat{\theta }^{a} \)
and the differential is defined by\[
\hat{d}\hat{f}=\hat{\theta }^{a}\hat{e}_{a}\hat{f}.\]
The components of the frame may be defined by\[
\hat{e}_{a}\hat{x}^{\alpha }=\hat{e}_{a}{}^{\alpha }.\]
Since the \( \hat{e}_{a} \) are derivations it is consistend to let
the \( \hat{\theta }^{a} \) commute with all generators of the algebra\[
\hat{x}^{a}\hat{\theta }^{b}=\hat{\theta }^{b}\hat{x}^{a}\]
The \( \hat{\theta }^{a} \) form a commuting frame for the algebra.
The differential \( \hat{d} \) and the forms \( \hat{\theta }^{a} \)
constitute a first order differential calculus on the algebra. To
construct an analog to the exterior algebra a higher order calculus
is necessary. As we have seen relations for the \( \hat{\theta }^{a} \)
among themselves and \( \hat{d}\hat{\theta }^{a} \)in terms of two
forms have to be given in a consistent way.

\bigskip

A very important structure for physical applications is a metric on
the manifolds which turns it into a (pseudo-) Riemannian manifold.
It can be shown that there always exists a dual frame\[
\theta ^{a}=e^{a}{}_{\mu }(x)dx^{\mu }\]
for which the metric is constant\[
g_{\mu \nu }dx^{\mu }dx^{\nu }=\eta _{ab}\theta ^{a}\theta ^{b}=\eta _{ab}e^{a}{}_{\mu }e^{b}{}_{\nu }dx^{\mu }dx^{\nu }.\]
Note that there are many frames resulting in the same metric. If \( M^{a}{}_{b}(x) \)
is a local \( SO(n) \) gauge transformation the metric stays the
same if we use the transformed frame\begin{equation}
\label{local_frame_transformation}
\theta ^{\prime a}=M^{a}{}_{b}(x)\theta ^{b}.
\end{equation}
 With the above construction it is very easy to generalize this to
the noncommutative case. We simply assume that the frame is always
adapted to the metric\[
\hat{g}=\eta _{ab}\hat{\theta }^{a}\hat{\theta }^{b}.\]
If the derivations are all inner derivations 

\begin{equation}
\label{frame_of_inner_derivations}
\hat{e}_{a}\hat{f}=[\hat{\lambda }_{a},\hat{f}],
\end{equation}
the algebra has to have a trivial center, if the module of one forms
should have the same number of generators as in the commutative case.
Otherwise one is not able to find enough linear independent derivations.
We will call the \( \hat{\lambda }_{a} \) {}``momentum maps''.
The components of the frame are now commutators\[
\hat{e}_{a}\hat{x}^{\alpha }=[\hat{\lambda }_{a},\hat{x}^{\alpha }]=\hat{e}_{a}{}^{\alpha }\]
and the differential may be written as a commutator with a one form\[
d\hat{f}=\hat{\theta }^{a}\hat{e}_{a}\hat{f}=[\hat{\theta }^{a}\hat{\lambda }_{a},\hat{f}].\]
We will call \( \hat{\theta }=\hat{\theta }^{a}\hat{\lambda }_{a} \)
the Dirac operator of the differential calculus. For a Dirac operator
in the sense of \cite{Connes:1994a} more conditions have to be fulfilled.
It is not clear how to generalize the notion of a local frame transformation
(\ref{local_frame_transformation}), since after that the frame will
not commute any more with functions.

\subsection{Semiclassical limit of \protect\( \star \protect \)-product representations}

Here we assume that the noncommutative frame consists of inner derivations
(\ref{frame_of_inner_derivations}). If we have represented the algebra
with a \( \star  \)-product then to first order the algebra relations
define a Poisson structure\[
[x^{\alpha }\stackrel{\star }{,}x^{\beta }]=h\{x^{\alpha },x^{\beta }\}+\cdots =h\Pi ^{\alpha \beta }(x)+\cdots .\]
 Further there are functions \( \lambda _{a} \) that correspond to
the momentum maps of the algebra. We now have\[
\{\lambda _{a},f\}=e_{a}{}^{\mu }\partial _{\mu }f\]
 and we can identify the functions \( e_{a}{}^{\mu } \) with the
coframe of the first section. In the semiclassical limit there is
direct correspondence between a frame and the momentum maps.

On the other hand one can ask the question if it is possible to construct
a Poisson structure and momentum maps that reproduce with the above
formalism a given frame. We know that\begin{equation}
\label{frame_from_inner_derivation}
\{\lambda _{a},x^{\mu }\}=\Pi ^{\alpha \mu }\partial _{\alpha }\lambda _{a}=e_{a}{}^{\mu }
\end{equation}
has to be fullfilled. If we introduce the closed symplectic form \( \omega =\Pi ^{-1} \)
we can translate the last equation into\[
\omega _{\alpha \beta }=-(\partial _{\alpha }\lambda _{a})e^{a}{}_{\beta }.\]
From this we derive two equations that have to be fullfilled for \( \lambda _{a} \)
and the frame \( e^{a} \). \( \omega  \) has to be antisymmetric
and closed \[
S_{\alpha \beta }=(\partial _{\alpha }\lambda _{a})e^{a}{}_{\beta }+(\partial _{\beta }\lambda _{a})e^{a}{}_{\alpha }=\omega _{\beta \alpha }+\omega _{\alpha \beta }=0,\]
\[
d\omega =0.\]
Since the algebra has trivial center it is neccessary that the dimension
of our space is even-dimensional \( N=2M \). The equation \( S=0 \)
has \( \frac{1}{2}N(N+1) \) and \( d\omega =0 \) has \( \left( \stackrel{N}{3}\right) =\frac{1}{6}N(N-1)(N-2) \)
components. Even in two dimensions these are 3 partial differential
equations for the 2 functions \( \lambda _{a} \) . We see that in
higher dimensions it will become very difficult to find a frame in
which the above system of equations may be solved. Further we are
free to make local frame transformations and coordinate transformations
on our classical manifold and there are no hints which frame to use
for quantizing the geometry.

\subsection{The flat metric}

First suppose we want to apply the formalism to the frame \( \theta ^{a}=\delta ^{a}{}_{\alpha }dx^{\alpha } \).
Then for \( S=0 \) we get \[
\partial _{\alpha }\lambda _{\beta }+\partial _{\beta }\lambda _{\alpha }=0.\]
 After some calculations one finds that \[
\lambda _{\alpha }=c_{\alpha \beta }x^{\beta }+\delta _{\alpha }\]
 is the most general solution. \( c \) is a constant antisymmetric
matrix and \( \delta _{\alpha } \) are some constants. For the inverse
of \( \Pi  \) this yields\[
\omega _{\alpha \beta }=c_{\alpha \beta },\]
which is clearly a closed form. We have reproduced the formalism with
constant invertible Poisson tensor.

Secondly we want to investigate the case of a holonomic frame \( \theta ^{a}=\partial _{\alpha }f^{a}dx^{\alpha } \).
After a coordinate transformation one sees immediatly that now\[
\lambda _{a}=c_{ab}f^{a}+\delta _{a}\]
 with \( c \) and \( \delta  \) again constant. Therefore\[
\omega _{\alpha \beta }=\partial _{\alpha }f^{a}\partial _{\beta }f^{b}c_{ab},\]
which is again closed. The first order formalism is invariant under
coordinate transformations, which we could have seen from the definitions
of \( S \) and \( \omega  \), too.

\subsection{\label{two_dimensional_examples}Two dimensional examples}

In the following we will apply the formalism to some two dimensional
examples. We will see that it works quite well in this case since
the equation \( d\omega =0 \) is fulfilled for all two forms in two
dimensions.

\subsubsection{Sphere}

The metric of the sphere in polar coordinates is \[
ds^{2}=d\vartheta ^{2}+\sin ^{2}\vartheta d\varphi ^{2}.\]
The most obvious frame is \begin{eqnarray*}
\theta ^{1}=d\vartheta , &  & \theta ^{2}=\sin \vartheta d\varphi .
\end{eqnarray*}
\( S=0 \) yields\begin{eqnarray*}
\partial _{\vartheta }\lambda _{1}+\partial _{\vartheta }\lambda _{1} & = & 0,\\
\partial _{\vartheta }\lambda _{2}\sin \vartheta +\partial _{\varphi }\lambda _{1} & = & 0,\\
\partial _{\varphi }\lambda _{2}+\partial _{\varphi }\lambda _{2} & = & 0.
\end{eqnarray*}
Therefore\begin{eqnarray*}
\lambda _{1}=-\frac{1}{h}\varphi +\delta _{1}, &  & \partial _{\vartheta }\lambda _{2}(\vartheta )=\frac{1}{h\sin \vartheta }
\end{eqnarray*}
and \( \omega  \) may be calculated\[
\omega _{\vartheta \varphi }=\frac{1}{h}.\]
 In two dimensions every two form is closed and therefore\[
\{\vartheta ,\varphi \}=h\]
fulfills the Jacobi-identities. An algebra having this Poisson structure
is the Heisenberg algebra \( [\hat{\vartheta },\hat{\varphi }]=h \)
in two dimensions. Since the second momentum map \( \lambda _{2} \)
is not a polynomial in the algebra generators a quantization of this
momentum map seems to be very unnatural.

\subsubsection{Constant curvature}

It is known that all two dimensional spaces with constant curvature
can be written in the following form (see e. g. \cite{Madore:2000aq})\[
ds^{2}=f^{2}(u,v)(du^{2}+dv^{2})\]
with\begin{eqnarray*}
f=\frac{1}{1+u^{2}+v^{2}} &  & \textrm{sphere},\\
f=\frac{1}{1-u^{2}-v^{2}} &  & \textrm{Poincare disk},\\
f=\frac{1}{v} &  & \textrm{Lobachewski plane}.
\end{eqnarray*}
For the sphere this are stereographic coordinates. We use the frame\begin{eqnarray*}
\theta ^{1}=fdu, &  & \theta ^{2}=fdv.
\end{eqnarray*}
\( S=0 \) yields \begin{eqnarray*}
\lambda _{1}=-\frac{1}{h}v+\delta _{1}, &  & \lambda _{2}=\frac{1}{h}u+\delta _{2}
\end{eqnarray*}
and we can calculate\[
\{u,v\}=\frac{h}{f}.\]
\bigskip

This Poisson bracket easily may be generalized to algebra relations.
All the momentum maps are linear in the coordinates. Therefore they
correspond to the algebra generators, no ordering ambiguity is present.
For the sphere in stereographic coordinates we get \[
[\hat{u},\hat{v}]=h(1+\hat{u}^{2}+\hat{v}^{2}).\]
Since we have started from the stereographic projection of the sphere
the resulting algebra for the sphere makes no reference to the different
topology. The resulting algebra is a noncommutative sphere with a
hole at the south pole and in this sense a noncommutative plane with
a non constant metric. Similar we get for the Poincare disk\[
\{\hat{u},\hat{v}\}=h(1-\hat{u}^{2}-\hat{v}^{2})\]
 and for the Lobachewski plane \[
\{\hat{u},\hat{v}\}=h\hat{v}.\]
 The resulting noncommutative Lobacheweski plane is known to the literature
\cite{Aghamohammadi:1993}.

\subsubsection{\label{t-r-slice Schwarzschild}Metric with one translational symmetry}

We start with the rather general ansatz \[
ds^{2}=\pm e^{2\psi (r)}dt^{2}+e^{2\phi (r)}dr^{2},\]
which is invariant under translations \( t\rightarrow t+c \) in the
\( t \) direction. We use the frame\begin{eqnarray*}
\theta ^{t}=e^{\psi }dt, &  & \theta ^{r}=e^{\phi }dr.
\end{eqnarray*}
\( S=0 \) yields\begin{eqnarray*}
\lambda _{r}=\frac{1}{h}t, &  & \partial _{r}\lambda _{t}(r)=-\frac{1}{h}e^{\phi -\psi }
\end{eqnarray*}
and the Poisson structure becomes\[
\{t,r\}=he^{-\phi }.\]

\subsubsection{Two dimensional Schwarzschild}

We specialise now to the case of the \( t \)-\( r \)-slice through
the Schwarzschild metric. Here \begin{eqnarray*}
e^{2\psi }=1-\frac{r_{0}}{r}, &  & e^{2\phi }=\frac{1}{1-\frac{r_{0}}{r}}
\end{eqnarray*}
and we get\begin{equation}
\label{2d_schwarzschild_poisson}
\{t,r\}=h\sqrt{1-\frac{r_{0}}{r}}.
\end{equation}
This is well defined if we restrict the manifold to \( t\in \mathbb {R} \)
and \( r\geq r_{0} \). In the limit \( r\rightarrow \infty  \) this
Poisson structure tends to the constant one. The momentum maps are

\[
\lambda _{r}=\frac{1}{h}t,\]
\[
\lambda _{t}=-\frac{1}{h}(1+r_{0}\ln (r-r_{0})).\]
We can write down an algebra which has this Poisson bracket as semiclassical
limit\begin{equation}
\label{2d_schwarzschild_algebra}
[\hat{t},\hat{r}]=h\sqrt{1-\frac{r_{o}}{\hat{r}}},
\end{equation}
where the square root is considered as a Taylor series in \( \hat{r} \).
This algebra may be represented with a \( \star  \)-product constructed
out of the Poisson bracket (\ref{2d_schwarzschild_poisson}). We will
use the resulting algebra in (\ref{4D_rotational_inv_frame}) to construct
a noncommutative frame for the four dimensional Schwarzschild metric.

\subsubsection{Higher order differential calculus}

We want to construct a higher order differential calculus for the
algebra (\ref{2d_schwarzschild_algebra}) for\[
\hat{d}\hat{f}=\hat{\theta }^{r}[\hat{\lambda }_{r},\hat{f}]+\hat{\theta }^{t}[\lambda _{t},\hat{f}]\]
with\begin{eqnarray*}
\hat{\lambda }_{r}=\frac{1}{h}\hat{t}, &  & \hat{\lambda }_{t}=-\frac{1}{h}(1+r_{0}\ln (\hat{r}-r_{0})).
\end{eqnarray*}
First we calculate\begin{eqnarray*}
\hat{d}\hat{t}=\hat{\theta }^{t}\frac{1}{\sqrt{1-\frac{r_{0}}{\hat{r}}}}, &  & \hat{d}\hat{r}=\hat{\theta }^{r}\sqrt{1-\frac{r_{0}}{\hat{r}}}.
\end{eqnarray*}
With this we get (\( \rho (\hat{r})=\frac{r_{0}}{2\hat{r}^{2}}\frac{1}{\sqrt{1-\frac{r_{0}}{\hat{r}}}} \))
\begin{eqnarray*}
{}[\hat{t},\hat{d}\hat{t}]=h\, \hat{d}\hat{t}\, \rho (\hat{r}), &  & {}[\hat{r},\hat{d}\hat{t}]=0,\\
{}[\hat{t},\hat{d}\hat{r}]=h\, \hat{d}\hat{r}\, \rho (\hat{r}), &  & {}[\hat{r},\hat{d}\hat{r}]=0.
\end{eqnarray*}
From this we derive\[
2\hat{d}\hat{t}\, \hat{d}\hat{t}=4\hat{d}\hat{t}\, \hat{d}\hat{r}\, \rho ^{\prime }(\hat{r}),\]
\[
\hat{d}\hat{r}\, \hat{d}\hat{t}+\hat{d}\hat{t}\, \hat{d}\hat{r}=0,\]
\[
\hat{d}\hat{r}\, \hat{d}\hat{t}+\hat{d}\hat{t}\, \hat{d}\hat{r}=\hat{d}\hat{r}\, \hat{d}\hat{r}\, \rho (\hat{r}),\]
\[
\hat{d}\hat{r}\, \hat{d}\hat{r}=0.\]
The first condition implies\[
0=[\hat{t},\hat{d}\hat{t}\, \hat{d}\hat{t}]=4\hat{d}\hat{t}\, \hat{d}\hat{r}\, [\hat{t},\rho ^{\prime }(\hat{r})].\]
Therefore\[
\hat{d}\hat{t}\, \hat{d}\hat{r}=0,\]
which has not the desired classical limit.

\subsection{Metrics in four dimensions}

Although the formalism works well in two dimensions we will see that
this is not the case in four dimensions. We tried to solve the system
of partial differential equations for diverse physically interesting
frames but we never were really successful.

\subsubsection{Schwarzschild metric}

The best known form of the Schwarzshild metric is\[
ds^{2}=-(1-\frac{r_{0}}{r})dt^{2}+\frac{1}{1-\frac{r_{0}}{r}}dr^{2}+r^{2}d\Omega ^{2}.\]
Here the most obvious frame is\begin{eqnarray*}
\theta ^{t}=e^{\psi }dt, &  & \theta ^{r}=e^{-\psi }dt,\\
\theta ^{\vartheta }=d\vartheta , &  & \theta ^{\varphi }=\sin \vartheta d\varphi ,
\end{eqnarray*}
with \( e^{\psi }=\sqrt{1-\frac{r_{o}}{r}} \). Here the \( S=0 \)
equations have no solution for arbitrary \( \psi  \) except \( \psi =0 \).
In another coordinate sytem the Schwarzschild metric becomes \cite{Nurowski:1999}
\[
ds^{2}=-dt^{2}+(dx^{i}-x^{i}\sqrt{\frac{2m}{r^{3}}}dt)^{2}.\]
A more general frame ist \begin{eqnarray*}
\theta ^{0}=dt, &  & \theta ^{i}=dx^{i}-x^{i}f(r)dt,
\end{eqnarray*}
with \( f=\sqrt{\frac{2m}{r^{3}}} \). For general \( f \) again
the \( S=0 \) equations imply \( f=0 \).

\subsubsection{Kasner metric}

One form of the Kasner metric is \cite{Nurowski:1999} \[
ds^{2}=-dt^{2}+(dx^{i}-p^{i}_{j}\frac{x^{j}}{t}dt)^{2}.\]
To be more flexible we start with the following frame\begin{eqnarray*}
\theta ^{0}=dt, &  & \theta ^{i}=dx^{i}+P^{i}{}_{j}(t)x^{j}dt.
\end{eqnarray*}
The \( S=0 \) equations become\begin{eqnarray*}
\partial _{0}\lambda _{0}+\partial _{0}\lambda _{i}P^{i}{}_{j}x^{j} & = & 0,\\
\partial _{0}\lambda _{i}+\partial _{i}\lambda _{l}P^{l}{}_{m}x^{m}+\partial _{i}\lambda _{0} & = & 0,\\
\partial _{i}\lambda _{j}+\partial _{j}\lambda _{i} & = & 0.
\end{eqnarray*}
From the last equation we deduce that\[
\lambda _{i}=c_{jk}(t)x^{k}+\delta _{i}(t).\]
We make the ansatz\begin{eqnarray*}
\lambda _{x}=-\alpha (t)y & \lambda _{y}=\alpha (t)x & \lambda _{z}=\beta (t),
\end{eqnarray*}
\[
P(t)=\left( \begin{array}{ccc}
p(t) & 0 & 0\\
 & p(t) & 0\\
 &  & q(t)
\end{array}\right) \]
and the \( S=0 \) equations reduce to\begin{eqnarray*}
\dot{\alpha }=p\alpha , &  & \ddot{\beta }=-q\dot{\beta },
\end{eqnarray*}
with \[
\lambda _{0}=-\dot{\beta }z+\gamma .\]
The two form \( \omega  \) becomes \begin{eqnarray*}
\omega  & = & -\dot{\alpha }(ydx-xdy)\wedge dt-\alpha dx\wedge dy+\dot{\beta }dz\wedge dt\\
 & = & (xdy-ydx)\wedge d\alpha -\alpha dx\wedge dy+dz\wedge d\beta .
\end{eqnarray*}
It is not closed. To cure this problem we make a slight modification\[
\omega =(axdy-(1-a)ydx)\wedge \alpha -\alpha dx\wedge dy+dz\wedge d\beta \]
with \( a \) some constant. We get now\begin{eqnarray*}
\{z,x\}=a\frac{\dot{\alpha }}{\alpha \dot{\beta }}x, &  & \{z,t\}=-\frac{1}{\dot{\beta }},\\
\{z,y\}=a\frac{\dot{\alpha }}{\alpha \dot{\beta }}y, &  & \{x,y\}=\frac{1}{\alpha }.
\end{eqnarray*}
If we use the definitions for the \( \lambda _{a} \) from above,
we can calculate the coframe\[
\{\lambda _{a},x^{\alpha }\}=\left( \begin{array}{cccc}
1 & -a\frac{\dot{\alpha }}{\alpha }x & -(1-a)\frac{\dot{\alpha }}{\alpha }y & -\frac{\ddot{\beta }}{\dot{\beta }}z\\
0 & 1 & 0 & -a\frac{\dot{\alpha }}{\dot{\beta }}y\\
0 & 0 & 1 & (1-a)\frac{\dot{\alpha }}{\dot{\beta }}x\\
0 & 0 & 0 & 1
\end{array}\right) .\]
The frame becomes \begin{eqnarray*}
\theta ^{0} & = & dt,\\
\theta ^{x} & = & dx+a\frac{\dot{\alpha }}{\alpha }x\, dt,\\
\theta ^{y} & = & dy+(1-a)\frac{\dot{\alpha }}{\alpha }y\, dt,\\
\theta ^{z} & = & dz+(\frac{\ddot{\beta }}{\dot{\beta }}-(a^{2}-(1-a)^{2})\frac{\dot{\alpha }^{2}}{\alpha \dot{\beta }}xy)dt\\
 &  & +\frac{\dot{\alpha }}{\dot{\beta }}(aydx-(1-a)xdy.
\end{eqnarray*}
 We see that with the commuting frame formalism it is impossible to
construct a frame, becoming in the classical limit the above given
frame of the Kasner metric.

\bigskip

In \cite{Maceda:2003xr} a noncommutative version of the Kasner algebra
was constructed using the commuting frame formalism we started from.
To relate our example to the one there we now further assume\begin{eqnarray*}
a\frac{\dot{\alpha }}{\alpha }=\frac{p_{1}}{t}, &  & (1-a)\frac{\dot{\alpha }}{\alpha }=\frac{p_{2}}{t}.
\end{eqnarray*}
Therefore\begin{eqnarray*}
\frac{p_{1}}{a}=\frac{p_{2}}{1-a} &  & \alpha =t^{\frac{p_{1}}{a}}.
\end{eqnarray*}
Further we set \[
\frac{\ddot{\beta }}{\dot{\beta }}=\frac{p_{3}}{t},\]
so\[
\dot{\beta }=t^{p_{3}}.\]
Now\[
\frac{\dot{\alpha }}{\dot{\beta }}=\frac{p_{1}}{a}t^{\frac{p_{1}}{a}-p_{3}-1}\]
and the frame becomes\begin{eqnarray*}
\theta ^{0} & = & dt,\\
\theta ^{x} & = & dx+p_{1}\frac{x}{t}dt,\\
\theta ^{y} & = & dy+p_{1}\frac{1-a}{a}\frac{y}{t}dt,\\
\theta ^{z} & = & dz+p_{3}\frac{z}{t}dt+(a^{2}-(1-a)^{2})\frac{p_{1}^{2}}{a^{2}}t^{\frac{p_{1}}{a}-p_{3}}xydt\\
 &  & +\frac{p_{1}}{a}t^{\frac{p_{1}}{a}-p_{3}-1}(aydx-(1-a)xdy).
\end{eqnarray*}
 The last term of \( \theta ^{z} \) has the same order of magnitude
as the ordinary deviation from the flat metric in \( \theta ^{x} \)
and \( \theta ^{y} \). 

With this solution the commutation relations become\begin{eqnarray*}
\{z,x\}=p_{1}t^{-p_{3}-1}x, &  & \{z,t\}=t^{-p_{3}},\\
\{z,y\}=\frac{1-a}{a}p_{1}t^{-p_{3}-1}y, &  & \{x,y\}=t^{-\frac{p_{1}}{a}}.
\end{eqnarray*}
If one sets the parameters \( p_{1} \) and \( p_{3} \) to zero the
above relations become constant commutator relations between the coordinates.
If we let \( t \) go to infinity we can fit the parameters \( p_{1} \),\( p_{3} \)
and \( a \) in such a way that all relations vanish.

\section{\protect\( SO(3)\times T\protect \) invariant Poisson structures
and algebras}

In this section we try to construct algebras having the same symmetries
as the Schwarzschild metric. Meaning invariance under rotations and
time translations. For this we first construct non-degenerate Poisson
structures with these properties. Since we know that every Poisson
structure may be quantized by a \( \star  \)-product we are able
to write down all possible algebras with trivial center. We will see
that these are quite unique. With the help of one of these algebras,
we propose a non-commuting frame, that becomes in the classical limit
a frame for the Schwarzschild metric.

\subsection{The Poisson structures}

We start with following Ansatz\begin{eqnarray*}
\{x^{i},x^{j}\} & = & \beta (r,t)\epsilon ^{ij}{}_{k}x^{k},\\
\{t,x^{i}\} & = & \alpha (r,t)x^{i}.
\end{eqnarray*}
where \( i=x,y,z \) and \( r=\sqrt{x^{2}+y^{2}+z^{2}} \). These
equations are obviously covariant under rotations. It would be invariant
under translations in the \( t \) direction if \( \alpha  \) and
\( \beta  \) do not depend on \( t \), but we keep them in this
form to be more general. The bracket with a function \( f \) is\begin{eqnarray*}
\{x^{i},f\} & = & -\alpha x^{i}\partial _{t}f+\beta \epsilon ^{ij}{}_{k}\partial _{j}f,\\
\{t,f\} & = & \alpha x^{i}\partial _{i}f.
\end{eqnarray*}
 With this the Jacobi identities are\begin{eqnarray*}
\{x^{i},\{x^{j},x^{k}\}\}+\textrm{cyc}. & = & -\alpha \partial _{t}\beta \epsilon ^{ijk}r^{2},\\
\{t,\{x^{j},x^{k}\}\}+\textrm{cyc}. & = & \alpha (r\partial _{r}\beta -\beta )\epsilon ^{ij}{}_{k}x^{k}.
\end{eqnarray*}
The brackets become a Poisson structure if the right hand side of
the above equations vanishes. This is the case if either\begin{eqnarray*}
\alpha =0, &  & \beta =\beta (r,t)
\end{eqnarray*}
or\begin{eqnarray*}
\alpha =\alpha (r,t), &  & \beta =br,
\end{eqnarray*}
where \( b \) is a constant. In the first case \( t \) commutes
with all functions. Only in the second case, there is the possibility
for all derivations to be inner derivations. Therefore we will later
restrict us to this case.

\subsection{\label{rotational_invariant_algebra}Algebras and isomorphisms}

After quantization the Poisson structures become algebras. Note that
in both cases \[
\{r,x^{i}\}=0\]
 and there is no ordering problem on the right hand side of the algebra
relations if \( \alpha  \) and \( \beta  \) do not depend on \( t \).
We will assume from now on that \( \beta  \) does not depend on \( \hat{t} \)
. The first case is now\begin{eqnarray*}
{[\hat{x}^{i},\hat{x}^{j}]}=\beta (\hat{r})\epsilon ^{ij}{}_{k}\hat{x}^{k}, &  & {[\hat{t},\hat{x}^{i}]}=0
\end{eqnarray*}
with \( \beta \neq b\hat{r} \). If we define \( \hat{s}^{i}=\beta ^{-1}(\hat{r})\hat{x}^{i} \)
these relations become\begin{eqnarray*}
{[\hat{s}^{i},\hat{s}^{j}]}=\epsilon ^{ij}{}_{k}\hat{s}^{k}, &  & {[x^{0},\hat{s}^{i}]}=0.
\end{eqnarray*}
Therefore this algebra is isomorphic to \( U(su(2))\times \mathbb {C} \).

The second case is\begin{eqnarray*}
{[\hat{x}^{i},\hat{x}^{j}]}=b\hat{r}\epsilon ^{ij}{}_{k}\hat{x}^{k}, &  & {[\hat{t},\hat{x}^{i}]}=\alpha (\hat{r},\hat{t}).
\end{eqnarray*}
If we again define \( \hat{s}^{i}=\hat{r}^{-1}\hat{x}^{i} \) we get
the constraint \( \hat{s}^{i}\hat{s}_{i}=1 \). The relations become\begin{eqnarray*}
{[\hat{s}^{i},\hat{s}^{j}]}=b\epsilon ^{ij}{}_{k}\hat{s}^{k}, &  & {[\hat{s}^{i},\hat{r}]}=0,\\
{[\hat{t},\hat{s}^{i}]}=0, &  & {[\hat{t},\hat{r}]}=\alpha (\hat{r},\hat{t})\hat{r}.
\end{eqnarray*}
This algebra is isomorphic to \( S_{b}\times \mathbb {C}^{2} \) for
\( \alpha =0 \) or \( S_{b}\times A \) otherwise. For \( b=\sqrt{\frac{4}{N^{2}-1}} \)
and \( N \) an integer \( S_{b} \) is a fuzzy sphere with deformation
parameter \( b \) and has finite dimensional representations. \( A \)
can be any two dimensional algebra.

\subsection{\label{4D_rotational_inv_frame}Rotational invariant metrics with
a minimal noncommuting frame}

In the last section we have seen that the second algebra decomposes
into a three dimensional rotational covariant algebra and a two dimensional
algebra for which we can now use the algebra from Section \ref{t-r-slice Schwarzschild}.
The relations become now (\( \hat{x}^{i}=\hat{r}\hat{s}^{i} \), \( \delta _{ij}\hat{s}^{i}\hat{s}^{j}=1 \))
\begin{eqnarray*}
{[\hat{s}^{i},\hat{s}^{j}]}=b\epsilon ^{ij}{}_{k}\hat{s}^{k}, &  & {[\hat{s}^{i},\hat{r}]}=0,\\
{[\hat{t},\hat{s}^{i}]}=0, &  & {[\hat{t},\hat{r}]}=he^{-\phi (\hat{r})}.
\end{eqnarray*}
We now use five inner derivations to define a frame\begin{eqnarray*}
\hat{e}^{0}_{i} & = & {}[\frac{\hat{s}^{i}}{b},\cdot ]\rightarrow -\epsilon _{ij}{}^{k}x^{j}\partial _{k},\\
\hat{e}_{r} & = & {}[\hat{\lambda }_{r},\cdot ]={}[\frac{\hat{t}}{h},\cdot ]\rightarrow e^{-\phi }\partial _{r},\\
\hat{e}_{t} & = & {}[\hat{\lambda }_{t},\cdot ]\rightarrow e^{-\psi }\partial _{t}.
\end{eqnarray*}
The arrows indicate the classical limit. \( \hat{\lambda }_{t} \)
is defined in the classical limit via \[
\partial _{r}\lambda _{t}(r)=-\frac{1}{h}e^{\phi -\psi }.\]
In the classical limit \( x^{i}e_{i}=0 \) and the derivations are
linearly dependent. The dual one forms to the \( \hat{e}_{i} \) are
well known, they form the differential calculus on the fuzzy sphere.
The dual forms of \( \hat{e}_{r} \) and \( \hat{e}_{t} \) are easily
constructed. The classical limit of these forms is \begin{eqnarray*}
\hat{\theta }_{0}^{i} & \rightarrow  & -\frac{1}{r^{2}}\epsilon ^{i}{}_{jk}x^{j}dx^{k},\\
\hat{\theta }^{r} & \rightarrow  & e^{\phi }dr,\\
\hat{\theta }^{t} & \rightarrow  & e^{\psi }dt.
\end{eqnarray*}
We know \[
\delta _{ij}\hat{\theta }_{0}^{i}\hat{\theta }_{0}^{j}\rightarrow d\Omega ^{2}.\]
We define now some forms that do not commute with functions\[
\hat{\theta }^{i}=\hat{r}\hat{\theta }_{0}^{i},\]
\[
\hat{f}\hat{\theta }^{i}=\hat{\theta }^{i}\hat{r}^{-1}\hat{f}\hat{r}.\]
Only \( \hat{t} \) does not commute with \( \hat{\theta }^{i} \).
With these forms we can now construct a noncommutative version of
the four dimensional metric\[
ds^{2}=-(\hat{\theta }^{t})^{2}+(\hat{\theta }^{r})^{2}+\delta _{ij}\hat{\theta }^{i}\hat{\theta }^{j}\rightarrow -e^{2\psi }dt^{2}+e^{2\phi }dr^{2}+r^{2}d\Omega ^{2}.\]
Note that dual to the \( \hat{\theta }^{i} \) are the following deformed
derivations\[
\hat{e}_{i}={}\hat{r}^{-1}[\frac{\hat{s}^{i}}{b},\cdot ]\rightarrow \frac{1}{r}\epsilon _{ij}{}^{k}x^{j}\partial _{k}\]
with \[
\hat{e}_{i}(\hat{f}\cdot \hat{g})=\hat{e}_{i}\hat{f}\cdot \hat{g}+\hat{r}^{-1}\hat{f}\hat{r}\cdot \hat{e}_{i}\hat{g}.\]
The inner isomorphism defined by \( \hat{r} \) is\[
\hat{r}^{-1}\hat{t}\hat{r}=\hat{t}+h\frac{e^{-\phi (\hat{r})}}{\hat{r}}.\]
In the case of the Schwarzschild metric this becomes\[
\hat{r}^{-1}\hat{t}\hat{r}=\hat{t}+\frac{h}{\hat{r}}\sqrt{1-\frac{r_{o}}{\hat{r}}}.\]
Again this only makes sense if the spectrum of \( \hat{r} \) has
no values smaller than \( r_{o} \). In the limit \( r\rightarrow \infty  \)
this tends to the identity.

\bigskip

If we define the one form\[
\hat{\theta }=\hat{x}^{i}\hat{\theta }^{i}+\hat{\lambda }_{r}\hat{\theta }^{r}+\hat{\lambda }_{t}\hat{\theta }^{t}\]
it follows that (\( a=r,t \))\[
[\hat{\theta },\hat{f}]=\hat{\theta }^{i}\hat{e}_{i}\hat{f}+\hat{\theta }^{a}\hat{e}_{a}\hat{f}=\hat{d}\hat{f}\]
\( \hat{\theta } \) is the Dirac operator of the differential calculus.

\bigskip

It would be nice to have a higher order differential calculus for
the first order one. To construct this we note that\begin{eqnarray*}
{}[\hat{e}_{i},\hat{e}_{j}] & = & \frac{1}{\hat{r}^{2}}\epsilon _{ij}{}^{k}\hat{e}_{k},\\
{}[\hat{e}_{t},\hat{e}_{i}] & = & 0,\\
{}[\hat{e}_{r},\hat{e}_{i}] & = & (\hat{e}_{r}\frac{1}{\hat{r}})\hat{r}\hat{e}_{i}=-\frac{1}{\hat{r}}\hat{e}_{i},\\
{}[\hat{e}_{r},\hat{e}_{t}] & = & -\frac{1}{h}[e^{-\psi (\hat{r})},\cdot ].
\end{eqnarray*}
It is consistent to define\begin{eqnarray*}
\hat{\theta }^{i}\hat{\theta }^{j} & = & -\hat{\theta }^{j}\hat{\theta }^{i},\\
\hat{\theta }^{a}\hat{\theta }^{i} & = & -\hat{\theta }^{i}\hat{\theta }^{a}
\end{eqnarray*}
and\begin{eqnarray*}
\hat{d}\hat{\theta }^{k} & = & \frac{1}{2\hat{r}^{2}}\epsilon _{ij}{}^{k}\hat{\theta }^{i}\hat{\theta }^{j}+\frac{1}{\hat{r}}\hat{\theta }^{r}\hat{\theta }^{k},\\
\hat{d}\hat{\theta }^{r} & = & 0.
\end{eqnarray*}
The claim \( \hat{d}^{2}=0 \) reduces to\[
d\hat{\theta }^{t}\hat{e}_{t}\hat{f}-\hat{\theta }^{a}\hat{\theta }^{b}\hat{e}_{b}\hat{e}_{a}\hat{f}=0.\]
In (1.5.4) we have shown that this implies \( (\hat{\theta }^{r})^{2}=(\hat{\theta }^{t})^{2}=\hat{\theta }^{r}\hat{\theta }^{t}=\hat{\theta }^{t}\hat{\theta }^{r}=\hat{d}\hat{\theta }^{t}=0 \).
Again we are not able to extend the first order differential calculus
to a differnential calculus of higher order.

\cleardoublepage

\chapter{Derivations of \protect\( \star \protect \)-products}

We have seen that if we want to make noncommutative geometry in the
\( \star  \)-product formulation we have been very successful interpreting
derivations as a noncommutative analog of frames. In the last chapter
we used invertible Poisson structures, the resulting algebras therefore
had trivial center and all derivations have been inner derivations.
To be more general we now relax our restrictions and take degenerated
Poisson structures in consideration. Consequently the algebras will
have central elements. This is due to the fact that in the classical
limit all vector fields formed from the inner derivations will commute
with the functions associated to these central elements. We are forced
to deal with outer derivations and in this chapter we first will examine
derivations of \( \star  \)-products without referring to any abstract
algebraic construction. 

In the beginning we will introduce Kontsevich's Formality map \cite{Kontsevich:1997vb}
to make some statements about derivations on quantized Poisson manifolds.
Then we will calculate the general form of derivations on the Weyl
ordered \( \star  \)-product. Knowing the restrictions and the form
of the derivations we are able to construct an interesting differential
calculus on the \( \star  \)-product algebra. This will be used later
to make contact with Seiberg-Witten gauge theory. After some considerations
how to construct traces for \( \star  \)-product algebras we are
able to write down consistent actions for noncommutative theories
becoming non abelian gauge theories on curved manifolds in the classical
limit.

\section{\label{Formality map}The Formality Map}

Kontsevich's Formality map \cite{Kontsevich:1997vb} is a very useful
tool for studying the relations between Poisson tensors and \( \star  \)-products.
To make use of the Formality map we first want to recall some definitions.
A polyvector field is a skew-symmetric tensor in the sense of differential
geometry. Every \( n \)-polyvector field \( \alpha  \) may locally
be written as\[
\alpha =\alpha ^{i_{1}\dots i_{n}}\, \partial _{i_{1}}\wedge \dots \wedge \partial _{i_{n}}.\]
 We see that the space of polyvector fields can be endowed with a
grading \( n \). For polyvector fields there is a grading respecting
bracket that in a natural way generalizes the Lie bracket \( [\cdot ,\cdot ]_{L} \)
of two vector fields, the Schouten-Nijenhuis bracket (see \ref{Schouten-Nijenhuis bracket}).
If \( \pi  \) is a Poisson tensor, the Hamiltonian vector field \( H_{f} \)
for a function \( f \) is\[
H_{f}={[\pi ,f]_{S}}=-\pi ^{ij}\partial _{i}f\partial _{j}.\]
Note that \( [\pi ,\pi ]_{S}=0 \) is the Jakobi identity of a Poisson
tensor. \bigskip

On the other hand a \( n \)-polydifferential operator is a multilinear
map that maps \( n \) functions to a function. For example, we may
write a \( 1 \)-polydifferential operator \( D \) as\[
D(f)=D_{0}f+D^{i}_{1}\partial _{i}f+D^{ij}_{2}\partial _{i}\partial _{j}f+\dots .\]
The ordinary multiplication \( \cdot  \) is a \( 2 \)-polydifferential
operator. It maps two functions to one function. Again the number
\( n \) is a grading on the space of polydifferential operators.
Now the Gerstenhaber bracket (see \ref{Gerstenhaber bracket}) is
natural and respects the grading. 

The Formality map is a collection of skew-symmetric multilinear maps
\( U_{n} \), \( n=0,1,\dots  \), that maps \( n \) polyvector fields
to a \( m \)-differential operator. To be more specific let \( \alpha _{1},\dots ,\alpha _{n} \)
be polyvector fields of grade \( k_{1},\dots ,k_{n} \). Then \( U_{n}(\alpha _{1},\dots ,\alpha _{n}) \)
is a polydifferential operator of grade\[
m=2-2n+\sum _{i}k_{i}.\]
 In particular the map \( U_{1} \) is a map from a \( k \)-vectorfield
to a \( k \)-differential operator. It is defined by\[
U_{1}(\alpha ^{i_{1}\dots i_{n}}\partial _{i_{1}}\wedge \dots \wedge \partial _{i_{n}})(f_{1},\dots ,f_{n})=\alpha ^{i_{1}\dots i_{n}}\partial _{i_{1}}f_{1}\cdot \dots \cdot \partial _{i_{n}}f_{n}.\]
The formality maps \( U_{n} \) fulfill the formality condition \cite{Kontsevich:1997vb,Arnal:2000hy}

\begin{equation}
\label{formality cond}
Q'_{1}U_{n}(\alpha _{1},\ldots ,\alpha _{n})+\frac{1}{2}\sum _{{I\sqcup J=\{1,\ldots ,n\}\atop I,J\neq \emptyset }}\epsilon (I,J)Q'_{2}(U_{|I|}(\alpha _{I}),U_{|J|}(\alpha _{J}))
\end{equation}
\[
=\frac{1}{2}\sum _{i\neq j}\epsilon (i,j,\ldots ,\hat{i},\ldots ,\hat{j},\ldots ,n)U_{n-1}(Q_{2}(\alpha _{i},\alpha _{j}),\alpha _{1},\ldots ,\widehat{\alpha }_{i},\ldots ,\widehat{\alpha }_{j},\ldots ,\alpha _{n}).\]
The hats stand for omitted symbols, \( Q'_{1}(\Upsilon )=[\Upsilon ,\mu ] \)
with \( \mu  \) being ordinary multiplication and \( Q'_{2}(\Upsilon _{1},\Upsilon _{2})=(-1)^{(|\Upsilon _{1}|-1)|\Upsilon _{2}|}[\Upsilon _{1},\Upsilon _{2}]_{G} \)
with \( |\Upsilon _{s}| \) being the degree of the polydifferential
operator \( \Upsilon _{s} \), i.e. the number of functions it is
acting on. For polyvectorfields \( \alpha ^{i_{1}\ldots i_{k_{s}}}_{s}\partial _{i_{1}}\wedge \ldots \wedge \partial _{i_{k_{s}}} \)
of degree \( k_{s} \) we have \( Q_{2}(\alpha _{1},\alpha _{2})=-(-1)^{(k_{1}-1)k_{2}}[\alpha _{2},\alpha _{1}]_{S} \). 

For a bivectorfield \( \pi  \) we can now define a bidifferential
operator

\[
\star =\sum ^{\infty }_{n=0}\frac{1}{n!}U_{n}(\pi ,\, \ldots ,\pi )\]
i.e.

\[
f\star g=\sum ^{\infty }_{n=0}\frac{1}{n!}U_{n}(\pi ,\, \ldots ,\pi )(f,g).\]
We further define the special polydifferential operators

\begin{eqnarray*}
\Phi (\alpha ) & = & \sum ^{\infty }_{n=1}\frac{1}{(n-1)!}U_{n}(\alpha ,\pi ,\, \ldots ,\pi ),\\
\Psi (\alpha _{1},\alpha _{2}) & = & \sum ^{\infty }_{n=2}\frac{1}{(n-2)!}U_{n}(\alpha _{1},\alpha _{2},\pi ,\ldots ,\pi ).
\end{eqnarray*}
For \( g \) a function, \( X \) and \( Y \) vectorfields and \( \pi  \)
a bivectorfield we see that\[
\delta _{X}=\Phi (X)\]
is a 1-differential operator and that both \( \phi (g) \) and \( \Psi (X,Y) \)
are functions. 

We now use the formality condition (\ref{formality cond}) to calculate\begin{eqnarray}
{[\star ,\star ]_{G}} & = & \Phi ([\pi ,\pi ]_{S}),\label{comm star star} \\
{[\Phi (f),\star ]_{G}} & = & -\Phi ([f,\pi ]_{S}),\label{comm phi f star} \\
{[\delta _{X},\star ]_{G}} & = & \Phi ([X,\pi ]_{S}),\label{comm delta X star} 
\end{eqnarray}
 \[
{[\delta _{X},\delta _{Y}]_{G}+[\Psi (X,Y),\star ]_{G}}\, \, \, \, \, \, \, \, \, \, \, \, \, \, \, \, \, \, \, \, \, \, \, \, \, \, \, \, \, \, \, \, \]
\begin{equation}
\label{comm delta X delta Y}
\, \, \, \, \, \, \, \, \, \, \, \, \, \, \, \, \, \, \, \, \, \, \, \, \, \, \, \, \, \, \, \, =\delta _{[X,Y]_{S}}+\Psi ([\theta ,Y]_{S},X)-\Psi ([\theta ,X]_{S},Y),
\end{equation}
\[
{}[\Phi (\pi ),\Phi (g)]_{G}+[\Psi (\pi ,g),\star ]_{G}\, \, \, \, \, \, \, \, \, \, \, \, \, \, \, \, \, \, \, \, \, \, \, \, \, \, \, \, \, \, \, \, \]
\[
\, \, \, \, \, \, \, \, \, \, \, \, \, \, \, \, \, \, \, \, \, \, \, \, \, \, \, \, \, \, \, \, =-\delta _{[\pi ,g]_{S}}-\Psi ([\theta ,g]_{S},\pi )-\Psi ([\theta ,\pi ]_{S},g),\]
\[
{}[\delta _{X},\Phi (g)]_{G}\, \, \, \, \, \, \, \, \, \, \, \, \, \, \, \, \, \, \, \, \, \, \, \, \, \, \, \, \, \, \, \, \, \, \, \, \, \, \, \, \, \, \, \, \, \, \, \, \, \, \, \]
\begin{equation}
\label{comm delta X Phi g}
\, \, \, \, \, \, \, \, \, \, \, \, \, \, \, \, \, \, \, \, \, \, \, \, \, \, =\phi ([X,g]_{S})-\Psi ([\theta ,g]_{S},X)-\Psi ([\theta ,X]_{S},g).
\end{equation}
If \( \pi  \) is Poisson, i. e. \( [\pi ,\pi ]_{S}=0 \) and if \( X \)
and \( Y \) are Poisson vector fields, i. e. \( [X,\pi ]_{S}=[Y,\pi ]_{S}=0 \),
the relations (\ref{comm star star}) to (\ref{comm delta X delta Y})
become \begin{eqnarray}
f\star (g\star h) & = & (f\star g)\star h,\nonumber \\
\delta _{H_{f}}(g) & = & -[\Phi (f)\stackrel{\star }{,}g],\nonumber \\
\delta _{X}(f\star g) & = & \delta _{X}(f)\star g+f\star \delta _{X}(g),\label{formality star derivation} \\
({[\delta _{X},\delta _{Y}]-\delta _{[X,Y]_{L}}})(g) & = & [\Psi (X,Y)\stackrel{\star }{,}g].\nonumber \label{formality star derivation} 
\end{eqnarray}
when evaluated on functions. \( [\cdot ,\cdot ] \) are now ordinary
brackets. \( \star  \) defines an associative product, the Hamiltonian
vector fields are mapped to inner derivations and Poisson vector fields
are mapped to outer derivations of the \( \star  \)-product. Additionally
the map \( \delta  \) preserves the bracket up to an inner derivation.
The last equation can be cast into a form which we will use in the
definition of deformed forms in (\ref{nc_forms}) \[
{[\delta _{X},\delta _{Y}]}=\delta _{[X,Y]_{\star }}\]
with\[
{[X,Y]_{\star }}=[X,Y]_{L}+H_{\Phi ^{-1}\Psi (X,Y)}.\]
For every Poisson manifold there not only exists a quantization with
the Formality \( \star  \)-product, but additionally there is a deformation
of the Lie bracket going with the derivations of this \( \star  \)-product.

\section{\label{weyl derivations}Weyl-ordered \protect\( \star \protect \)-products}

The formality \( \star  \)-product is the obvious choice if we start
from a Poisson manifold and therefore if we only need a \( \star  \)-product
that reproduces the Poisson structure to first order. But starting
from an algebra, we need a \( \star  \)-product that reproduces the
whole algebra, not just the Poisson structure. If we extract a Poisson
structure from an algebra generated by noncommutative coordinates
fulfilling certain commutation relations, there's no way of knowing
if the formality \( \star  \)-product built from this Poisson structure
will reproduce the commutation relations or not. For this purpose
the Weyl ordered \( \star  \)-product (\ref{wely_star_construction})
is more suitable. In the following we will calculate the derivations
for this type of \( \star  \)-product.

We have shown that for the Formality \( \star  \)-product there exists
a map \( \delta  \) from the derivations of the Poisson manifold
\( T_{\pi }M=\{X\in TM|[X,\pi ]_{S}=0\} \) to the derivations of
the \( \star  \)-product \( T_{\star }M=\{\delta \in T_{poly}|[\delta ,\star ]_{G}=0\} \).
Since an arbitrary \( \star  \)-product is equivalent to the Formality
\( \star  \)-product, we can expect that such a map exists for every
\( \star  \)-product. Here we state some facts, that we can say about
such a map in general. For this we expand it on a local patch in terms
of partial derivatives\[
\delta _{X}=\delta ^{i}_{X}\partial _{i}+\delta ^{ij}_{X}\partial _{i}\partial _{j}+\cdots .\]
Due to its property to be a derivation, it is now easy to see that
\( \delta _{X} \) is completely determined by the first term \( \delta ^{i}_{X}\partial _{i} \).
This means that if the first term is zero, the other terms have to
vanish, too. If further \( e \) is an arbitrary derivation of the
\( \star  \)-product there must exist a vector field \( X_{e} \)
such that\[
\delta _{X_{e}}=e.\]
If \( X,Y\in T_{\pi }M \), then \( [\delta _{X},\delta _{Y}] \)
is again a derivation of the \( \star  \)-product and we can conclude
that \begin{equation}
\label{star_bracket}
[\delta _{X},\delta _{Y}]=\delta _{[X,Y]_{\star }},
\end{equation}
where \( [X,Y]_{\star } \) is a deformation of the ordinary Lie bracket
of vector fields. Obviously it is linear, antisymmetric and fulfills
the Jacobi identity.

\bigskip

We will now calculate \( \delta  \) and \( [\cdot ,\cdot ]_{\star } \)
up to second order for the Weyl ordered \( \star  \)-product. We
assume that \( \delta _{X} \) can be expanded in the following way\[
\delta _{X}=X^{i}\partial _{i}+\delta ^{ij}_{X}\partial _{i}\partial _{j}+\delta _{X}^{ijk}\partial _{i}\partial _{j}\partial _{k}+\cdots .\]
 Expanding the equation \[
\delta _{X}(f\star g)=\delta _{X}(f)\star g+f\star \delta _{X}(g)\]
 order by order and using \( [X,c]_{S}=0 \) we find that\begin{eqnarray}
\delta _{X} & = & X^{i}\partial _{i}-\frac{1}{12}c^{lk}\partial _{k}c^{im}\partial _{l}\partial _{m}X^{j}\partial _{i}\partial _{j}\label{weyl_derivations} \\
 &  & +\frac{1}{24}c^{lk}c^{im}\partial _{l}\partial _{i}X^{j}\partial _{k}\partial _{m}\partial _{j}+\mathcal{O}(3).\nonumber 
\end{eqnarray}
For \( [\cdot ,\cdot ]_{\star } \) we simply calculate \( [\delta _{X},\delta _{Y}] \)
and get\begin{eqnarray*}
[X,Y]_{\star } & = & [X,Y]_{L}\nonumber \\
 &  & -\frac{1}{12}(c^{lk}\partial _{k}c^{im}\partial _{l}\partial _{m}X^{j}\partial _{i}\partial _{j}Y^{n}-c^{lk}\partial _{k}c^{im}\partial _{l}\partial _{m}Y^{j}\partial _{i}\partial _{j}X^{n})\partial _{n}\nonumber \\
 &  & +\frac{1}{24}(c^{lk}c^{im}\partial _{l}\partial _{i}X^{j}\partial _{k}\partial _{m}\partial _{j}Y^{n}-c^{lk}c^{im}\partial _{l}\partial _{i}Y^{j}\partial _{k}\partial _{m}\partial _{j}X^{n})\partial _{n}\\
 &  & +\mathcal{O}(3).\nonumber 
\end{eqnarray*}

\section{\label{nc_forms}Forms}

Now we want to introduce noncommutative forms, which will later be
used in Seiberg-Witten gauge theory (\ref{Seiberg_Witten_gauge_theory}).
If we have a map \( \delta  \) from the Poisson vector fields to
the derivations of the \( \star  \)-product, we have seen that there
is a natural Lie-algebra structure \( [\cdot ,\cdot ]_{\star } \)
(\ref{star_bracket}) over the space of Poisson vector fields, the
space of derivations of the Poisson structure. On the space of derivations
of the \( \star  \)-product we can easily construct the Chevalley
cohomology. Further, again with the map \( \delta  \), we can lift
derivations of the Poisson structure to derivations of the \( \star  \)-product.
Therefore it should be possible to pull back the Chevalley cohomology
from the space of derivations to vector fields. This will be done
in the following.

\bigskip

A deformed \( k \)-form is defined to map \( k \) Poisson vector
fields to a function and has to be skew-symmetric and linear over
\( \mathbb {C} \). This is a generalization of the undeformed case,
where a form has to be linear over the algebra of functions. Functions
are defined to be \( 0 \)-forms. The space of forms \( \Omega _{\star }M \)
is now a \( \star  \)-bimodule via \begin{equation}
\label{bimodule_structure_forms}
(f\star \omega \star g)(X_{1},\dots ,X_{k})=f\star \omega (X_{1},\dots ,X_{k})\star g.
\end{equation}
As expected, the exterior differential is defined with the help of
the map \( \delta  \).\[
\delta \omega (X_{0},\dots ,X_{k})=\sum ^{k}_{i=0}(-1)^{i}\, \delta _{X_{i}}\omega (X_{0},\dots ,\hat{X}_{i},\dots ,X_{k})\]
\begin{equation}
\label{nc_differential}
+\sum _{0\leq i<j\leq k}(-1)^{i+j}\omega ([X_{i},X_{j}]_{\star },X_{0},\dots ,\hat{X}_{i},\dots ,\hat{X}_{j},\dots ,X_{k}).
\end{equation}
With the properties of \( \delta  \) and \( [\cdot ,\cdot ]_{\star } \)
it follows that \[
\delta ^{2}\omega =0.\]
To be more explicit we give formulas for a function \( f \), a one
form \( A \) and a two form \( F \)\begin{eqnarray*}
\delta f(X) & = & \delta _{X}f,\nonumber \\
\delta A(X,Y) & = & \delta _{X}A_{Y}-\delta _{Y}A_{X}-A_{[X,Y]_{\star }},\nonumber \\
\delta F(X,Y,Z) & = & \delta _{X}F_{Y,Z}-\delta _{Y}F_{X,Z}+\delta _{Z}F_{X,Y},\\
 &  & -F_{[X,Y]_{\star },Z}+F_{[X,Z]_{\star },Y}-F_{[Y,Z]_{\star },X}.\nonumber 
\end{eqnarray*}
A wedge product may be defined \[
\omega _{1}\wedge \omega _{2}(X_{1},\dots ,X_{p+q})=\frac{1}{p!q!}\sum _{I,J}\varepsilon (I,J)\, \omega _{1}(X_{i_{1}},\dots ,X_{i_{p}})\star \omega _{2}(X_{j_{1}},\dots ,X_{j_{q}})\]
where \( (I,J) \) is a partition of \( (1,\dots ,p+q) \) and \( \varepsilon (I,J) \)
is the sign of the corresponding permutation. The wedge product is
linear and associative and generalizes the bimodule structure (\ref{bimodule_structure_forms}).
We note that it is no more graded commutative. We again give some
formulas.\begin{eqnarray*}
(f\wedge a)_{X} & = & f\star a_{X},\nonumber \\
(a\wedge f)_{X} & = & a_{X}\star f,\\
(a\wedge b)_{X,Y} & = & a_{X}\star b_{Y}-a_{Y}\star b_{X}.\nonumber 
\end{eqnarray*}
The differential (\ref{nc_differential}) fulfills the graded Leibniz
rule\[
\delta (\omega _{1}\wedge \omega _{2})=\delta \omega _{1}\wedge \omega _{2}+(-1)^{k_{2}}\, \omega _{1}\wedge \delta \omega _{2}.\]

\section{Construction of actions}

All field theories in physics may be formulated by an action principle.
For field theories on a curved manifold the action is of the form\[
\mathcal{S}=\int d^{n}x\, \sqrt{g}\, \mathcal{L}(g^{ij},\phi ,\partial _{i}\phi )\]
with \( g \) the determinant of the metric. \( \mathcal{L} \) is
a local Lagrange density depending on the fields and its partial derivatives.
A simple examples is a single scalar field \( \phi  \). In this case
we have \( \mathcal{L}=g^{ij}\partial _{i}\phi \partial _{j}\phi +p(\phi ) \)
where \( p \) is a polynomial in the field \( \phi  \). We can formulate
the action in the language of frames and in this case\[
\mathcal{S}=\int d^{n}x\, e\, \mathcal{L}(\phi ,e_{a}\phi )\]
where \( e_{a} \) are the vectorfields of the frame as defined in
(\ref{definition_comoving_frame}). Now \( e \) is the determinant
of the frame and in the example of the single scalar field the Lagrange
density becomes \( \mathcal{L}=\eta ^{ab}e_{a}\phi e_{b}\phi +p(\phi ) \).

If we do not want to give up the action principle we have to generalize
the notion of an action functional to noncommutative geometry. The
generalization of the Lagrange density is quite clear. We only have
to replace the fields by algebra elements and if we believe that the
commuting frame formalism is the right one we can use derivations
of the algebra as a substitute for the frame of derivations . 

What to use for the integral and the measure function \( e \) is
not so clear in the beginning. As an action maps fields to numbers
from an algebraic point of view, the most obvious candidate is a trace
on a representation of the algebra. A trace is cyclic with respect
to the product of algebra elements\[
Tr\, \hat{f}\hat{g}=Tr\, \hat{g}\hat{f}.\]
Therefore it is possible to do partial integration with inner derivations\[
Tr\, [\hat{\lambda },\hat{f}]\hat{g}=-Tr\, \hat{f}[\hat{\lambda },\hat{g}].\]
Another argument for using a trace comes from noncommutative gauge
theory (\ref{nc_gauge_theory}): Suppose we have a field invariant
under the following transformation\[
\delta _{\hat{\alpha }}\hat{\phi }=i[\hat{\alpha },\hat{\phi }],\]
then the trace of a polynomial in \( \hat{\phi } \) is invariant
under this type of gauge transformations. We will see in the following
that the use of a trace in the \( \star  \)-product representation
will solve the problem with the measure function in a quite remarkable
way.

\subsection{Traces for \protect\( \star \protect \)-products}

If the algebra is in the \( \star  \)-product representation, the
algebra elements are functions on some manifold and we are able to
integrate them. But the pure integral is not cyclic with respect to
the \( \star  \)-product. To cure this we introduce a measure function
\( \Omega  \) and make the following ansatz for the trace of the
\( \star  \)-product \[
tr\, f=\int d^{n}x\, \Omega (x)\, f(x).\]
If we expand the equation of cyclicity \[
\int d^{n}x\, \Omega (x)\, [f(x)\stackrel{\star }{,}g(x)]=0\]
up to first order we see that \( \Omega  \) has to fulfill\begin{equation}
\label{equation_for_measure_function}
\partial _{i}(\Omega \Pi ^{ij})=0,
\end{equation}
 where \( \Pi ^{ij} \) is the Poisson structure corresponding to
the \( \star  \)-product \( [f\stackrel{\star }{,}g]=h\Pi ^{ij}\partial _{i}f\, \partial _{j}g+\cdots  \).
It is known \cite{Felder:2000hy} that there is always a gauge equivalent
\( \star  \)-product in the sense of (\ref{linear_transformation_on_star_product})
for which cyclicity is guaranteed to all orders.

\bigskip

If the Poisson structure \( \Pi ^{ij} \) is invertible then a solution
to the equation (\ref{equation_for_measure_function}) can be given.
In this case the inverse of the Pfaffian\[
\frac{1}{\Omega }=Pf(\Pi )=\sqrt{det(\Pi )}=\frac{1}{2^{n}n!}\epsilon _{i_{1}i_{2}\cdots i_{2n}}\Pi ^{i_{1}i_{2}}\cdots \Pi ^{i_{2n-1}i_{2n}}\]
is the measure function.

\subsection{Commuting frames from inner derivations}

Form (\ref{frame_from_inner_derivation}) we know that in the commuting
frame formalism with inner derivations \( e_{a}{}^{i}=\Pi ^{li}\partial _{l}\lambda _{a} \).
In two dimensions we have

\[
e_{a}{}^{i}=\Pi ^{12}\left( \begin{array}{cc}
-\partial _{2}\lambda _{1} & -\partial _{1}\lambda _{1}\\
\partial _{2}\lambda _{2} & \partial _{1}\lambda _{2}
\end{array}\right) ,\]
\[
e^{-1}=\det (e_{a}{}^{i})=(\Pi ^{12})^{2}(\partial _{1}\lambda _{1}\, \partial _{2}\lambda _{2}-\partial _{2}\lambda _{1}\, \partial _{1}\lambda _{2}).\]
\( e^{-1} \) is the inverse of the measure function induced by the
metric. On the other hand the inverse of the measure function induced
by the trace is\[
\Omega ^{-1}=\frac{1}{2}\epsilon _{ij}\Pi ^{ij}=\Pi ^{12}.\]
If we want these two measure functions to be equal, \begin{equation}
\label{constraint_on_the_lambdas}
\partial _{1}\lambda _{1}\, \partial _{2}\lambda _{2}-\partial _{2}\lambda _{1}\, \partial _{1}\lambda _{2}=\frac{1}{\Pi ^{12}}
\end{equation}
has to be fulfilled. This is not the case in any of the examples from
(\ref{two_dimensional_examples}). 

\bigskip

In four dimensons the measure function induced by the trace is \[
\Omega ^{-1}=\frac{1}{8}\epsilon _{ijkl}\Pi ^{ij}\Pi ^{kl}=\Pi ^{12}\Pi ^{34}-\Pi ^{13}\Pi ^{24}+\Pi ^{14}\Pi ^{23},\]
which is quadratic in the elements of \( \Pi  \). Due to (\ref{frame_from_inner_derivation})
the measure function induced by the metric contains monomials of order
four. Again there are constraints of the form (\ref{constraint_on_the_lambdas})
if want the two functions to be equal. This is also the case in higher
dimensions.

\subsection{Commuting frames and quantum spaces}

We will now propose another method how to find Poisson structures
with compatible frames. On several quantum spaces deformed derivations
have been constructed \cite{Wess:1991vh,Lorek:1997eh,Cerchiai:1998ef}.
In many cases the deformed Leibniz rule may be written in the following
form\[
\hat{\partial }_{\mu }(\hat{f}\hat{g})=\hat{\partial }_{\mu }\hat{f}\hat{g}+\hat{T}_{\mu }{}^{\nu }(\hat{f})\hat{\partial }_{\nu }\hat{g},\]
where \( \hat{T} \) is an algebra morphism from the quantum space
to its matrix ring\[
\hat{T}_{\mu }{}^{\nu }(\hat{f}\hat{g})=\hat{T}_{\mu }{}^{\alpha }(\hat{f})\hat{T}_{\alpha }{}^{\nu }(\hat{g}).\]
Again in some cases it is possible to implement this morphism with
some kind of inner morphism\[
\hat{T}_{\mu }{}^{\nu }(\hat{f})=\hat{e}_{\mu }{}^{a}\hat{f}\hat{e}_{a}{}^{\nu },\]
where \( \hat{e}_{a}{}^{\mu } \) is an invertible matrix with entries
from the quantum space. If we define\[
\hat{e}_{a}=\hat{e}_{a}{}^{\mu }\hat{\partial }_{\mu },\]
the \( \hat{e}_{a} \) are derivations\[
\hat{e}_{a}(\hat{f}\hat{g})=\hat{e}_{a}(\hat{f})\hat{g}+\hat{f}\hat{e}_{a}(\hat{g}).\]
The dual formulation of this with covariant differential calculi on
quantum spaces is the formalism with commuting frames investigated
for example in \cite{Dimakis:1996,Madore:1999bi,Cerchiai:2000qi,Madore:2000aq}. 

We can now represent the quantum space with the help of a \( \star  \)-product.
For example, we can use the Weyl-ordered \( \star  \)-product we
have constructed in section \ref{wely_star_construction}. Further
we can calculate the action of the operators \( \hat{e}_{a} \) on
functions. Since these are now derivations of a \( \star  \)-product,
their classical limits are necessarily a Poisson vector fields \( e_{a} \)
for the Poisson structure of the \( \star  \)-product and with (\ref{weyl_derivations})
the derivations are represented by \[
\hat{e}_{a}=\delta _{e_{a}}.\]

\subsection{Example: \protect\( M(so_{a}(n))\protect \)}

Now we continue the example (\ref{example_so_a(n)}). As a special
frame we take the deformed commuting derivations acting on the coordinates
like\begin{eqnarray*}
\hat{\partial }_{o}\hat{x}^{0} & = & 1+\hat{x}^{0}\hat{\partial }_{o},\nonumber \\
\hat{\partial }_{o}\hat{x}^{i} & = & \hat{x}^{i}\hat{\partial }_{o},\\
\hat{\partial }_{i}\hat{x}^{j} & = & \delta _{i}^{j}+\hat{x}^{j}\hat{\partial }_{i},\nonumber \\
\hat{\partial }_{i}\hat{x}^{0} & = & (\hat{x}^{0}+ia)\hat{\partial }_{i}.\nonumber 
\end{eqnarray*}
 Note that the \( \hat{\partial }_{i} \) are not derivations on the
quantum space. But we can apply the procedure described previously.
If we define \[
\hat{\rho }=\sqrt{\sum _{i}(\hat{x}^{i})^{2}}\]
 and assume that it is invertible, we can write the above formulas
for \( \hat{\partial }_{i} \) in another way\[
\hat{\partial }_{i}\hat{f}\hat{g}=\hat{\partial }_{i}\hat{f}\cdot \hat{g}+\hat{\rho }^{-1}\hat{f}\hat{\rho }\cdot \hat{\partial }_{i}\hat{g},\]
since\begin{eqnarray*}
\hat{\rho }^{-1}\hat{x}^{0}\hat{\rho }= & \hat{x}^{0}+ia, & \hat{\rho }^{-1}\hat{x}^{i}\hat{\rho }=\hat{x}^{i}.
\end{eqnarray*}
Therefore as we have seen \begin{eqnarray*}
\hat{e}_{o}=\hat{\partial }_{0}, &  & \hat{e}_{i}=\hat{\rho }\hat{\partial }_{i}
\end{eqnarray*}
is a frame on the quantum space. The classical limit of this is obviously\begin{eqnarray*}
e_{o}=\partial _{0}, &  & e_{i}=\rho \partial _{i}.
\end{eqnarray*}
The derivations (\ref{weyl_derivations}) going with the Weyl ordered
\( \star  \)-product are identical up to third order \begin{eqnarray*}
\delta _{0} & = & \partial _{o}+\mathcal{O}(a^{3}),\\
\delta _{i} & = & \rho \partial _{i}+\mathcal{O}(a^{3}).
\end{eqnarray*}

\bigskip 

In the classical limit we have \( n \) linear independent derivations
and we can apply the commutating frame formalism. The forms dual to
the derivations are\begin{eqnarray*}
\theta ^{0}=dt, &  & \theta ^{i}=\frac{1}{\rho }dx^{i},
\end{eqnarray*}
and the classical metric (with \( \eta _{ab}=diag(1,-1,-1,\cdots ) \)
) becomes\[
g=\eta _{ab}\theta ^{a}\theta ^{b}=(dx^{0})^{2}-\rho ^{-2}((dx^{1})^{2}+\cdots +(dx^{n-1})^{2}).\]
We know that we can write\[
(dx^{1})^{2}+\cdots +(dx^{n-1})^{2}=d\rho ^{2}+\rho ^{2}d\Omega ^{2}_{n-2},\]
where \( d\Omega ^{2}_{n-2} \) is the metric of the \( n-2 \) dimensional
sphere. Therefore in this new coordinate system \[
g=(dx^{0})^{2}-(d\ln \rho )^{2}+d\Omega ^{2}_{n-2}\]
and we see that the classical space is a cross product of two dimensional
Euclidean space and a \( (n-2) \)-sphere. Therefore it is a space
of constant non vanishing curvature. Further we calculate that\[
\sqrt{\det g}=\rho ^{-(n-1)}\]
fulfills the equation for the measure function (\ref{equation_for_measure_function})
of the \( \star  \)-product trace. Here we are lucky and are able
to write down actions for field theories on this special quantum space
with the correct classical limit. For example\begin{eqnarray*}
S & = & Tr\, (\eta ^{\alpha \beta }\hat{e}_{\alpha }\hat{\phi }\, \hat{e}_{b}\hat{\phi }+m^{2}\hat{\phi }^{2}+a\hat{\phi }^{4})\\
 & = & \int \frac{d^{n}x}{\rho ^{n-1}}\, \delta _{o}\phi \star \delta _{o}\phi -\delta _{i}\phi \star \delta _{i}\phi +m^{2}\phi \star \phi +a\phi \star \phi \star \phi \star \phi 
\end{eqnarray*}
is a well defined action with the \( \star  \)-product (\ref{SO_a(n)_star_product})
and reduces in the classical limit to \( \phi ^{4} \)-theory on above
described manifold. We will continue this example at (\ref{example_so_a(n)_and_sw_map}),
where we will have explicit formulas for the Seiberg-Witten-maps and
we are able to do gauge theory.

\cleardoublepage

\chapter{Gauge theory}

In this chapter we will investigate noncommutative gauge theory formulated
in the \( \star  \)-product formalism, where it is possible to formulate
general nonabelian gauge theories on noncommutative spacetime. Nonexpanded
theories can in general only deal with \( U(n) \)-gauge groups, but
using Seiberg-Witten-maps relating noncommutative quantities with
their commutative counterparts makes it possible to consider arbitrary
nonabelian gauge groups \cite{Seiberg:1999vs,Madore:2000en,Jurco:2001rq}.

The case of an algebra with constant commutator has been extensively
studied. This theory reduces in the classical limit to a theory on
a flat spacetime. Therefore it is necessary to develop concepts working
with more general algebras, since one would expect that curved backgrounds
are related to algebras with nonconstant commutation relations. We
present here a method using derivations of \( \star  \)-products
to build covariant derivatives for Seiberg-Witten gauge theory. Further
we are able to write down a noncommutative action by linking the derivations
to a frame field induced by a nonconstant metric as explaned in the
last chapter. An example is given where the action reduces in the
classical limit to scalar electrodynamics on a curved background.

\section{\label{classical gauge theory}Classical gauge theory}

First let us recall some properties of a general classical gauge theory.
A non-abelian gauge theory is based on a Lie group with Lie algebra\[
[T^{a},T^{b}]=i\, f^{ab}{}_{c}T^{c}.\]
Matter fields transform under a Lie algebra valued infinitesimal parameter\begin{equation}
\label{com_g_field}
\delta _{\alpha }\psi =i\alpha \psi ,\, \, \, \, \, \, \, \, \, \alpha =\alpha _{a}T^{a}
\end{equation}
in the fundamental representation. It follows that\begin{equation}
\label{com_consitency_condition}
(\delta _{\alpha }\delta _{\beta }-\delta _{\beta }\delta _{\alpha })\psi =\delta _{-i[\alpha ,\beta ]}\psi .
\end{equation}
The commutator of two consecutive infinitesimal gauge transformation
closes into an infinitesimal gauge transformation. Further a Lie algebra
valued gauge potential is introduced with the transformation property\begin{eqnarray}
a_{i} & = & a_{ia}T^{a},\nonumber \\
\delta _{\alpha }a_{i} & = & \partial _{i}\alpha +i[\alpha ,a_{i}].\label{com_gt_a} 
\end{eqnarray}
With this the covariant derivative of a field is\[
D_{i}\psi =\partial _{i}\psi -ia_{i}\psi .\]
The field strength of the gauge potential is defined to be the commutator
of two covariant derivatives\[
iF_{ij}=[D_{i},D_{j}]=\partial _{i}a_{j}-\partial _{j}a_{i}-i[a_{i},a_{j}].\]
The last equations can all be stated in the language of forms. For
this a connection one form is introduced\[
a=a_{ia}T^{a}dx^{i}.\]
The covariant derivative now acts as\[
D\psi =d\psi -ia\psi .\]
The field strength becomes a two form\[
F=da-ia\wedge a.\]

\bigskip

We note that all this may be formulated with finite gauge transformations
\( g \). They are related to the infinitesimal gauge parameter by
\[
g=e^{i\alpha }=e^{i\alpha _{a}(x)t^{a}}.\]
 With this finite gauge transformations for the covariant derivative
and a field in the fundamental representation are\begin{eqnarray*}
T_{g}D_{\mu } & = & gD_{\mu }g^{-1},\\
T_{g}a_{\mu } & = & gag^{-1}+ig\partial _{\mu }g^{-1},\\
T_{g}\psi  & = & g\psi .
\end{eqnarray*}

\section{\label{nc_gauge_theory}Noncommutative gauge theory}

In a gauge theory on a noncommutative space, fields should again transform
like (\ref{com_g_field})\begin{equation}
\label{nc_gt_field}
\hat{\delta }_{\hat{\Lambda }}\hat{\Psi }=i\hat{\Lambda }\hat{\Psi }.
\end{equation}
It follows again that\begin{equation}
\label{nc_consitency_condition}
(\hat{\delta }_{\hat{\Lambda }}\hat{\delta }_{\hat{\Gamma }}-\hat{\delta }_{\hat{\Gamma }}\hat{\delta }_{\hat{\Lambda }})\hat{\Psi }=\hat{\delta }_{-i[\hat{\Lambda },\hat{\Gamma }]}\hat{\Psi }.
\end{equation}

Since multiplication of a function with a field is not again a covariant
operation we are forced to introduce a covariantizer with the transformation
property \begin{equation}
\label{nc_gt_D}
\hat{\delta }_{\hat{\Lambda }}D(\hat{f})=i[\hat{\Lambda },D(\hat{f})].
\end{equation}
From this it follows that\begin{equation}
\label{nc_gt_field}
\hat{\delta }_{\hat{\Lambda }}(D(\hat{f})\hat{\Psi })=i\hat{\Lambda }D(\hat{f})\hat{\Psi }.
\end{equation}
If we covariantize the coordinate functions \( \hat{x}^{i} \) we
get covariant coordinates \begin{equation}
\label{definition_covariant_coordinates}
\hat{X}^{i}=D(\hat{x}^{i})=\hat{x}^{i}+\hat{A},^{i}
\end{equation}
where the gauge field now transforms according to \begin{equation}
\label{nc_gt_covcord}
\hat{\delta }_{\hat{\Lambda }}\hat{A}^{i}=-i[\hat{x}^{i},\hat{\Lambda }]+i[\hat{\Lambda },\hat{A}^{i}].
\end{equation}

Unluckily, this does not have a meaninful commutative limit, a problem
that can only be fixed for the canonical case (i.e. \( [\widehat{x}^{i},\widehat{x}^{j}]=i\theta ^{ij} \)
with \( \theta  \) a constant) and invertible \( \theta  \). 

For noncommutative algebras where we already have derivatives with
a commutative limit, it therefore seems natural to gauge these. But
due to their nontrivial coproduct the resulting gauge field would
have to be derivative-valued to match the rather awkward behaviour
under gauge transformations. The physical reason for this might be
the following: The noncommutative derivatives are in general built
to reduce to derivatives on flat spacetime, which might not be the
correct commutative limit.

We therefore advocate a solution using derivations that will later
on (see section \ref{constrcution_of_gauge_inv_actions}) be linked
to derivatives on curved spacetime:

If we have a derivation \( \hat{\partial } \), i. e. a map with the
property \[
\hat{\partial }(\hat{f}\hat{g})=(\hat{\partial }\hat{f})\hat{g}+\hat{f}(\hat{\partial }\hat{g})\]
for arbitrary elements \( \hat{f} \) and \( \hat{g} \) of the algebra,
we can introduce a noncommutative gauge parameter \( \hat{A}_{\hat{\partial }} \)
and demand that the covariant derivative (or covariant derivation)
of a field \[
\hat{D}\hat{\Psi }=(\hat{\partial }-i\hat{A}_{\hat{\partial }})\hat{\Psi }\]
again transforms like a field \[
\hat{\delta }_{\hat{\Lambda }}\hat{D}\hat{\Psi }=i\hat{\Lambda }\hat{D}\hat{\Psi }.\]
From this it follows that \( \hat{A}_{\hat{\partial }} \) has to
transform like \begin{equation}
\label{nc_gt_A}
\hat{\delta }_{\hat{\Lambda }}\hat{A}_{\hat{\partial }}=\hat{\partial }\hat{A}_{\hat{\partial }}+i[\hat{\Lambda },\hat{A}_{\hat{\partial }}].
\end{equation}
This is the transformation property we would expect a noncommutative
gauge potential to have, and in the next section we will show that
for this object we can indeed construct a Seiberg-Witten map in a
natural way. If we have an involution on the algebra, we can demand
that the gauge potential is hermitian \( \hat{A}_{\hat{\partial }}=\overline{\hat{A}_{\hat{\partial }}} \)
. Additionally the field \( \overline{\hat{\Psi }} \) transforms
on the right hand side. In this case expressions of the form\begin{eqnarray*}
\overline{\hat{\Psi }}\hat{\Psi } & \mbox {and} & \overline{\hat{D}\hat{\Psi }}\hat{D}\hat{\Psi }
\end{eqnarray*}
become gauge invariant quantities.

\section{\label{Seiberg_Witten_gauge_theory}Seiberg-Witten gauge theory}

In \cite{Jurco:2001rq} a method how to construct noncommutative non
abelian gauge theories using Seiberg-Witten-maps was presented. In
the case of constant Poisson structure treated there, it is possible
to introduce the momenta via covariant coordinates: \( \partial _{i}=[\theta _{ij}^{-1}x^{j},\cdot ] \).
In general this approach does not yield the desired classical limit.
The momentum operators have to be introduced in another way. We will
approach the problem by considering derivations of \( \star  \)-products.

\subsection{Gauge transformations and derivations}

If we have a look at (\ref{nc_consitency_condition}), we see that
the commutator of two gauge transformations only closes into the Lie
algebra in the fundamental representation of \( U(n) \). For non
abelian gauge groups, we are forced to go to the enveloping algebra,
giving us infinitely many degrees of freedom. But this problem can
be solved using Seiberg-Witten maps \cite{Seiberg:1999vs,Jurco:2001rq}.

The noncommutative gauge parameter and the noncommutative gauge potential
will be enveloping algebra valued, but they will only depend on their
commutative counterparts, therefore preserving the right number of
degrees of freedom. These Seiberg-Witten maps \( \Lambda  \), \( \Psi  \)
and \( D \) are functionals of their classical counterparts and additionally
of the gauge potential \( a_{i} \). Their transformation properties
(\ref{nc_gt_D}) and (\ref{nc_gt_field}) should be induced by the
classical ones (\ref{com_g_field}) and (\ref{com_gt_a}) like\begin{eqnarray*}
\widehat{\Lambda }_{\beta }[a]+\widehat{\delta }_{\alpha }\widehat{\Lambda }_{\beta }[a] & = & \widehat{\Lambda }_{\beta }[a+\delta _{\alpha }a],\\
\widehat{\Psi }_{\psi }[a]+\widehat{\delta }_{\alpha }\widehat{\Psi }_{\psi }[a] & = & \widehat{\Psi }_{\psi +\delta _{\alpha }\psi }[a+\delta _{\alpha }a],\\
\widehat{A}[a]+\widehat{\delta }_{\alpha }\widehat{A}[a] & = & \widehat{A}[a+\delta _{\alpha }a].
\end{eqnarray*}

The Seiberg-Witten maps can be found order by order using a \( \star  \)-product
to represent the algebra on a space of functions. Translated into
this language we get for the fields \cite{Jurco:2001rq} \begin{equation}
\label{sp_gt_field}
\delta _{\alpha }\Psi _{\psi }[a]=i\Lambda _{\alpha }[a]\star \Psi _{\psi }[a].
\end{equation}

From (\ref{sp_gt_field}) we can derive a consistency condition for
the noncommuative gauge parameter \cite{Jurco:2000ja}. Insertion
into (\ref{nc_consitency_condition}) and the use of (\ref{com_consitency_condition})
yields\begin{equation}
\label{st_consitency_condition}
i\delta _{\alpha }\Lambda _{\beta }-i\delta _{\beta }\Lambda _{\alpha }+[\Lambda _{\alpha }\stackrel{\star }{,}\Lambda _{\beta }]=i\Lambda _{-i[\alpha ,\beta ]}.
\end{equation}
The transformation law for the covariantizer is now\begin{equation}
\label{sp_gt_cov}
\delta _{\alpha }(D[a](f))=i[\Lambda _{\alpha }[a]\stackrel{\star }{,}D[a](f)].
\end{equation}
We now want to extend the Seiberg-Witten-map to derivations of the
\( \star  \)-product. In the next section we will see that we are
able to identify derivations of a \( \star  \)-products with Poisson
vector fields of the Poisson structure associated with the \( \star  \)-product.
To be more explicit, let us assume that \( X \) is a Poisson vector
field\[
X^{i}\partial _{i}\{f,g\}=\{X^{i}\partial _{i}f,g\}+\{f,X^{i}\partial _{i}g\},\]
 then we know that there exists a polydifferential operator \( \delta _{X} \)
with the following property (see chapter \ref{Formality map} esp.
(\ref{formality star derivation}) and \ref{weyl derivations})\[
\delta _{X}(f\star g)=\delta _{X}f\star g+f\star \delta _{X}g.\]
It is easy to see that all derivations of this kind exhaust the space
of derivations of the \( \star  \)-product. Since the commutator
of two derivations is again a derivation we have concluded that there
has to be a deformed Lie bracket \( [\cdot ,\cdot ]_{\star } \) with
the following property\[
[\delta _{X},\delta _{Y}]=\delta _{[X,Y]_{\star }}.\]

With help of the operator \( \delta _{X} \) we can now introduce
the covariant derivative of a field and the gauge potential in the
following way\begin{equation}
\label{sp_gt_cov_derivative}
D_{X}[a]\Psi _{\psi }[a]=\delta _{X}\Psi _{\psi }[a]-iA_{X}[a]\star \Psi _{\psi }[a].
\end{equation}
It follows that the gauge potential has to transform like\begin{equation}
\label{sp_gt_gauge_pot}
\delta _{\alpha }A_{X}[a]=\delta _{X}\Lambda _{\alpha }[a]+i[\Lambda _{\alpha }[a]\stackrel{\star }{,}A_{X}[a]].
\end{equation}
A field strength may be defined\begin{equation}
\label{sp_field_strength}
iF_{X,Y}[a]=[D_{X}[a]\stackrel{\star }{,}D_{Y}[a]]-D_{[X,Y]_{\star }}[a].
\end{equation}
The properties of \( \delta _{\cdot } \) and \( [\cdot ,\cdot ]_{\star } \)
ensure that this is really a function and not a polydifferential operator\[
F_{X,Y}[a]=\delta _{X}A_{Y}[a]-\delta _{Y}A_{X}[a]-i[A_{X}[a]\stackrel{\star }{,}A_{Y}[a]]-A_{[X,Y]_{\star }}[a].\]

\bigskip

We are able to translate Seiberg-Witten gauge theory into the language
of the forms introduced in (\ref{nc_forms}). \( A_{X} \) is the
connection one form \( A \) evaluated on the vector field \( X \).
It transforms like\[
\delta _{\alpha }A=\delta \Lambda _{\alpha }+i\Lambda _{\alpha }\wedge A-A\wedge \Lambda _{\alpha }.\]
The covariant derivative of a field \( \Psi  \) is now\[
D\Psi =\delta \Psi -iA\wedge \Psi ,\]
 and the field strength becomes\[
F=DF=\delta A-iA\wedge A.\]
One easily can show that the field strength is covariant constant\[
DF=\delta F-iA\wedge F=0.\]

\subsection{Finite Seiberg-Witten gauge transformations}

Ii is interesting that noncommutative infinitesimal and finite gauge
transformations may be related like in the classical case. To show
this let us first define noncommutative finite gauge transformations
similar to the classical case \cite{Jurco:2001kp}\[
T_{g}\Psi _{\psi }[a]=\Psi _{T_{g}\psi }[T_{g}a]=G_{g}[a]\star \Psi _{\psi }[a],\]
\[
T_{g}D_{X}[a]=D_{X}[T_{g}a]=G_{g}[a]\star D_{X}[a]\star G_{g}[a]^{-1}.\]
If we apply two consecutive gauge transformations on a field \[
T_{g_{2}}T_{g_{1}}\Psi _{\psi }[a]=G_{g_{1}}[T_{g2}a]\star (G_{g_{2}}[a]\star \Psi _{\psi }[a]),\]
 we can derive a consistency condition for finite Seiberg-Witten gauge
transfomations\[
G_{g_{1}g_{2}}[a]=G_{g_{1}}[T_{g_{2}}a]\star G_{g_{2}}[a].\]
Now we are able to relate finite and infinitesimal gauge transformations.
In the classical case \[
T_{g}\psi =e^{i\alpha }\psi =e^{\delta _{\alpha }}\psi \]
where \( \delta _{\alpha } \) is the action of the infinitesimal
gauge transformation on the field \( \psi  \). We can use the same
fomula to define the noncommutative gauge transformations\[
G_{e^{i\alpha }}[a]\star \Psi _{\psi }[a]=e^{\delta _{\alpha }}_{\star }\star \Psi _{\psi }[a].\]
 To get an explicit formula we note that\[
(\delta _{\alpha }-i\Lambda _{\alpha }[a])\star \Psi _{\alpha }[a]=0\]
and calculate\begin{eqnarray*}
e^{\delta _{\alpha }}_{\star }\star \Psi _{\psi }[a] & = & e^{\delta _{\alpha }}_{\star }\star e^{-\delta _{\alpha }+i\Lambda _{\alpha }[a]}_{\star }\star \Psi _{\psi }[a]\\
 & = & e_{\star }^{i\Lambda _{\alpha }[a]+\frac{i}{2}\delta _{\alpha }\Lambda _{\alpha }[a]+\cdots }\star \Psi _{\psi }[a]
\end{eqnarray*}
where we have used the Baker-Campbell-Hausdorff formula.

\subsection{Enveloping algebra valued gauge transformations}

Gauge theories on noncommutative spaces cannot be formulated with
Lie algebra valued infinitesimal transformations and therefore not
with Lie algebra valued gauge fields. To see this we assume that the
noncommutative gauge parameter is Lie algebra valued\[
\hat{\alpha }=\hat{\alpha }_{a}T^{a},\]
where the \( \hat{\alpha }_{a} \) are elements of the algebra describing
the noncommutative space and the \( T^{a} \) are generators of a
Lie algebra with \( [T^{a},T^{b}]=if^{ab}{}_{c}T^{c} \). We have
seen (\ref{nc_consitency_condition}) that due to consistency the
commutator of two gauge parameters again has to be a gauge parameter,
but now\[
[\hat{\alpha },\hat{\beta }]=\frac{i}{2}\{\hat{\alpha }_{a}\hat{\beta }_{b}\}f^{ab}{}_{c}T^{a}+\frac{1}{2}[\hat{\alpha }_{a},\hat{\beta }_{b}]\{T^{a},T^{b}\}\]
where \( \{,\} \) denotes the anticommutator. All higher powers of
the generators \( T^{a} \) of the gauge group may be created in this
way. Thus the enveloping algebra of the Lie algebra seems to be a
proper setting for nonabelian noncommutative gauge theory. In general
this is not very attractive because the enveloping algebra is infinite
dimensional and consequently requires an infinite number of gauge
parameters and gauge potentials. 

In Seiberg-Witten gauge theory, however, it is possible to restrict
the number of infinitesimal enveloping algebra valued gauge parameters
to the usual ones \cite{Jurco:2000ja}. In this case the gauge parameter
depends on the Lie algebra valued parameters and its derivatives of
the corresponding commutative gauge theory. The construction of this
kind of enveloping valued gauge parameter is based on the Seiberg-Witten
map. We have seen that in Seiberg-Witten gauge theory the gauge parameter
\( \Lambda _{\alpha }[a] \) is a functional of the classical one
\( \alpha =\alpha _{a}T^{a} \) and the classical potential \( a_{i}=a_{ia}T^{a} \).
We expand it order by order in the expansion parameter \begin{equation}
\label{expanded_SW_gauge_paramter}
\Lambda _{\alpha }[a]=\Lambda ^{0}_{\alpha }[a]+\Lambda ^{1}_{\alpha }[a]+\cdots .
\end{equation}
Further it has to fulfill the consistency condition (\ref{st_consitency_condition}).
If we plug (\ref{expanded_SW_gauge_paramter}) into this equation
we get to zeroth and first order\[
i\delta _{\alpha }\Lambda ^{0}_{\beta }-i\delta _{\beta }\Lambda ^{0}_{\alpha }+[\Lambda ^{0}_{\alpha },\Lambda ^{0}_{\beta }]=i\Lambda ^{0}_{-i[\alpha ,\beta ]},\]
\[
i\delta _{\alpha }\Lambda ^{1}_{\beta }-i\delta _{\beta }\Lambda ^{1}_{\alpha }+[\Lambda ^{0}_{\alpha },\Lambda ^{1}_{\beta }]-[\Lambda ^{0}_{\beta },\Lambda ^{1}_{\alpha }]-i\Lambda ^{1}_{-i[\alpha ,\beta ]}=-\frac{1}{2}c^{ij}[\partial _{i}\Lambda ^{0}_{\alpha },\partial _{j}\Lambda ^{0}_{\beta }]\]
where we have assumed that we use as usual \( f\star g=fg+\frac{1}{2}c^{ij}\partial _{i}f\partial _{j}g+\cdots  \).
The first equation is fulfilled by the commutative gauge parameter.
Since this yields additionally the correct classical limit we set
\[
\Lambda ^{0}_{\alpha }=\alpha .\]
With this, a solution to the first order equation is\[
\Lambda ^{1}_{\alpha }=-\frac{i}{4}c^{ij}\{\partial _{i}\alpha ,a_{j}\}=-\frac{i}{4}c^{ij}\partial _{i}\alpha _{a}\, a_{jb}\{T^{a},T^{b}\}.\]
This is now obviously enveloping algebra valued. The solution is not
unique, since a solution to the homogeneous part of the first order
equation may be added. 

\bigskip

We have made the above considerations only for the noncommutative
gauge parameter. It should be clear that the method can be extended
to the gauge potential with help of (\ref{sp_gt_gauge_pot}), to fields
with (\ref{sp_gt_field}) and to the covariantizer with (\ref{sp_gt_cov}).
This will be done in (\ref{SW_map_weyl_ordered}) for the special
case of the Weyl-ordered \( \star  \)-product. Again all these solutions
are not unique, due to cohomologies induced by the homogeneous parts
of the equations. Other methods have to be used to restrict the possible
solutions. In \cite{Bichl:2001cq} this is done for the constant case
by demanding that the resulting action is renormalizable up to all
orders.

\subsection{\label{SW_map_weyl_ordered}Seiberg-Witten map for Weyl-ordered \protect\( \star \protect \)-product}

With the methods developed in the last section we will now present
a consistent solution for the Seiberg-Witten maps up to second order
for the Weyl ordered \( \star  \)-product and non-abelian classical
gauge transformations. The solutions have been chosen in such a way
that they reproduce the ones obtained in \cite{Jurco:2001rq} for
the constant case. In the following we will use the Weyl-ordered \( \star  \)-product
expanded order by order\[
f\star g=fg+f\star _{1}g+f\star _{2}g+\cdots \]
with e. g. \[
f\star _{1}g=\frac{1}{2}c^{ij}\partial _{i}f\, \partial _{j}g.\]

\subsubsection{Noncommutative gauge parameter}

As we have seen we have to expand \( \Lambda  \) in terms of the
deformation parameter\[
\Lambda _{\alpha }[a]=\alpha +\Lambda ^{1}_{\alpha }[a]+\Lambda ^{2}_{\alpha }[a]+\cdots .\]
To zeroth order the consistency condition (\ref{st_consitency_condition})
is equal to the commutative one (\ref{com_consitency_condition}).
Therefore, we already have set \( \Lambda ^{0}_{\alpha }=\alpha  \).

To first order we obtain\[
i\delta _{\alpha }\Lambda ^{1}_{\beta }-i\delta _{\beta }\Lambda ^{1}_{\alpha }+[\alpha ,\Lambda ^{1}_{\beta }]-[\beta ,\Lambda ^{1}_{\alpha }]-i\Lambda ^{1}_{-i[\alpha ,\beta ]}=-[\alpha \stackrel{\star _{1}}{,}\beta ]\]
\[
=-\frac{1}{2}c^{ij}[\partial _{i}\alpha ,\partial _{j}\beta ]\]
and to second order\[
i\delta _{\alpha }\Lambda ^{2}_{\beta }-i\delta _{\beta }\Lambda ^{2}_{\alpha }+[\alpha \, \Lambda ^{2}_{\beta }]-[\beta ,\Lambda ^{2}_{\alpha }]-i\Lambda ^{2}_{-i[\alpha ,\beta ]}\]
\[
=-[\alpha \stackrel{\star _{1}}{,}\Lambda ^{1}_{\beta }]-[\beta \stackrel{\star _{1}}{,}\Lambda ^{1}_{\alpha }]-[\Lambda ^{1}_{\alpha },\Lambda ^{1}_{\beta }]-[\alpha \stackrel{\star _{2}}{,}\beta ]\]
\[
=-\frac{1}{2}c^{ij}[\partial _{i}\alpha ,\partial _{j}\Lambda ^{1}_{\beta }]-\frac{1}{2}c^{ij}[\partial _{i}\beta ,\partial _{j}\Lambda ^{1}_{\alpha }]-[\Lambda ^{1}_{\alpha },\Lambda ^{1}_{\beta }]\]
\[
-\frac{1}{8}c^{mn}c^{ij}[\partial _{m}\partial _{i}\alpha ,\partial _{n}\partial _{j}\beta ]-\frac{1}{12}c^{ml}\partial _{l}c^{ij}([\partial _{m}\partial _{i}\alpha ,\partial _{j}\beta ]-[\partial _{i}\alpha ,\partial _{m}\partial _{j}\beta ]).\]
A solution to this is \begin{eqnarray*}
\Lambda _{\alpha }[a] & = & \alpha -\frac{i}{4}c^{ij}\{\partial _{i}\alpha ,a_{j}\}\nonumber \\
 &  & +\frac{1}{32}c^{ij}c^{kl}\Big {(}4\{\partial _{i}\alpha ,\{a_{k},\partial _{l}a_{j}\}\}-2i[\partial _{i}\partial _{k}\alpha ,\partial _{j}a_{l}]\nonumber \\
 &  & \; \; \; \; \; \; +2[\partial _{j}a_{l},[\partial _{i}\alpha ,a_{k}]]-2i[[a_{j},a_{l}],[\partial _{i}\alpha ,a_{k}]]\\
 &  & \; \; \; \; \; \; +i\{\partial _{i}\alpha ,\{a_{k},[a_{j},a_{l}]\}\}+\{a_{j},\{a_{l},[\partial _{i}\alpha ,a_{k}]\}\}\Big {)}\nonumber \\
 &  & +\frac{1}{24}c^{kl}\partial _{l}c^{ij}\Big {(}\{\partial _{i}\alpha ,\{a_{k},a_{j}\}\}-2i[\partial _{i}\partial _{k}\alpha ,a_{j}]\Big {)}+\mathcal{O}(3).\nonumber 
\end{eqnarray*}

\subsubsection{Noncommutative matter field}

Now we derive formulas for fields that transforms according to (\ref{sp_gt_field}).
We again expand the noncommuting field in terms of the deformation
parameter\[
\Psi _{\psi }[a]=\psi +\Psi ^{1}_{\psi }[a]+\Psi ^{2}_{\psi }[a]+\cdots .\]
To first order (\ref{sp_gt_field}) reduces again to the classical
transformation law (\ref{com_g_field}). To first order we obtain\[
\delta _{\alpha }\Psi _{\psi }^{1}-i\alpha \Psi _{\psi }^{1}=i\left( \alpha \star _{1}\psi +\Lambda _{\alpha }^{1}\psi \right) =i\left( \frac{1}{2}c^{ij}\partial _{i}\alpha \partial _{j}\psi +\Lambda _{\alpha }^{1}\psi \right) \]
and to second order\[
\delta _{\alpha }\Psi _{\psi }^{2}-i\alpha \Psi _{\psi }^{2}=i\left( \alpha \star _{2}\psi +\alpha \star _{1}\Psi _{\psi }^{1}+\Lambda _{\alpha }^{1}\star _{1}\psi +\Lambda _{\alpha }^{1}\Psi _{\psi }^{1}+\Lambda _{\alpha }^{2}\psi \right) \]
\[
=i\Big {(}\frac{1}{8}c^{mn}c^{ij}\partial _{m}\partial _{i}\alpha ,\partial _{n}\partial _{j}\psi +\frac{1}{12}c^{ml}\partial _{l}c^{ij}(\partial _{m}\partial _{i}\alpha \partial _{j}\psi -\partial _{i}\alpha \partial _{m}\partial _{j}\psi )\]
\[
+\frac{1}{2}c^{ij}\partial _{i}\alpha \partial _{j}\Psi _{\psi }^{1}+\frac{1}{2}c^{ij}\partial _{i}\Lambda _{\alpha }^{1}\partial _{j}\psi +\Lambda _{\alpha }^{1}\Psi _{\psi }^{1}+\Lambda _{\alpha }^{2}\psi \Big {)}.\]
A solution for the noncommutative field is \begin{eqnarray*}
\Psi _{\psi }[a] & = & \psi +\frac{1}{4}c^{ij}\Big {(}2ia_{i}\partial _{j}\psi +a_{i}a_{j}\psi \Big {)}\nonumber \\
 &  & +\frac{1}{32}c^{ij}c^{kl}\Big {(}4i\partial _{i}a_{k}\partial _{j}\partial _{l}\psi -4a_{i}a_{k}\partial _{j}\partial _{l}\psi -8a_{i}\partial _{j}a_{k}\partial _{l}\psi \nonumber \\
 &  & \; \; \; \; \; \; +4a_{i}\partial _{k}a_{j}\partial _{l}\psi +4ia_{i}a_{j}a_{k}\partial _{l}\psi -4ia_{k}a_{j}a_{i}\partial _{l}\psi \nonumber \\
 &  & \; \; \; \; \; \; +4ia_{j}a_{k}a_{i}\partial _{l}\psi -4\partial _{j}a_{k}a_{i}\partial _{l}\psi +2\partial _{i}a_{k}\partial _{j}a_{l}\psi \nonumber \\
 &  & \; \; \; \; \; \; -4ia_{i}a_{l}\partial _{k}a_{j}\psi -4ia_{i}\partial _{k}a_{j}a_{l}\psi +4ia_{i}\partial _{j}a_{k}a_{l}\psi \\
 &  & \; \; \; \; \; \; -3a_{i}a_{j}a_{l}a_{k}\psi -4a_{i}a_{k}a_{j}a_{l}\psi -2a_{i}a_{l}a_{k}a_{j}\psi \Big {)}\nonumber \\
 &  & +\frac{1}{24}c^{kl}\partial _{l}c^{ij}\Big {(}2ia_{j}\partial _{k}\partial _{i}\psi +2i\partial _{k}a_{i}\partial _{j}\psi +2\partial _{k}a_{i}a_{j}\psi \nonumber \\
 &  & \; \; \; \; \; \; -a_{k}a_{i}\partial _{j}\psi -3a_{i}a_{k}\partial _{j}\psi -2ia_{j}a_{k}a_{i}\psi \Big {)}+\mathcal{O}(3).\nonumber 
\end{eqnarray*}

\subsubsection{Covariantizer}

As in the preceeding cases we again expand the covariantizer in terms
of the deformation parameter\[
D(f)=f+D^{1}(f)+D^{2}(f)+\cdots .\]
Since the transformation law (\ref{sp_gt_cov}) to zeroth order is
trivial we can assume that \( D \) starts with the identity. To first
order we get\[
\delta _{\alpha }D^{1}(f)=i[\alpha \stackrel{\star _{1}}{,}f]+i[\alpha ,D^{1}(f)]=\frac{i}{2}c^{ij}[\partial _{i}\alpha ,\partial _{j}f]+i[\alpha ,D^{1}(f)]\]
and to second order\[
\delta _{\alpha }D^{2}(f)=i[\alpha \stackrel{\star _{2}}{,}f]+i[\alpha \stackrel{\star _{1}}{,}D^{1}(f)]+i[\Lambda _{\alpha }^{1}\stackrel{\star _{1}}{,}f]+i[\alpha ,D^{2}(f)]+i[\Lambda ^{1}_{\alpha },D^{2}(f)]\]
\[
=\frac{i}{8}c^{mn}c^{ij}[\partial _{m}\partial _{i}\alpha ,\partial _{n}\partial _{j}f]+\frac{i}{12}c^{ml}\partial _{l}c^{ij}([\partial _{m}\partial _{i}\alpha ,\partial _{j}f]-[\partial _{i}\alpha ,\partial _{m}\partial _{j}f])\]
\[
+\frac{i}{2}c^{ij}[\partial _{i}\alpha ,\partial _{j}D^{1}(f)]+\frac{i}{2}c^{ij}[\partial _{i}\Lambda _{\alpha }^{1},\partial _{j}f]+i[\alpha ,D^{2}(f)]+i[\Lambda ^{1}_{\alpha },D^{2}(f)].\]
A solution to this is

\[
\begin{array}{rcl}
D[a](f) & = & f+ic^{ij}a_{i}\partial _{j}f\\
 &  & +\frac{1}{4}c^{ij}c^{kl}\Big {(}-2\{a_{i},\partial _{j}a_{k}\}\partial _{l}f+\{a_{i},\partial _{k}a_{j}\}\partial _{l}f\\
 &  & \; \; \; \; \; \; +i\{a_{i},[a_{j},a_{k}]\}\partial _{l}f-\{a_{i},a_{k}\}\partial _{j}\partial _{l}f\Big {)}\\
 &  & +\frac{1}{4}c^{il}\partial _{l}c^{jk}\{a_{i},a_{k}\}\partial _{j}f+\mathcal{O}(3).
\end{array}\]

\subsubsection{Noncommutative gauge potential}

Again we expand the noncommutative gauge potential, starting with
the usual one\[
A_{X}=X^{n}a_{n}+A^{1}_{X}+A^{2}_{X}+\cdots .\]
 Since it is a noncommutative form in the sense of (\ref{nc_forms}),
we have to evaluate it on a Poisson vector field \( X \). We again
expand the equation (\ref{sp_gt_gauge_pot}) and obtain to first order\[
\delta _{\alpha }A^{1}_{X}=X^{i}\partial _{i}\Lambda ^{1}_{\alpha }+\delta ^{1}_{X}\alpha +i[\alpha \stackrel{\star _{1}}{,}X^{n}a_{n}]+i[\alpha ,A^{1}_{X}]+i[\Lambda ^{1}_{\alpha },X^{n}a_{n}]\]
\[
=X^{i}\partial _{i}\Lambda ^{1}_{\alpha }+\frac{i}{2}c^{ij}[\partial _{i}\alpha ,\partial _{j}(X^{n}a_{n})]+i[\alpha ,A^{1}_{X}]+i[\Lambda ^{1}_{\alpha },X^{n}a_{n}].\]
For the second order we get \[
\delta _{\alpha }A^{2}_{X}=i[\alpha \stackrel{\star _{2}}{,}X^{n}a_{n}]+i[\alpha \stackrel{\star _{1}}{,}A^{1}_{X}]+i[\Lambda ^{1}_{\alpha }\stackrel{\star _{1}}{,}X^{n}a_{n}]\]
\[
+X^{i}\partial _{i}\Lambda ^{2}_{\alpha }+\delta ^{2}_{X}\alpha +\delta ^{1}_{X}\Lambda ^{1}_{\alpha }+i[\alpha ,A^{2}_{X}]+i[\Lambda ^{1}_{\alpha },A^{1}_{X}]+i[\Lambda ^{2}_{\alpha },X^{n}a_{n}]\]
\[
=\frac{i}{8}c^{mn}c^{ij}[\partial _{m}\partial _{i}\alpha ,\partial _{n}\partial _{j}(X^{n}a_{n})]\]
\[
+\frac{i}{12}c^{ml}\partial _{l}c^{ij}([\partial _{m}\partial _{i}\alpha ,\partial _{j}(X^{n}a_{n})]-[\partial _{i}\alpha ,\partial _{m}\partial _{j}(X^{n}a_{n})])\]
\[
+\frac{i}{2}c^{ij}[\partial _{i}\alpha ,\partial _{j}A^{1}_{X}]+\frac{i}{2}c^{ij}[\partial _{i}\Lambda ^{1}_{\alpha },\partial _{j}(X^{n}a_{n})]\]
\[
-\frac{1}{12}c^{lk}\partial _{k}c^{im}\partial _{l}\partial _{m}X^{j}\partial _{i}\partial _{j}\alpha +\frac{1}{24}c^{lk}c^{im}\partial _{l}\partial _{i}X^{j}\partial _{k}\partial _{m}\partial _{j}\alpha \]
\[
+X^{i}\partial _{i}\Lambda ^{2}_{\alpha }+\delta ^{1}_{X}\Lambda ^{1}_{\alpha }+i[\alpha ,A^{2}_{X}]+i[\Lambda ^{1}_{\alpha },A^{1}_{X}]+i[\Lambda ^{2}_{\alpha },X^{n}a_{n}].\]
We found the following solution to the noncommutative gauge potential
\begin{eqnarray*}
A_{X} & = & X^{n}a_{n}+\frac{i}{4}c^{kl}X^{n}\{a_{k},\partial _{l}a_{n}+f_{ln}\}+\frac{i}{4}c^{kl}\partial _{l}X^{n}\{a_{k},a_{n}\}\\
 &  & +\frac{1}{32}c^{kl}c^{ij}X^{n}\Big {(}-4i[\partial _{k}\partial _{i}a_{n},\partial _{l}a_{j}]\\
 &  & \; \; \; \; \; \; \; \; \; +2i[\partial _{k}\partial _{n}a_{i},\partial _{l}a_{j}]-4\{a_{k},\{a_{i},\partial _{j}f_{ln}\}\}\\
 &  & \; \; \; \; \; \; \; \; \; -2[[\partial _{k}a_{i},a_{n}],\partial _{l}a_{j}]+4\{\partial _{l}a_{n},\{\partial _{i}a_{k},a_{j}\}\}\\
 &  & \; \; \; \; \; \; \; \; \; -4\{a_{k},\{f_{li},f_{jn}\}\}+i\{\partial _{n}a_{j},\{a_{l},[a_{i},a_{k}]\}\}\\
 &  & \; \; \; \; \; \; \; \; \; +i\{a_{i},\{a_{k},[\partial _{n}a_{j},a_{l}]\}\}-4i[[a_{i},a_{l}],[a_{k},\partial _{j}a_{n}]]\\
 &  & \; \; \; \; \; \; \; \; \; +2i[[a_{i},a_{l}],[a_{k},\partial _{n}a_{j}]]+\{a_{i},\{a_{k},[a_{l},[a_{j},a_{n}]]\}\}\\
 &  & \; \; \; \; \; \; \; \; \; -\{a_{k},\{[a_{l},a_{i}],[a_{j},a_{n}]\}\}-[[a_{i},a_{l}],[a_{k},[a_{j},a_{n}]]]\Big {)}\\
 &  & +\frac{1}{32}c^{kl}c^{ij}\partial _{j}X^{n}\Big {(}2i[\partial _{k}a_{i},\partial _{l}a_{n}]+2i[\partial _{i}a_{k},\partial _{l}a_{n}]\\
 &  & \; \; \; \; \; \; \; \; \; +2i[\partial _{i}a_{k},\partial _{l}a_{n}-\partial _{n}a_{l}]+4\{a_{n},\{a_{l},\partial _{k}a_{i}\}\}\\
 &  & \; \; \; \; \; \; \; \; \; +4\{a_{k},\{a_{i},\partial _{n}a_{l}-\partial _{l}a_{n}\}\}-2i\{a_{k},\{a_{i},[a_{n},a_{l}]\}\}\\
 &  & \; \; \; \; \; \; \; \; \; +i\{a_{i},\{a_{l},[a_{n},a_{k}]\}\}+i\{a_{n},\{a_{l},[a_{i},a_{k}]\}\}\Big {)}\\
 &  & +\frac{1}{24}c^{kl}c^{ij}\partial _{l}\partial _{j}X^{n}\Big {(}\partial _{i}\partial _{k}a_{n}-2i[a_{i},\partial _{k}a_{n}]-\{a_{n},\{a_{k},a_{i}\}\}\Big {)}\\
 &  & +\frac{1}{24}c^{kl}\partial _{l}c^{ij}X^{n}\Big {(}2i[a_{j},\partial _{k}\partial _{i}a_{n}]+2i[\partial _{k}a_{i},f_{jn}]\\
 &  & \; \; \; \; \; \; \; \; \; -\{\partial _{j}a_{n},\{a_{k},a_{i}\}\}+2\{a_{i},\{a_{k},f_{nj}\}\}\Big {)}\\
 &  & +\frac{1}{24}c^{kl}\partial _{l}c^{ij}\partial _{j}X^{n}\Big {(}-4i[a_{i},\partial _{k}a_{n}]+2i[a_{k},\partial _{i}a_{n}]-\{a_{n},\{a_{k},a_{i}\}\}\Big {)}\\
 &  & -\frac{1}{12}c^{kl}\partial _{l}c^{ij}\partial _{j}\partial _{k}X^{n}\partial _{i}a_{n}+\mathcal{O}(3).
\end{eqnarray*}

\subsection{Seiberg-Witten map for Formality \protect\( \star \protect \)-products}

We will now apply the formalism of Seiberg-Witten gauge theory to
Kontsevich's formality \( \star  \)-product. Here we are able to
calculate the abelian Seiberg-Witten map up to all orders. We have
seen that derivations for this \( \star  \)-product are easily obtained
from Poisson vector fields. With them we have all the key ingredients
to do noncommutative gauge theory on any Poisson manifold. To relate
the noncommutative theory to commutative gauge theory, we need the
Seiberg-Witten maps for the formality \( \star  \)-product. In \cite{Jurco:2000fs}
and \cite{Jurco:2001my} the Seiberg-Witten maps for the noncommutative
gauge parameter and the covariantizer were already constructed to
all orders in \( \theta  \) for abelian gauge theory. We will extend
the method developed there to the Seiberg-Witten map for covariant
derivations.

\subsubsection{Semi-classical construction}

We will first do the construction in the semi-classical limit, where
the star commutator is replaced by the Poisson bracket. As in \cite{Jurco:2000fs}
and \cite{Jurco:2001my}, we define, with the help of the Poisson
tensor \( \theta =\frac{1}{2}\theta ^{kl}\partial _{k}\wedge \partial _{l} \)
\[
d_{\theta }=-[\cdot ,\theta ]\]
and (locally) \[
a_{\theta }=\theta ^{ij}a_{j}\partial _{i}.\]
Note that the bracket used in the definition of \( d_{\theta } \)
is not the Schouten-Nijenhuis bracket (\ref{Schouten-Nijenhuis bracket}).
For polyvectorfields \( \pi _{1} \) and \( \pi _{2} \) it is \[
{}[\pi _{1},\pi _{2}]=-[\pi _{2},\pi _{1}]_{S},\]
giving an extra minus sign for \( \pi _{1} \) and \( \pi _{2} \)
both even (see \ref{Quantum commutators}). Especially, we get for
\( d_{\theta } \) acting on a function \( g \)

\[
d_{\theta }g=-[g,\theta ]=[g,\theta ]_{S}=\theta ^{kl}\partial _{l}g\partial _{k}.\]
Now a parameter \( t \) and \( t \)-dependent \( \theta _{t}=\frac{1}{2}\theta _{t}^{kl}\partial _{k}\wedge \partial _{l} \)
and \( X_{t}=X^{k}_{t}\partial _{k} \) are introduced, fulfilling

\[
\partial _{t}\theta _{t}=f_{\theta }=-\theta _{t}f\theta _{t}\; \; \; \; \mbox {and}\; \; \; \; \partial _{t}X_{t}=-X_{t}f\theta _{t},\]
where the multiplication is ordinary matrix multiplication. Given
the Poisson tensor \( \theta  \) and the Poisson vectorfield \( X \),
the formal solutions are\[
\theta _{t}=\theta \sum _{n=0}^{\infty }(-t\; f\theta )^{n}=\frac{1}{2}(\theta ^{kl}-t\theta ^{ki}f_{ij}\theta ^{jl}_{}+\ldots )\partial _{k}\wedge \partial _{l}\]
and

\[
X_{t}=X\sum _{n=0}^{\infty }(-t\; f\theta )^{n}=X^{k}\partial _{k}-tX^{i}f_{ij}\theta ^{jk}\partial _{k}+\ldots .\]
\( \theta _{t} \) is still a Poisson tensor and \( X_{t} \) is still
a Poisson vectorfield, i.e.

\[
[\theta _{t},\theta _{t}]=0\; \; \; \; \; \; \; \mbox {and}\; \; \; \; \; \; \; \; [X_{t},\theta _{t}]=0.\]
 For the proof see \ref{Commutator theta theta and theta X}.

With this we calculate\begin{equation}
\label{f theta}
f_{\theta }=\partial _{t}\theta _{t}=-\theta _{t}f\theta _{t}=-[a_{\theta },\theta ]=d_{\theta }a_{\theta }.
\end{equation}
We now get the following commutation relations

\begin{eqnarray}
{[}a_{\theta _{t}}+\partial _{t},d_{\theta _{t}}(g)] & = & d_{\theta _{t}}((a_{\theta _{t}}+\partial _{t})(g)),\label{comm rel 1} \\
{[}a_{\theta _{t}}+\partial _{t},X_{t}] & = & -d_{\theta _{t}}(X_{t}^{k}a_{k}),\label{comm rel} 
\end{eqnarray}
where \( g \) is some function which might also depend on \( t \)
(see \ref{Semi-classical commutators}).

To construct the Seiberg-Witten map for the gauge potential \( A_{X} \),
we first define

\[
K_{t}=\sum ^{\infty }_{n=0}\frac{1}{(n+1)!}(a_{\theta _{t}}+\partial _{t})^{n}.\]
With this, the semi-classical gauge parameter reads \cite{Jurco:2000fs,Jurco:2001my}
\[
\Lambda _{\lambda }[a]=K_{t}(\lambda )\Big {|}_{t=0}.\]
To see that this has indeed the right transformation properties under
gauge transformations, we first note that the transformation properties
of \( a_{\theta _{t}} \) and \( X^{k}_{t}a_{k} \) are\begin{equation}
\label{delta a theta}
\delta _{\lambda }a_{\theta _{t}}=\theta ^{kl}_{t}\partial _{l}\lambda \partial _{k}=d_{\theta _{t}}\lambda 
\end{equation}
and\begin{equation}
\label{delta X}
\delta _{\lambda }(X_{t}^{k}a_{k})=X_{t}^{k}\partial _{k}\lambda =[X_{t},\lambda ].
\end{equation}
Using (\ref{delta a theta}), (\ref{delta X}) and the commutation
relations (\ref{comm rel 1}), (\ref{comm rel}), a rather tedious
calculation (see \ref{Transformation properties of K}) shows that
\[
\delta _{\lambda }K_{t}(X_{t}^{k}a_{k})=X_{t}^{k}\partial _{k}K_{t}(\lambda )+d_{\theta _{t}}(K_{t}(\lambda ))K_{t}(X_{t}^{k}a_{k}).\]
Therefore, the semi-classical gauge potential is

\[
A_{X}[a]=K_{t}(X_{t}^{k}a_{k})\Big {|}_{t=0}.\]

\subsubsection{Quantum construction}

We can now use the Kontsevich formality map to quantise the semi-classical
construction. All the semi-classical expressions can be mapped to
their counterparts in the \( \star  \)-product formalism without
loosing the properties necessary for the construction. One higher
order term will appear, fixing the transformation properties for the
quantum objects.

The star-product we will use is \[
\star =\sum ^{\infty }_{n=0}\frac{1}{n!}U_{n}(\theta _{t},\, \ldots ,\theta _{t}).\]
 We define 

\[
d_{\star }=-[\cdot ,\star ]_{G\: \: },\]
which for functions \( f \) and \( g \) reads

\[
d_{\star }(g)\: f=[f\stackrel{\star }{,}g].\]
The bracket used in the definition of \( d_{\star } \) is the Gerstenhaber
bracket (\ref{Gerstenhaber bracket}). We now calculate the commutators
(\ref{comm rel 1}) and (\ref{comm rel}) in the new setting (see
\ref{Quantum commutators}). We get

\begin{eqnarray*}
{[}\Phi (a_{\theta _{t}})+\partial _{t},d_{\star }(\Phi (g))] & = & d_{\star }((\Phi (a_{\theta _{t}})+\partial _{t})\Phi (f)),\\
{[}\Phi (a_{\theta _{t}})+\partial _{t},\Phi (X_{t})] & = & -d_{\star }(\Phi (X_{t}^{k}a_{k})-\Psi (a_{\theta _{t}},X_{t})).
\end{eqnarray*}
The higher order term \( \Psi (a_{\theta _{t}},X_{t}) \) has appeared,
but looking at the gauge transformation properties of the quantum
objects we see that it is actually necessary. We get \[
\delta _{\lambda }\Phi (a_{\theta _{t}})=\Phi (d_{\theta _{t}}\lambda )=d_{\star }\Phi (\lambda )\]
with (\ref{formality star derivation}) and (\ref{delta a theta})
and\begin{eqnarray*}
\delta _{\lambda }(\Phi (X_{t}^{k}a_{k})-\Psi (a_{\theta },X_{t})) & = & \Phi ([X_{t},\lambda ])-\Psi (d_{\theta }\lambda ,X_{t})\\
 & = & [\Phi (X_{t}),\Phi (\lambda )]-\Psi ([\theta _{t},\lambda ],X_{t})\\
 &  & +\Psi ([\theta _{t},X_{t}],\lambda )-\Psi (d_{\theta }\lambda ,X_{t})\\
 & = & [\Phi (X_{t}),\Phi (\lambda )]\\
 & = & \delta _{X_{t}}\Phi (\lambda ),
\end{eqnarray*}
where the addition of the new term preserves the correct transformation
property. With

\[
K_{t}^{\star }=\sum _{n=0}^{\infty }\frac{1}{(n+1)!}(\Phi (a_{\theta _{t}})+\partial _{t})^{n},\]
a calculation analogous to the semi-classical case gives \begin{eqnarray*}
\delta _{\lambda }(K_{t}^{\star }(\Phi (X_{t}^{k}a_{k})-\Psi (a_{\theta _{t}},X_{t}))) & = & \delta _{X_{t}}K_{t}^{\star }(\Phi (\lambda ))\\
 &  & +d_{\star }(K_{t}^{\star }(\Phi (\lambda )))K_{t}^{\star }(\Phi (X_{t}^{k}a_{k})-\Psi (a_{\theta _{t}},X_{t})).
\end{eqnarray*}
As in \cite{Jurco:2000fs,Jurco:2001my}, the noncommutative gauge
parameter is

\[
\Lambda _{\lambda }[a]=K_{t}^{\star }(\Phi (\lambda ))\Big {|}_{t=0},\]
and we therefore get for the noncommutative gauge potential

\[
A_{X}[a]=K_{t}^{\star }(\Phi (X_{t}^{k}a_{k})-\Psi (a_{\theta _{t}},X_{t}))\Big {|}_{t=0},\]
transforming with

\[
\delta _{\lambda }A_{X}=\delta _{X}\Lambda _{\lambda }-[\Lambda _{\lambda }\stackrel{\star }{,}A_{X}].\]

\subsection{\label{constrcution_of_gauge_inv_actions}Construction of gauge invariant
actions}

We have seen that in the noncommutative realm the integral may be
replaced by a trace on a representation of the algebra describing
the noncommutative space. In the \( \star  \)-product formalism this
has been the ordinary integral together with a measure function (\ref{equation_for_measure_function}).
With the covariantizer \( D[a](f) \) (\ref{sp_gt_cov}) for functions
at hand it is now easy to construct actions invariant under noncommutative
gauge transformations that reduce in the classical limit to gauge
theory on a flat space. For example the measure function can be compensated
by \( D[a](\Omega ^{-1}(x)) \). But from the point of view of noncommutative
gauge theory this looks quite unnatural. To make contact with the
commuting frame formalism we will have to go another way.

\bigskip

First we want to translate classical gauge theory (\ref{classical gauge theory})
into the language of frames. Since forms are dual to vector fields,
they may be evaluated on a frame. In the special case of the connection
one form this yields\[
a_{a}=a(e_{a})=a_{i}dx^{i}(e_{a})=a_{i}e^{i}{}_{a}.\]
The same we can do with the covariant derivate \[
(D\psi )(e_{a})=e_{a}\psi -ia_{a}\psi .\]
The field strength becomes\[
f(e_{a},e_{b})=f_{ab}=e_{a}a_{b}-e_{b}a_{a}-a([e_{a},e_{b}])-i[a_{a},a_{b}].\]
Since in scalar electrodynamics we do not need a spin connection,
it is simple to write down its action on an curved manifold with the
frame formalism\[
\mathcal{S}=\int d^{n}x\, e\, (-\frac{1}{4}\eta ^{ab}\eta ^{cd}f_{ac}f_{bd}+\eta ^{ab}D_{a}\overline{\phi }D_{b}\phi +m^{2}\overline{\phi }\phi ).\]
 Here again \[
e=(\det \, e_{a}{}^{\mu })^{-1}=\sqrt{\det \, (g_{\mu \nu })}\]
is the measure function for the curved manifold.

\bigskip

The considerations above can be generalized to a curved noncommutative
space, i.e. a noncommutative space with a Poisson structure that is
compatible with a frame \( e_{a} \). For a curved noncommutative
space we are now able to mimic the previous classical constructions
and evaluate the noncommutative covariant derivative (\ref{sp_gt_cov_derivative})
and field strength (\ref{sp_field_strength}) on it\[
D_{a}\Phi =\delta _{e_{a}}\Phi -iA_{e_{a}}\star \Phi ,\]
\[
F_{ab}=F(e_{a},e_{b}).\]
Using the measure function and our noncommutative versions of field
strength and covariant derivative we end up with the following action
\begin{equation}
\label{noncommutative_gauge_action}
\mathcal{S}=tr\, \int d^{n}x\, \Omega \, (-\frac{1}{4}\eta ^{ab}\eta ^{cd}F_{ac}\star F_{bd}+\eta ^{ab}D_{a}\overline{\Phi }\star D_{b}\Phi -m^{2}\overline{\Phi }\star \Phi ).
\end{equation}
 \( tr \) is the trace of the Lie algebra representation. By construction
this action is invariant under noncommutative gauge transformations\[
\delta _{\alpha }S=0.\]
To lowest order we obtain\[
\mathcal{S}_{0}=tr\, \int d^{n}x\, \Omega \, (-\frac{1}{4}g^{\alpha \beta }g^{\gamma \delta }f_{\alpha \gamma }f_{\beta \delta }+g^{\alpha \beta }D_{\alpha }\bar{\phi }D_{\beta }\phi -m^{2}\bar{\phi }\phi ),\]
with \( g_{\alpha \beta } \) the metric induced by the frame. If
\( g=\Omega  \), the commuting frame formalism yields the desired
classical limit.

\subsection{\label{example_so_a(n)_and_sw_map}Example: \protect\( M(so_{a}(n))\protect \)}

We have seen that the components of the frame (\( e_{\alpha }=X_{\alpha }{}^{\mu }\partial _{u} \))
are \begin{eqnarray*}
X^{\mu }_{0} & = & \delta _{0}^{\mu },\\
X^{\mu }_{i} & = & \rho \delta ^{\mu }_{i}.
\end{eqnarray*}
These we can plug into our solution of the Seiberg-Witten map and
the derivation corresponding to the Weyl-ordered \( \star  \)-product
and get\begin{eqnarray*}
\Lambda _{\lambda }[a] & = & \lambda +\frac{a}{4}x^{i}\{\partial _{0}\lambda ,a_{i}\}-\frac{a}{4}x^{i}\{\partial _{i}\lambda ,a_{0}\}+\mathcal{O}(a^{2}),\\
\Phi _{\phi }[a] & = & \phi -\frac{a}{2}x^{i}a_{0}\partial _{i}\phi +\frac{a}{2}x^{i}a_{i}\partial _{0}\phi +\frac{ia}{4}x^{i}[a_{0},a_{i}]\phi +\mathcal{O}(a^{2}),\\
A_{X_{0}} & = & a_{0}-\frac{a}{4}x^{i}\{a_{0},\partial _{i}a_{0}+f_{i0}\}+\frac{a}{4}x^{i}\{a_{i},\partial _{0}a_{0}\}+\mathcal{O}(a^{2}),\\
A_{X_{j}} & = & \rho a_{j}-\frac{a}{4}\rho \{a_{j},a_{0}\}-\frac{a}{4}\rho x^{i}\{a_{0},\partial _{i}a_{j}+f_{ij}\}\\
 &  & +\frac{a}{4}\rho x^{i}\{a_{i},\partial _{0}a_{j}+f_{0j}\}+\mathcal{O}(a^{2}),\\
\delta _{X_{\mu }} & = & X^{\nu }_{\mu }\partial _{\nu }+\mathcal{O}(a^{2}).
\end{eqnarray*}
 The action (\ref{noncommutative_gauge_action}) becomes up to first
order\begin{eqnarray*}
S=\int d^{n}x\, \Big {(} &  & -\frac{1}{2}\rho ^{3-n}\eta ^{00}\eta ^{ij}Tr(f_{0i}f_{0j})-\frac{1}{4}\rho ^{5-n}\eta ^{kl}\eta ^{ij}Tr(f_{ki}f_{lj})\\
 &  & +\rho ^{1-n}\eta ^{00}\overline{D_{0}\phi }D_{0}\phi +\rho ^{3-n}\eta ^{kl}\overline{D_{k}\phi }D_{l}\phi \\
 &  & -\frac{a}{2}\rho ^{3-n}\eta ^{00}\eta ^{ij}x^{p}Tr(f_{0p}f_{0i}f_{0j})\\
 &  & +\frac{a}{4}\rho ^{5-n}\eta ^{kl}\eta ^{ij}x^{p}Tr(f_{0p}f_{ki}f_{lj})\\
 &  & -\frac{a}{2}\rho ^{5-n}\eta ^{kl}\eta ^{ij}x^{p}Tr(f_{jp}\{f_{ki},f_{l0}\})\\
 &  & -\frac{a}{2}\rho ^{3-n}\eta ^{kl}x^{i}\overline{D_{k}\phi }f_{l0}D_{i}\phi +\frac{a}{2}\rho ^{3-n}\eta ^{kl}x^{i}\overline{D_{k}\phi }f_{li}D_{0}\phi \\
 &  & -\frac{a}{2}\rho ^{3-n}\eta ^{kl}x^{i}\overline{D_{i}\phi }f_{l0}D_{k}\phi +\frac{a}{2}\rho ^{3-n}\eta ^{kl}x^{i}\overline{D_{0}\phi }f_{li}D_{k}\phi \\
 &  & -a\rho ^{3-n}\eta ^{kl}x^{i}\overline{D_{k}\phi }f_{0i}D_{l}\phi \, \Big {)}+\mathcal{O}(a^{2}).
\end{eqnarray*}
We know that in the classical limit \( a\rightarrow 0 \) the action
reduces to scalar electrodynamics on a curved background or its nonabelian
generalization, respectively.

\section{Observables}

In the previous sections we have seen that gauge theory on noncommutative
spaces is a very interesting and fruitful subject. Nevertheless we
need a method to extract physical predictions from the theory. Since
a gauge transformation should not affect the predictions we make,
we have to find gauge invariant objects. Such observables are not
easy to find if we want them to have a sensible classical limit.

A second reason for studying observables is the similarity between
noncommutative gauge theory and gravity in view of the gauge structure.
The equations of general relativity transform covariantly under coordinate
transformations. Therefore the group of local diffemorphisms is part
of the gauge group. Now take a gauge transformation in the \( \star  \)-product
representation of a noncommutative \( U(1) \)-gauge theory. Then
we have seen that the coordiantes are not invariant under gauge transformations
\begin{equation}
\label{finitel_gauge_transformation_expanded}
x^{i}\rightarrow e^{i\alpha (x)}_{\star }\star x^{i}\star e^{-i\alpha (x)}_{\star }=x^{i}+\theta ^{ij}\partial _{j}\alpha +\cdots 
\end{equation}
In the whole section we will assume that the \( \star  \)-product
looks up to first order like \begin{equation}
\label{DEF_STAR_PRODUCT}
f\star g=fg+\frac{i}{2}\theta ^{ij}\partial _{i}f\partial _{j}g+\cdots ,
\end{equation}
where \( \theta ^{ij} \) is antisymmetric and fulfills the Poisson
equation. In the semiclassical limit the transformations (\ref{finitel_gauge_transformation_expanded})
become the Hamiltonian flows of the Poisson manifold. In a sense the
gauge group of noncommutative gauge theories contains a large class
of diffeomorphisms. Since it is not easy to find a full set of meaningful
observables in general relativity (see e. g. \cite{Perez:2001gj}
and for a more general review \cite{Rovelli:1999hz}), the study of
nonncommutative gauge theory will perhaps give new insights into this
subject.

In the case of constant commutator so called open Wilson lines \cite{Ishibashi:1999hs}
have been introduced as observables of noncommutative gauge theory.
We will use covariant coordinates (\ref{definition_covariant_coordinates})
to generalize this construction to general \( \star  \)-products.
In \cite{Okawa:2001mv} they were used to give an exact formula for
the inverse Seiberg-Witten map. We will generalize this construction
for \( \star  \)-products with invertible Poisson structure \( \theta ^{ij} \).

\subsection{Classical Wilson lines}

Let us first recall some aspects of the commutative gauge theory.
For this let \( a_{\mu } \) be a gauge field. Then an infinitesimal
parallel transporter (infinitesimal wilson line) may be defined via\begin{eqnarray*}
U(x,x+l) & = & 1+il^{\mu }a_{\mu }(x)\\
 & = & e^{il^{\mu }a_{\mu }(x)}+\mathcal{O}(l^{2}),
\end{eqnarray*}
where \( l^{\mu } \) is an infinitesimal constant vector. The infinitesimal
Wilson line transforms like\[
T_{g}U(x,x+l)=g(x)U(x,x+l)g^{-1}(x+l)+\mathcal{O}(l^{2}).\]
Now let \( \Gamma _{yz} \) a Path connecting the points \( a \)
and \( b \). And let \( \{x_{i}\}_{i\in 0\dots N} \) be a partition
of this Path. Then we define\begin{eqnarray*}
U_{N}[\Gamma _{yz}] & = & \prod ^{N}_{i=1}U(x_{i-1},x_{i})\\
 & = & \prod ^{N}_{i=1}\left( 1+i(x^{\mu }_{i}-x^{\mu }_{i-1})a_{\mu }(x_{i})\right) \\
 & = & \prod ^{N}_{i=1}\left( 1+il_{i}^{\mu }a_{\mu }(y+\sum ^{i}_{j=1}l_{i})\right) .
\end{eqnarray*}
 \( U_{N} \) transforms in the following way\[
T_{g}U_{N}[\Gamma _{xy}]=g(x)U_{N}[\Gamma _{xy}]g^{-1}(y)+\mathcal{O}(l_{i}^{2}).\]
Further the Wilson line of the Path \( \Gamma _{xy} \) is the continuum
limit of the \( U_{N} \)\begin{eqnarray*}
U[\Gamma _{xy}] & = & \lim _{N\rightarrow \infty }U_{N}[\Gamma _{xy}]\\
 & = & P\exp (i\int dx^{\mu }a_{\mu }),
\end{eqnarray*}
where \( P \) denotes path ordering of the exponential. If one acts
with a gauge transformation on the Wilson line\[
T_{g}U[\Gamma _{xy}]=g(x)U[\Gamma _{xy}]g^{-1}(y),\]
one sees that it transforms only at its endpoints.

\subsection{Noncommutative Wilson lines }

In the case \( \theta ^{ij}=const. \) the basic observation was that
translations in space are gauge transformations \cite{Ishibashi:1999hs}.
They are realized by \[
T_{l}x^{j}=x^{j}+l_{i}\theta ^{ij}=e^{il_{i}x^{i}}\star f\star e^{-il_{i}x^{i}}.\]
Now one can pose the question what happens if one uses covariant coordinates
\cite{Behr:2003hg}. In this case the inner automorphism\[
f\rightarrow e^{il_{i}X^{i}}\star f\star e^{-il_{i}X^{i}}\]
should consist of a translation and a gauge transformation dependent
of the translation. If we subtract the translation again only the
gauge transformation remains and the resulting object \[
W_{l}=e^{il_{i}X^{i}}\star e^{-il_{i}x^{i}}\]
has a very interesting transformation behavior under a gauge transformation\[
W_{l}^{\prime }(x)=g(x)\star W_{l}(x)\star g^{-1}(x+l_{i}\theta ^{ij}).\]
It transforms like a Wilson line starting at \( x \) and ending at
\( x+l\theta  \).

As in the constant case we can start with \[
W_{l}=e^{il_{i}X^{i}}_{\star }\star e^{-il_{i}x^{i}}_{\star },\]
where now \( e_{\star } \) is the \( \star  \)-exponential. Every
multiplication in its Taylor series is replaced by the \( \star  \)-product.
In contrast to the constant case, \( e^{l_{i}x^{i}}_{\star }=e^{l_{i}x^{i}} \)
isn't true any more. The transformation property of \( W_{l} \) is
now\[
W^{\prime }_{l}(x)=g(x)\star W_{l}(x)\star g^{-1}(T_{l}x),\]
where \[
T_{l}x^{j}=e^{il_{i}x^{i}}\star x^{j}\star e^{-il_{i}x^{i}}\]
is an inner automorphism of the algebra, which can be interpreted
as a quantized coordinate transformation. If we replace commutators
by Poisson brackets, the classical limit of this coordinate transformation
may be calculated\[
T_{l}x^{k}=e^{il_{i}[x^{i}\stackrel{\star }{,}\cdot ]}x^{k}\approx e^{-l_{i}\{x^{i},\cdot \}}x^{j}=e^{-l_{i}\theta ^{ij}\partial _{j}}x^{k},\]
the formula becoming exact for \( \theta ^{ij} \) constant or linear
in \( x \). We see that the classical coordinate transformation is
the flow induced by the Hamiltonian vector field \( -l_{i}\theta ^{ij}\partial _{j} \).
At the end we may expand \( W_{l} \) in terms of \( \theta  \) and
get\[
W_{l}=e^{il_{i}\theta ^{ij}a_{j}}+\mathcal{O}(\theta ^{2}),\]
where we have replaced \( A^{i} \) by its Seiberg-Witten expansion.
We see that for \( l \) small this really is a Wilson line starting
at \( x \) and ending at \( x+l\theta  \).

\subsection{Observables}

Now we are able to write down a large class of observables for the
above defined noncommutative gauge theory, namely\[
U_{l}=\int d^{n}x\, \Omega (x)\, W_{l}(x)\star e_{\star }^{il_{i}x^{i}}=\int d^{n}x\, \Omega (x)\, e^{il_{i}X^{i}(x)}_{\star }\]
or more general\[
f_{l}=\int d^{n}x\, \Omega (x)\, f(X^{i})\star e^{il_{i}X^{i}(x)}_{\star }\]
with \( f \) an arbitrary function of the covariant coordinates.
Obviously they are invariant under gauge transformations.

\subsection{Inverse Seiberg-Witten-map}

As an application of the above constructed observables we generalize
\cite{Okawa:2001mv} to arbitrary \( \star  \)-products, i. e. we
give a formula for the inverse Seiberg-Witten map for \( \star  \)-products
with invertible Poisson structure. In order to map noncommutative
gauge theory to its commutative counterpart we need a functional \( f_{ij}[X] \)
fulfilling 

\[
f_{ij}[g\star X\star g^{-1}]=f_{ij}[X],\]
\[
df=0\]
and\[
f_{ij}=\partial _{i}a_{j}-\partial _{j}a_{i}+\mathcal{O}(\theta ^{2}).\]
 \( f \) is a classical field strength and reduces in the limit \( \theta \rightarrow 0 \)
to the correct expression.

To proove the first and the second property we will only use the algebra
properties of the \( \star  \)-product and the cyclicity of the trace.
All quantities with a hat will be elements of an algebra. With this
let \( \hat{X}^{i} \) be covariant coordinates in an algebra, transforming
under gauge transformations like\[
\hat{X}^{i\prime }=\hat{g}\hat{X}^{i}\hat{g}^{-1}\]
with \( \hat{g} \) an invertible element of the algebra. Now define
\[
\hat{F}^{ij}=-i[\hat{X}^{i},\hat{X}^{j}]\]
and \[
(\hat{F}^{n-1})_{ij}=\frac{1}{2^{n-1}(n-1)!}\epsilon _{iji_{1}i_{2}\cdots i_{2n-2}}\hat{F}^{i_{1}i_{2}}\cdots \hat{F}^{i_{2n-3}i_{2n-2}}.\]
Since an antisymmetric matrix in odd dimensions is never invertible
we have assumed that the space is \( 2n \) dimensional. The expression\begin{equation}
\label{DEF_F}
\mathcal{F}_{ij}(k)=str_{\hat{F},\hat{X}}\left( (\hat{F}^{n-1})_{ij}e^{ik_{j}\hat{X}^{j}}\right) 
\end{equation}
clearly fulfills the first property due to the properties of the trace.
\( str \) is the symmetrized trace, every monomial in \( \hat{F}^{ij} \)and
\( \hat{X}^{k} \) should be symmetrized. For an exact definition
see \cite{Okawa:2001mv}. Note that symmetrization is only necessary
for spaces with dimension higher than 4 due to the cyclicity of the
trace. In dimensions 2 and 4 we may replace \( str \) by the ordinary
trace \( tr \). \( \mathcal{F}_{ij}(k) \) is the Fourier transform
of a closed form if \[
k_{[i}\mathcal{F}_{jk]}=0\]
or if the current \[
J^{i_{1}\cdots i_{2n-2}}=str_{\hat{F},X}\left( \hat{F}^{[i_{1}i_{2}}\cdots \hat{F}^{i_{2n-3}i_{2n-2}]}e^{ik_{j}\hat{F}^{j}}\right) \]
is conserved, respectively\[
k_{i}J^{i\cdots }=0.\]
This is easy to show, if one uses\[
str_{\hat{F},\hat{X}}\left( [k\hat{X},\hat{X}^{l}]e^{ik_{j}\hat{X}^{j}}\cdots \right) =str_{\hat{F},\hat{X}}\left( [\hat{X}^{l},e^{ik_{j}\hat{X}^{j}}]\cdots \right) =str_{\hat{F},\hat{X}}\left( e^{ik_{j}\hat{X}^{j}}[\hat{X}^{l},\cdots ]\right) ,\]
which can be calculated by simple algebra.

To show the last property we have to switch to the \( \star  \)-product
formalism and expand the formula in \( \theta ^{ij} \). The expression
(\ref{DEF_F}) now becomes \[
\mathcal{F}[X]_{ij}(k)=\int \frac{d^{2n}x}{Pf(\theta )}\left( (F_{\star }^{n-1})_{ij}\star e_{\star }^{ik_{j}X^{j}}\right) _{sym\, F,X}.\]
The expression in brackets has to be symmetrized in \( F^{ij} \)
and \( X^{i} \) for \( n>2 \). Up to third order in \( \theta ^{ij} \),
the commutator \( F^{ij} \) of two covariant coordinates is \[
F^{ij}=-i[X^{i}\stackrel{\star }{,}X^{j}]=\theta ^{ij}-\theta ^{ik}f_{kl}\theta ^{lj}-\theta ^{kl}\partial _{l}\theta ^{ij}a_{k}+\mathcal{O}(3)\]
with \( f_{ij}=\partial _{i}a_{j}-\partial _{j}a_{i} \) the ordinary
field strength. Furthermore we have\[
e^{ik_{i}X^{i}}_{\star }=e^{ik_{i}x^{i}}(1+ik_{i}\theta ^{ij}a_{j})+\mathcal{O}(2).\]
If we choose the antisymmetric \( \star  \)-product (\ref{DEF_STAR_PRODUCT}),
the symmetrization will annihilate all the first order terms of the
\( \star  \)-products between the \( F^{ij} \) and \( X^{i} \),
and therefore we get\[
-\mathcal{F}[X]_{ij}(k)=\]
\[
=-2n\int \frac{d^{2n}x}{\epsilon \theta ^{n}}\left( \epsilon _{ij}\theta ^{n-1}-(n-1)\epsilon _{ij}\theta ^{n-2}\theta f\theta -\theta ^{kl}\partial _{l}(\epsilon _{ij}\theta ^{n-1})a_{k}\right) e^{ik_{i}x^{i}}+\mathcal{O}(1)\]
\[
=-2n\int \frac{d^{2n}x}{\epsilon \theta ^{n}}\left( \epsilon _{ij}\theta ^{n-1}-(n-1)\epsilon _{ij}\theta ^{n-2}\theta f\theta -\frac{1}{2}\epsilon _{ij}\theta ^{n-1}f_{kl}\theta ^{kl}\right) e^{ik_{i}x^{i}}+\mathcal{O}(1)\]
\[
=d^{2n}x\, \left( \theta _{ij}^{-1}+2n(n-1)\frac{\epsilon _{ij}\theta ^{n-2}\theta f\theta }{\epsilon \theta ^{n}}-\frac{1}{2}\theta _{ij}^{-1}f_{kl}\theta ^{kl}\right) e^{ik_{i}x^{i}}+\mathcal{O}(1),\]
using partial integration and \( \partial _{i}(\epsilon \theta ^{n}\theta ^{ij})=0 \).
To simplify notation we introduced \[
\epsilon _{ij}\theta ^{n-1}=\epsilon _{iji_{1}j_{1}\cdots i_{n-1}j_{n-1}}\theta ^{i_{1}j_{1}}\cdots \theta ^{i_{n-1}j_{n-1}}\]
 etc. In the last line we have used\[
\theta _{ij}^{-1}=-\frac{(\theta ^{n-1})_{ij}}{Pf(\theta )}=-2n\frac{\epsilon _{ij}\theta ^{n-1}}{\epsilon \theta ^{n}}.\]
We will now have a closer look at the second term, noting that\[
\theta ^{ij}\frac{\epsilon _{ij}\theta ^{n-2}\theta f\theta }{\epsilon \theta ^{n}}=-\frac{1}{2n}\theta ^{-1}_{kl}\theta ^{kr}f_{rs}\theta ^{sl}=-\frac{1}{2n}f_{rs}\theta ^{rs}\]
and therefore\begin{equation}
\label{defining formula for a b}
\frac{\epsilon _{ij}\theta ^{n-2}\theta f\theta }{\epsilon \theta ^{n}}=a\frac{\epsilon _{ij}\theta ^{n-1}}{\epsilon \theta ^{n}}f_{rs}\theta ^{rs}+bf_{ij}
\end{equation}
with \( a+b=-\frac{1}{2n} \). Taking e. g. \( i=1,j=2 \) we see
that\[
\epsilon _{12\cdots kl}\theta ^{n-2}\theta ^{kr}f_{rs}\theta ^{sl}=\epsilon _{12\cdots kl}\theta ^{n-2}(\theta ^{k1}\theta ^{2l}-\theta ^{k2}\theta ^{1l})f_{12}+\mbox {terms\, without}\: f_{12}.\]
 Especially there are no terms involving \( f_{12}\theta ^{12} \)
and we get for the two terms on the right hand side of (\ref{defining formula for a b})\[
2a\epsilon _{12}\theta ^{n-1}f_{12}\theta ^{12}=-2nb\epsilon _{12}\theta ^{12}\theta ^{n-1}f_{12}\]
 and therefore \( b=-\frac{a}{n} \). This has the solution \[
a=-\frac{1}{2(n-1)}\: \: \: \mbox {and}\: \: \: b=\frac{1}{2n(n-1)}.\]
With the resulting 

\[
2n(n-1)\frac{\epsilon _{ij}\theta ^{n-2}\theta f\theta }{\epsilon \theta ^{n}}=\frac{1}{2}\theta _{ij}^{-1}f_{kl}\theta ^{kl}+f_{ij}\]
we finally get \[
-\mathcal{F}[X]_{ij}(k)=\int d^{2n}x\, \left( \theta _{ij}^{-1}+f_{ij}\right) e^{ik_{i}x^{i}}+\mathcal{O}(1).\]
Therefore\[
f[X]_{ij}=\mathcal{F}[X]_{ij}(k)-\mathcal{F}[x]_{ij}(k)\]
is a closed form that reduces in the classical limit to the classical
Abelian field strength. We have found an expression for the inverse
Abelian Seiberg-Witten map.

\cleardoublepage

\begin{appendix}

\chapter{Definitions and calculations}

\section{\label{Schouten-Nijenhuis bracket}The Schouten-Nijenhuis bracket}

The Schouten-Nijenhuis bracket for multivector fields \( \pi ^{i_{1}\ldots i_{k_{s}}}_{s}\partial _{i_{1}}\wedge \ldots \wedge \partial _{i_{k_{s}}} \)
can be written as (\cite{Arnal:2000hy}, IV.2.1):

\[
{}[\pi _{1},\pi _{2}]_{S}=(-1)^{k_{1}-1}\pi _{1}\bullet \pi _{2}-(-1)^{k_{1}(k_{2}-1)}\pi _{2}\bullet \pi _{1},\]

\[
\pi _{1}\bullet \pi _{2}=\sum ^{k_{1}}_{l=1}(-1)^{l-1}\pi _{1}^{i_{1}\ldots i_{k_{1}}}\partial _{l}\pi _{2}^{j_{1}\ldots j_{k_{2}}}\partial _{i_{1}}\wedge \ldots \wedge \widehat{{\partial _{i_{l}}}}\wedge \ldots \wedge \partial _{i_{k_{1}}}\wedge \partial _{j_{1}}\wedge \ldots \wedge \partial _{j_{k_{2}}},\]
 where the hat marks an omitted derivative.

For a function \( g \), vectorfields \( X=X^{k}\partial _{k} \)
and \( Y=Y^{k}\partial _{k} \) and a bivectorfield \( \pi =\frac{1}{2}\pi ^{kl}\partial _{k}\wedge \partial _{l} \)
we get:

\begin{eqnarray*}
{}[X,g]_{S} & = & X^{k}\partial _{k}g,\\
{}[\pi ,g]_{S} & = & -\pi ^{kl}\partial _{k}g\partial _{l},\\
{}[X,\pi ]_{S} & = & \frac{1}{2}(X^{k}\partial _{k}\pi ^{ij}-\pi ^{ik}\partial _{k}X^{j}+\pi ^{jk}\partial _{k}X^{i})\partial _{i}\wedge \partial _{j},\\
{}[\pi ,\pi ]_{S} & = & \frac{1}{3}(\pi ^{kl}\partial _{l}\pi ^{ij}+\pi ^{il}\partial _{l}\pi ^{jk}+\pi ^{jl}\partial _{l}\pi ^{ki})\partial _{k}\wedge \partial _{i}\wedge \partial _{j}.
\end{eqnarray*}

\section{\label{Gerstenhaber bracket}The Gerstenhaber bracket}

The Gerstenhaber bracket for polydifferential operators \( A_{s} \)
can be written as (\cite{Arnal:2000hy}, IV.3):

\[
{}[A_{1},A_{2}]_{G}=A_{1}\circ A_{2}-(-1)^{(|A_{1}|-1)(|A_{2}|-1)}A_{2}\circ A_{1},\]
\[
(A_{1}\circ A_{2})(f_{1},\ldots \, f_{m_{1}+m_{2}-1})=\]
\[
\sum ^{m_{1}}_{j=1}(-1)^{(m_{2}-1)(j-1)}A_{1}(f_{1},\ldots \, f_{j-1},A_{2}(f_{j},\ldots ,f_{j+m_{2}-1}),f_{j+m_{2}},\ldots ,f_{m_{1}+m_{2}-1}),\]
where \( |A_{s}| \) is the degree of the polydifferential operator
\( A_{s} \), i.e. the number of functions it is acting on.

For functions \( g \) and \( f \), differential operators \( D_{1} \)and
\( D_{2} \) of degree one and \( P \) of degree two we get

\begin{eqnarray}
{}[D,g]_{G} & = & D(g),\nonumber \\
{}[P,g]_{G}(f) & = & P(g,f)-P(f,g),\nonumber \\
{}[D_{1},D_{2}]_{G}(g) & = & D_{1}(D_{2}(g))-D_{2}(D_{1}(g)),\nonumber \\
{}[P,D]_{G}(f,g) & = & P(D(f),g)+P(f,D(g))-D(P(f,g)).\label{gerst 2 1} 
\end{eqnarray}

\section{\label{Commutator theta theta and theta X}Calculation of \protect\( [\theta _{t},\theta _{t}]\protect \)
and \protect\( [\theta _{t},X_{t}]\protect \)}

We want to show that \( \theta _{t} \) is still a Poisson tensor
and that \( X_{t} \) still commutes with \( \theta _{t} \). For
this we first define \( \theta (n)^{k}_{l}=(\theta f)^{n}=\theta ^{ki}f_{ij}\ldots \theta ^{rs}f_{sl}=f_{li}\theta ^{ij}\ldots f_{rs}\theta ^{sk}=(f\theta )^{n} \)
and \( \theta (n)^{kl}=\theta (f\theta )^{n}=\theta ^{ki}f_{ij}\ldots f_{rs}\theta ^{sl} \).
In the calculations to follow we will sometimes drop the derivatives
of the polyvectorfields and associate \( \pi ^{k_{1}\ldots k_{n}} \)
with \( \pi ^{k_{1}\ldots k_{n\: \: }}\frac{1}{n}\partial _{k_{1}}\wedge \ldots \wedge \partial _{k_{n}} \)for
simplicity. All the calculations are done locally.

We evaluate

\begin{eqnarray*}
{[}\theta _{t},\theta _{t}]_{S} & = & \theta _{t}^{kl}\partial _{l}\theta ^{ij}_{t}+\mbox {c.p.}\; \mbox {in}\; (kij)\\
 & = & \sum ^{\infty }_{n,m=0}\sum ^{m}_{o=0}(-t)^{n+m}\theta (n)^{k}_{r}\theta (o)^{i}_{s}\theta (m-o)_{p}^{j}\theta ^{rl}\partial _{l}\theta ^{sp}+\mbox {c.p.}\; \mbox {in}\; (kij)\\
 &  & +\sum ^{\infty }_{n,m=0}\sum ^{m}_{o=0}(-t)^{n+m+1}\theta (n)^{kl}\theta (o)^{is}\theta (m-o)^{pj}\partial _{l}f_{sp}+\mbox {c.p.}\; \mbox {in}\; (kij)\\
 & = & \sum ^{\infty }_{n,m,o=0}(-t)^{n+m+o}\theta (n)^{k}_{r}\theta (o)^{i}_{s}\theta (m)_{p}^{j}\theta ^{rl}\partial _{l}\theta ^{sp}+\mbox {c.p.}\; \mbox {in}\; (kij)\\
 &  & -\sum ^{\infty }_{n,m,o=0}(-t)^{n+m+o+1}\theta (n)^{kl}\theta (o)^{is}\theta (m)^{jp}\partial _{l}f_{sp}+\mbox {c.p.}\; \mbox {in}\; (kij).
\end{eqnarray*}
The first part vanishes because \( \theta _{t} \) is a Poisson tensor,
i.e.\begin{equation}
\label{poisson}
{[}\theta ,\theta ]_{S}=\theta ^{kl}\partial _{l}\theta ^{ij}+\mbox {c.p.}\; \mbox {in}\; (kij)=0,
\end{equation}
the second part because of\begin{equation}
\label{cyclic f}
\partial _{k}f_{ij}+\mbox {c.p.}\; \mbox {in}\; (kij)=0.
\end{equation}
To prove that \( X_{t} \) still commutes with \( \theta _{t} \),
we first note that\[
X_{t}=X\sum _{n=0}^{\infty }(-tf\theta )=X(1-tf\theta _{t}).\]
With this we can write

\begin{eqnarray}
{[}X_{t},\theta _{t}] & = & [X,\theta _{t}]-t[Xf\theta _{t},\theta _{t}]\label{comm X theta} \\
 & = & X^{n}\partial _{n}\theta _{t}^{kl}-\theta _{t}^{kn}\partial _{n}X^{l}+\theta _{t}^{ln}\partial _{n}X^{k}\nonumber \\
 &  & -tX^{m}f_{mi}\theta _{t}^{in}\partial _{n}\theta _{t}^{kl}+t\theta _{t}^{kn}\partial _{n}(X^{m}f_{mi}\theta _{t}^{il})-t\theta _{t}^{ln}\partial _{n}(X^{m}f_{mi}\theta _{t}^{ik})\nonumber \\
 & = & X^{n}\partial _{n}\theta _{t}^{kl}-\theta _{t}^{kn}\partial _{n}X^{l}+\theta _{t}^{ln}\partial _{n}X^{k}\nonumber \\
 &  & +t\theta _{t}^{kn}\partial _{n}X^{m}f_{mi}\theta _{t}^{il}-t\theta _{t}^{ln}\partial _{n}X^{m}f_{mi}\theta _{t}^{ik}\nonumber \\
 &  & +t\theta _{t}^{kn}X^{m}\partial _{n}f_{mi}\theta _{t}^{il}-t\theta _{t}^{ln}X^{m}\partial _{n}f_{mi}\theta _{t}^{ik}.\nonumber 
\end{eqnarray}
In the last step we used (\ref{poisson}). To go on we note that

\[
t\theta _{t}^{kn}X^{m}\partial _{n}f_{mi}\theta _{t}^{il}-t\theta _{t}^{ln}X^{m}\partial _{n}f_{mi}\theta _{t}^{ik}=tX^{n}\theta _{t}^{km}\partial _{n}f_{mi}\theta _{t}^{il},\]
where we used (\ref{cyclic f}). Making use of the power series expansion
and the fact that \( X \) commutes with \( \theta , \) i.e. \[
{[}X,\theta ]=X^{n}\partial _{n}\theta ^{kl}-\theta ^{kn}\partial _{n}X^{l}+\theta ^{ln}\partial _{n}X^{k}=0,\]
 we further get

\begin{eqnarray*}
X^{n}\partial _{n}\theta _{t}^{kl}+tX^{n}\theta _{t}^{km}\partial _{n}f_{mi}\theta _{t}^{il} & = & \sum ^{\infty }_{r,s=0}(-t)^{r+s}\theta (r)^{k}_{i}X^{n}\partial _{n}\theta ^{ij}\theta (s)_{j}^{l}\\
 & = & \sum ^{\infty }_{r,s=0}(-t)^{r+s}\theta (r)^{k}_{i}\theta ^{in}\partial _{n}X^{j}\theta (s)_{j}^{l}\\
 &  & -\sum ^{\infty }_{r,s=0}(-t)^{r+s}\theta (r)^{k}_{i}\theta ^{jn}\partial _{n}X^{i}\theta (s)_{j}^{l}.
\end{eqnarray*}
Therefore (\ref{comm X theta}) reads

\begin{eqnarray*}
{[}X_{t},\theta _{t}] & = & \sum ^{\infty }_{r,s=0}(-t)^{r+s}\theta (r)^{k}_{i}\theta (s)_{j}^{l}\theta ^{in}\partial _{n}X^{j}-\sum ^{\infty }_{r,s=0}(-t)^{r+s}\theta (r)^{k}_{i}\theta (s)_{j}^{l}\theta ^{jn}\partial _{n}X^{i}\\
 &  & -\theta _{t}^{kn}\partial _{n}X^{l}+\theta _{t}^{ln}\partial _{n}X^{k}+t\theta _{t}^{kn}\partial _{n}X^{m}f_{mi}\theta _{t}^{il}-t\theta _{t}^{ln}\partial _{n}X^{m}f_{mi}\theta _{t}^{ik}\\
 & = & 0.
\end{eqnarray*}

\section{\label{Transformation properties of K}The transformation properties
of \protect\( K_{t}\protect \)}

To calculate the transformation properties of \( K_{t}(X_{t}^{k}a_{k}) \),
we first evaluate

\begin{eqnarray*}
\delta _{\lambda }((a_{\theta }+\partial _{t})^{n})X^{k}a_{k} & = & \sum ^{n-1}_{i=0}(a_{\theta }+\partial _{t})^{i}d_{\theta }(\lambda )(a_{\theta }+\partial _{t})^{n-1-i}X^{k}a_{k}\\
 & = & \sum ^{n-1}_{i=0}\sum ^{i}_{l=0}{i\choose l}d_{\theta }((a_{\theta }+\partial _{t})^{l}(\lambda ))(a_{\theta }+\partial _{t})^{n-1-l}X^{k}a_{k}
\end{eqnarray*}
and\\
\\
\\
\begin{eqnarray*}
\lefteqn {(a_{\theta }+\partial _{t})^{n}\delta _{\lambda }(X^{k}a_{k})} &  & \\
 & = & (a_{\theta }+\partial _{t})^{n}X^{k}\partial _{k}\lambda \\
 & = & X^{k}\partial _{k}(a_{\theta }+\partial _{t})^{n}-\sum ^{n-1}_{i=0}(a_{\theta }+\partial _{t})^{i}d_{\theta }(X^{k}a_{k})(a_{\theta }+\partial _{t})^{n-1-i}\lambda \\
 & = & X^{k}\partial _{k}(a_{\theta }+\partial _{t})^{n}\\
 &  & -\sum ^{n-1}_{i=0}\sum ^{n-1-i}_{j=0}{n-1-i\choose j}(-1)^{n-1-i-j}\\
 &  & \, \, \, \, \, \, \, \, \, \, \, \, \, \, \, \, \, \, \, \, \, \, \, \, \, \, \, \, \, \, \, \, \, \, \, \, \, \, (a_{\theta }+\partial _{t})^{i+j}d_{\theta }((a_{\theta }+\partial _{t})^{n-1-i-j}(X^{k}a_{k}))(\lambda )\\
 & = & X^{k}\partial _{k}(a_{\theta }+\partial _{t})^{n}\\
 &  & +\sum ^{n-1}_{i=0}\sum ^{n-1-i}_{j=0}{n-1-i\choose j}(-1)^{n-1-i-j}\\
 &  & \, \, \, \, \, \, \, \, \, \, \, \, \, \, \, \, \, \, \, \, \, \, \, \, \, \, \, \, \, \, \, \, \, \, \, \, \, \, (a_{\theta }+\partial _{t})^{i+j}d_{\theta }(\lambda )((a_{\theta }+\partial _{t})^{n-1-i-j}(X^{k}a_{k}))\\
 & = & X^{k}\partial _{k}(a_{\theta }+\partial _{t})^{n}\\
 &  & +\sum ^{n-1}_{i=0}\sum ^{n-1-i}_{j=0}\sum ^{i+j}_{l=0}{n-1-i\choose j}{i+j\choose l}(-1)^{n-1-i-j}\\
 &  & \, \, \, \, \, \, \, \, \, \, \, \, \, \, \, \, \, \, \, \, \, \, \, \, \, \, \, \, \, \, \, \, \, \, \, \, \, \, d_{\theta }((a_{\theta }+\partial _{t})^{l}(\lambda ))((a_{\theta }+\partial _{t})^{n-1-l}(X^{k}a_{k})).
\end{eqnarray*}

We go on by simplifying these expressions. Using\begin{equation}
\label{binomial}
{i\choose l}={i-1\choose l}+{i-1\choose l-1}\; \; \; \; \mbox {for}\; \; \; \; i>l,
\end{equation}
we get\[
\sum ^{n-1}_{m=l}\sum ^{m}_{i=0}{n-1-i\choose m-i}{m\choose l}(-1)^{n-1-m}=\sum ^{n-1}_{m=l}{n\choose m}{m\choose l}(-1)^{n-1-m}.\]
 Using (\ref{binomial}) again two times and then using induction
we go on to\[
\sum ^{n-1}_{m=l}{n\choose m}{m\choose l}(-1)^{n-1-m}=\sum ^{l}_{i=0}{n-1-i\choose n-1-l},\]
 giving, after using (\ref{binomial}) again\[
\sum ^{l}_{i=0}{n-1-i\choose n-1-l}={n\choose l}.\]
Together with\[
\sum ^{n-1}_{i=l}{i\choose l}={n\choose l+1}\]
these formulas add up to give\[
\sum ^{n-1}_{m=l}\sum ^{m}_{i=0}{n-1-i\choose m-i}{m\choose l}(-1)^{n-1-m}+\sum ^{n-1}_{i=l}{i\choose l}={n+1\choose l+1}\]
and therefore

\[
\delta _{\lambda }(K_{t}(X^{k}a_{k}))=X^{k}\partial _{k}(K_{t}(\lambda ))+d_{\theta }(K_{t}(\lambda ))K_{t}(X^{k}a_{k}).\]

\section{Calculation of the commutators}

\subsection{\label{Semi-classical commutators}Semi-classical construction}

We calculate the commutator (\ref{comm rel 1}) (see also \cite{Jurco:2001my}),
dropping the t-subscripts on \( \theta _{t} \) for simplicity and
using local expressions.

\begin{eqnarray*}
{[}a_{\theta },d_{\theta }(g)] & = & -\theta ^{ij}a_{j}\partial _{i}\theta ^{kl}\partial _{k}g\partial _{l}-\theta ^{ij}a_{j}\theta ^{kl}\partial _{i}\partial _{k}g\partial _{l}\\
 &  & +\theta ^{kl}\partial _{k}g\partial _{l}\theta ^{ij}a_{j}\partial _{i}+\theta ^{kl}\partial _{k}g\theta ^{ij}\partial _{l}a_{j}\partial _{i}\\
 & = & -\theta ^{kl}\partial _{k}\theta ^{ij}a_{j}\partial _{i}g\partial _{l}-\theta ^{kl}\theta ^{ij}a_{j}\partial _{k}\partial _{i}g\partial _{l}-\theta ^{kl}\theta ^{ij}\partial _{j}a_{k}\partial _{i}g\partial _{l}\\
 & = & +\theta ^{ij}f_{jk}\theta ^{kl}\partial _{i}g\partial _{l}-\theta ^{kl}\partial _{k}(\theta ^{ij}a_{j}\partial _{i}g)\partial _{l}\\
 & = & -d_{\theta f\theta }g+d_{\theta }(a_{\theta }(g))\\
 & = & -\partial _{t}(d_{\theta })g+d_{\theta }(a_{\theta }(g)).
\end{eqnarray*}
For (\ref{comm rel}) we get

\begin{eqnarray*}
{[}a_{\theta },X_{t}] & = & \theta ^{ij}a_{j}\partial _{i}X^{k}\partial _{k}-X^{k}\partial _{k}\theta ^{ij}a_{j}\partial _{i}-X^{k}\theta ^{ij}\partial _{k}a_{j}\partial _{i}\\
 & = & -\theta ^{ij}X^{k}\partial _{k}a_{j}\partial _{i}-\theta ^{ik}\partial _{k}X^{j}a_{j}\partial _{i}\\
 & = & X^{k}f_{ki}\theta ^{ij}\partial _{j}+\theta ^{ij}\partial _{i}(X^{k}a_{k})\partial _{j}\\
 & = & -\partial _{t}X-d_{\theta }(X^{k}a_{k}).
\end{eqnarray*}

\subsection{\label{Quantum commutators}Quantum construction}

In \cite{Manchon:2000hy}, (\ref{comm phi f star},\ref{comm delta X star},\ref{comm delta X Phi g})
have already been calculated, unluckily (and implicitly) using a different
sign convention for the brackets of polyvectorfields. In \cite{Jurco:2001my},
again a different sign convention is used, coinciding with the one
in \cite{Manchon:2000hy} in the relevant cases. In order to keep
our formulas consistent with the ones used in \cite{Manchon:2000hy,Jurco:2001my},
we define our bracket on polyvectorfields \( \pi _{1} \) and \( \pi _{2} \)
as in \cite{Manchon:2000hy} to be\[
{}[\pi _{1},\pi _{2}]=-[\pi _{2},\pi _{1}]_{S},\]
giving an extra minus sign for \( \pi _{1} \) and \( \pi _{2} \)
both even. The bracket on polydifferential operators is always the
Gerstenhaber bracket.

With these conventions and\[
d_{\star }=-[\cdot ,\star ],\]

we rewrite the formulas (\ref{comm delta X Phi g},\ref{comm delta X delta Y},\ref{comm phi f star},\ref{comm delta X star})
so we can use them in the following

\begin{eqnarray}
{}[\Phi (X),\Phi (g)]_{G} & = & \Phi ([X,g])+\Psi ([\theta ,g],X)-\Psi ([\theta ,X],g),\label{Phi X Phi g} \\
{}[\Phi (X),\Phi (Y)]_{G} & = & d_{\star }\Psi (X,Y)\label{Phi X, Phi Y} \\
 &  & +\Phi ([X,Y])+\Psi ([\theta ,Y],X)-\Psi ([\theta ,X],Y),\nonumber \\
d_{\star }\Phi (g) & = & \Phi (d_{\theta }(g)),\\
d_{\star }\Phi (X) & = & \Phi (d_{\theta }(X)).\label{d star x} 
\end{eqnarray}
For the calculation of the commutators of the quantum objects we first
define\[
a_{\star }=\Phi (a_{\theta _{t}})\]
and \[
f_{\star }=\Phi (f_{\theta _{t}}).\]
With (\ref{d star x}) we get the quantum version of (\ref{f theta})\[
f_{\star }=d_{\star }a_{\star }.\]
For functions \( f \) and \( g \) we get

\[
\partial _{t}(f\star g)=\sum ^{\infty }_{n=0}\frac{1}{n!}\partial _{t}U_{n}(\theta _{t},\, \ldots ,\theta _{t})(f,g)\]
\[
=\sum ^{\infty }_{n=1}\frac{1}{(n-1)!}U_{n}(f_{\theta },\, \ldots ,\theta _{t})(f,g)=f_{\star }(f,g).\]
With these two formulas we can now calculate the quantum version of
(\ref{comm rel 1}) as in \cite{Jurco:2001my}. On two functions \( f \)
and \( g \) we have

\begin{eqnarray*}
\partial _{t}(f\star g) & = & f_{\star }(f,g)\\
 & = & d_{\star }a_{\star }(f,g)\\
 & = & -[a_{\star },\star ](f,g)\\
 & = & -a_{\star }(f\star g)+a_{\star }(f)\star g+f\star a_{\star }(g),
\end{eqnarray*}
where we used (\ref{gerst 2 1}) in the last step. Therefore\begin{eqnarray*}
{}[a_{\star },d_{\star }(g)](f) & = & a_{\star }(d_{\star }(g)(f))-d_{\star }(g)(a_{\star }(f))\\
 & = & a_{\star }([f\stackrel{\star }{,}g])-[a_{\star }(f)\stackrel{\star }{,}g]\\
 & = & -\partial _{t}[f\stackrel{\star }{,}g]-[a_{\star }(g)\stackrel{\star }{,}f]\\
 & = & -\partial _{t}d_{\star }(g)(f)+d_{\star }(a_{\star }(g))(f).
\end{eqnarray*}
For a function \( g \) which might also depend on \( t \) the quantum
version of (\ref{comm rel 1}) now reads\[
{}[a_{\star }+\partial _{t},d_{\star }(g)]=d_{\star }(a_{\star }(g)).\]
We go on to calculate the quantum version of (\ref{comm rel}). We
first note that\[
\partial _{t}\Phi (X_{t})=\sum ^{\infty }_{n=1}\frac{1}{(n-1)!}\partial _{t}U_{n}(X_{t},\theta _{t},\, \ldots ,\theta _{t})=\Phi (\partial _{t}X_{t})+\Psi (f_{\theta },X_{t}).\]
With this we get\begin{eqnarray*}
{}[\Phi (a_{\theta }),\Phi (X_{t})] & = & d_{\star }\Psi (a_{\theta },X_{t})+\Phi ([a_{\theta },X_{t}])-\Psi ([\theta _{t}\, a_{\theta }])+\Psi ([\theta _{t},X_{t}],a_{\theta })\\
 & = & d_{\star }\Psi (a_{\theta },X_{t})+\Phi (-d_{\theta }(X_{t}^{k}a_{k}))+\Phi (-\partial _{t}X_{t})-\Psi (f_{\theta },X_{t})\\
 & = & -d_{\star }(\Phi (X_{t}^{k}a_{k})-\Psi (a_{\theta },X_{t}))-\partial _{t}\Phi (X_{t}),
\end{eqnarray*}
where we have used (\ref{Phi X, Phi Y}). 

\end{appendix}

\cleardoublepage

\bibliographystyle{diss}
\bibliography{mainbib}

\end{document}